\documentclass[aps,prd,twocolumn,preprintnumbers,showpacs,nofootinbib,amssymb]{revtex4}
\usepackage{graphicx}
\usepackage{amsmath}
\usepackage{amssymb}
\usepackage{amsfonts}
\usepackage{bm}
\usepackage{color}

\def\be{\begin{equation}}
\def\ee{\end{equation}}
\def\bea{\begin{eqnarray}}
\def\eea{\end{eqnarray}}

\begin{document}

\title{Collapse of a self-gravitating Bose-Einstein condensate with attractive
self-interaction}
\author{Pierre-Henri Chavanis}
\affiliation{Laboratoire de Physique Th\'eorique,
Universit\'e Paul Sabatier, 118 route de Narbonne  31062 Toulouse, France}

\begin{abstract}

We study the collapse of a self-gravitating Bose-Einstein condensate with
attractive self-interaction. Equilibrium states in which the gravitational
attraction and the attraction due to the self-interaction are counterbalanced by
the quantum pressure (Heisenberg uncertainty principle) exist only below a
maximum mass $M_{\rm
max}=1.012\hbar/\sqrt{Gm|a_s|}$ where $a_s<0$ is the scattering length of the
bosons and $m$ is their mass [Chavanis, Phys. Rev. D {\bf 84}, 043531 (2011)].
For $M>M_{\rm max}$ the
system is expected to collapse and form a black hole. We study the collapse
dynamics by making a Gaussian ansatz for the wave function and reducing the
problem to the study of the motion of a particle in an effective potential. We
find that the collapse time scales as $(M/M_{\rm max}-1)^{-1/4}$ for
$M\rightarrow M_{\rm max}^+$ and as $M^{-1/2}$ for $M\gg M_{\rm max}$. Other
analytical results are given above and below the critical point corresponding
to a saddle node bifurcation. We apply our results to standard axions with mass
$m=10^{-4}\,
{\rm eV}/c^2$ and scattering length
$a_s=-5.8\times 10^{-53}\, {\rm m}$  for
which $M_{\rm max}=6.5\times 10^{-14}M_{\odot}$ and $R=3.3\times
10^{-4}\, R_{\odot}$. We
confirm our previous claim that bosons with attractive
self-interaction, such as standard axions, may form low mass stars (axion
stars or dark matter stars) but cannot form dark matter halos of relevant mass
and size. These mini
axions stars could be the constituents of dark matter. They can collapse
into mini black holes of mass $\sim 
10^{-14}\, M_{\odot}$ in a few hours. In that case,  dark matter halos would be
made of mini
black holes.  We
also  apply our results to ultralight axions with mass  $m=1.93\times 10^{-20}\,
{\rm eV}/c^2$ and scattering length
$a_s=-8.29\times 10^{-60}\, {\rm fm}$  for
which $M_{\rm max}=0.39\times 10^6\, M_{\odot}$ and $R=33\, {\rm pc}$. These
ultralight axions could cluster into dark matter halos. Axionic dark matter
halos with attractive self-interaction can collapse into supermassive black
holes
of mass $\sim 
10^{6}\, M_{\odot}$ (similar to those reported at the center of galaxies) in
about one million years.

\end{abstract}


\maketitle

\section{Introduction}
\label{sec_introduction}

The nature of dark matter is still unknown. It has been proposed that dark
matter may be made of bosons in the form of Bose-Einstein condensates (BECs) at
absolute zero temperature
\cite{baldeschi,khlopov,membrado,sin,jisin,leekoh,schunckpreprint,
matosguzman,
guzmanmatos,hu,peebles,goodman,mu,arbey1,silverman1,matosall,silverman,
lesgourgues,arbey,bohmer,bmn,sikivie,mvm,lee09,ch1,lee,prd1,prd2,prd3,briscese,
harkocosmo,harko,abrilMNRAS,aacosmo,velten,pires,park,rmbec,rindler,lora2,
abrilJCAP,mhh,lensing,glgr1,ch2,ch3,shapiro,bettoni,lora,mlbec,madarassy,
abrilph,playa,stiff,guth,souza,freitas,alexandre,schroven,eby,cembranos,braaten,
davidson,fan,calabrese}
(see \cite{revueabril,revueshapiro,bookspringer} for reviews and the
Introduction of \cite{prd1} for a
short historical account of the
development of the BEC dark matter scenario). For dark matter halos, Newtonian
gravity can be used so the evolution of the wave function of the
self-gravitating BEC is governed by the Gross-Pitaevskii-Poisson (GPP) equations
(see, e.g., \cite{prd1,prd2,prd3}). By using the Madelung transformation
\cite{madelung}, these equations can be written in the form of hydrodynamic
equations, the so-called quantum Euler-Poisson (EP) equations. These
equations
are similar to the hydrodynamic equations of cold dark matter (CDM) except that
they include a quantum force arising from the Heisenberg uncertainty principle
and a pressure force due to the self-interaction of the bosons measured by their
scattering length $a_s$. Quantum mechanics may be a way to
solve the small scale
problems of the CDM model such as the cusp problem, the missing satellite
problem, and the ``too big to fail'' problem. In the BEC model, dark matter
halos are stable equilibrium solutions of the GPP, or quantum EP, equations.
They satisfy a condition of hydrostatic equilibrium corresponding to the balance
between the quantum force (Heisenberg uncertainty principle), the pressure due
to the
self-interaction (scattering), and the gravitational attraction. The mass-radius
relation of self-gravitating BECs at $T=0$ has been obtained numerically
(exactly) and analytically (approximately) in \cite{prd1,prd2} for any value of
the scattering length $a_s$ of the bosons. This study makes the link between the
noninteracting case $a_s=0$  and the Thomas-Fermi (TF)
limit $GM^2ma_s/\hbar^2\gg 1$ in which the
quantum potential can be neglected. It also treats the case of bosons
with negative
scattering lengths ($a_s<0$).

When $a_s\ge 0$, the short-range interaction between the bosons is repulsive (or
absent). In that case, there is a stable equilibrium state for any value of the
mass $M$.\footnote{This is true in the Newtonian regime appropriate to dark
matter halos.
In the general relativistic regime, corresponding to boson stars
(possibly mimicking massive black holes at the center of
galaxies), equilibrium
states exist only below a maximum mass. For noninteracting boson stars, the
maximum mass is
given by $M_{\rm max}=0.633 M_P^2/m$, where $M_P=(\hbar c/G)^{1/2}=2.18\times
10^{-8}\, {\rm kg}$ is the Planck mass. It is obtained  from the
Klein-Gordon-Einstein (KGE) equations \cite{kaup,rb}. For self-interacting boson
stars with a $\lambda\phi^4$ potential, the maximum mass is given
by $M_{\rm max}=0.0612 \sqrt{\lambda} M_P^3/m^2$, where  $\lambda=8\pi a_s m
c/\hbar$ is the dimensionless self-interaction constant
(see Appendix \ref{sec_d}). It can be  obtained
from the KGE equations \cite{colpi} or from their hydrodynamic representation 
\cite{chavharko}. In  \cite{chavharko} it is argued that, because
of their superfluid core, neutron stars could actually be BEC stars. Indeed, the
neutrons could form Cooper pairs and behave as bosons of mass $2m_n$ (where
$m_n=0.940\, {\rm GeV/c^2}$ is the mass of the neutron). By adjusting the
value of the self-interaction constant, the maximum mass of these BEC stars
could account for the abnormal mass (in the range $2-2.4\, M_{\odot}$) of
certain neutron stars \cite{Lat,Dem,black1,black2,black3,antoniadis} that
is much larger than the Oppenheimer-Volkoff  limit $M_{\rm OV}=0.376\, 
M_P^3/m^2=0.7\,  M_\odot$ based on the assumption
that neutron stars are ideal fermion stars \cite{ov}.} 
It corresponds to the balance
between the repulsive quantum force, the repulsive self-interaction (if any),
and the
attractive gravitational force.  When $a_s<0$, the short-range interaction
between the bosons is attractive. In that case, there is an equilibrium state
only
for $M\le M_{\rm max}$ with \cite{prd1,prd2}:
\begin{eqnarray}
\label{intro1}
M_{\rm max}^{\rm exact}=1.012\frac{\hbar}{\sqrt{Gm|a_s|}}.
\end{eqnarray}
The stable configurations have a radius $R_{99}\ge R_{99}^*$  with
\cite{prd1,prd2}:
\begin{eqnarray}
\label{intro2}
(R_{99}^*)^{\rm exact}=5.5\left (\frac{|a_s|\hbar^2}{Gm^3}\right )^{1/2},
\end{eqnarray}
where the subscript  $99$ means that $R_{99}$ is the radius containing $99\%$
of the mass (the density profile has not a compact support but extends to
infinity so the radius of the system is formally infinite). This equilibrium
state
corresponds to the balance between the repulsive quantum force, the attractive
force coming from the self-interaction of the bosons, and the attractive
gravitational force. For $M>M_{\rm max}$, there is no equilibrium state anymore
because the quantum force cannot balance the attractive self-interaction and the
gravitational attraction. In that case, nothing can prevent the collapse of the
BEC, not even quantum mechanics (Heisenberg uncertainty principle). Therefore,
the system is expected to collapse and form a black hole.

In the case of an attractive self-interaction, the maximum mass obtained in
\cite{prd1,prd2} is usually extremely small. This is because it can be rewritten
as \cite{prd1}:\footnote{This
scaling can be compared to the scaling $M_{\rm max}=0.376 M_P^3/m^2$ of the
maximum mass of
fermion stars \cite{ov}, to the scaling  $M_{\rm max}=0.633 M_P^2/m$ of
the maximum mass of noninteracting
boson stars \cite{kaup,rb} and to the scaling $M_{\rm max}=0.062 \sqrt{\lambda}
M_P^3/m^2$ of
the maximum mass of 
self-interacting boson stars \cite{colpi,chavharko}. We emphasize, however, that
the
maximum mass given by Eq.
(\ref{intro3}) is a Newtonian result contrary to the other limits that
come from general relativity. The usually very small value of the maximum mass
(or mass-radius ratio) of self-gravitating
BECs with attractive self-interaction justifies {\it a posteriori} why a
Newtonian treatment is sufficient to describe them (see Appendix
\ref{sec_vna} for more details).} 
\begin{eqnarray}
\label{intro3}
M_{\rm max}^{\rm
exact}=5.073\frac{M_P}{\sqrt{|\lambda|}}.
\end{eqnarray}
Similarly, the maximum radius
can be written as \cite{prd1}:
\begin{eqnarray}
\label{intro4}
(R_{99}^*)^{\rm
exact}=1.1\sqrt{|\lambda|}\frac{M_P}{m}
\lambda_c ,
\end{eqnarray}
where $\lambda_c=\hbar/mc$ is the Compton wavelength of the bosons. Unless the
self-interaction constant $\lambda$ is extraordinarily small, the
maximum mass
and the
corresponding radius of self-gravitating BECs with attractive self-interaction
are much smaller than the masses and radii of dark matter halos \cite{prd1}.

This remark has
important consequences. It has often been proposed, in connection to the BEC
dark matter model, that one possible dark matter candidate could be the axion
\cite{kc,marsh}.
One reason is that, unlike the Higgs boson, axions are sufficiently long-lived
to coalesce into dark matter halos which constitute the seeds of galaxy
formation. Axions
are hypothetical pseudo-Nambu-Goldstone bosons of the Peccei-Quinn
phase transition associated with a $U(1)$ symmetry that
solves the strong charge parity (CP) problem of quantum chromodynamics (QCD)
\cite{pq}. Axions also appear in string theory leading to the
notion of string axiverse \cite{axiverse}. The axion is a spin-$0$ particle
with a very small mass and an extremely weak self-interaction arising from
nonperturbative effects in QCD. Axions are extremely nonrelativistic and have
huge occupation numbers, so they can be described by a classical field.
Recently, it has been proposed that axionic dark matter can form a BEC during
the radiation-dominated era \cite{sikivie1,sikivie2}.   Axions can thus be
described by a
relativistic quantum field theory with a real scalar field $\phi$ whose
evolution is governed by the Klein-Gordon-Einstein (KGE) equations. In the
nonrelativistic
limit, they can
be described by an effective field theory with a complex scalar field $\psi$
whose evolution is governed  by the GPP equations. Therefore, axions seem to be
good candidates for the BEC dark matter scenario. However, the axion has a
negative scattering length corresponding to an attractive self-interaction.
Therefore, according to the results of \cite{prd1}, it is not obvious that
axions can form dark
matter halos of relevant mass and size. This can be
checked by a simple numerical application. Eqs. (\ref{intro1}) and
(\ref{intro2}) can be rewritten as
\begin{eqnarray}
\label{intro4b}
\frac{M_{\rm max}^{\rm exact}}{M_{\odot}}=1.56\times 10^{-34} \left (\frac{{\rm
eV}/c^2}{m}\right )^{1/2}  \left (\frac{\rm fm}{|a_s|}\right )^{1/2}, 
\end{eqnarray}
\begin{eqnarray}
\label{intro4c}
\frac{(R_{99}^*)^{\rm exact}}{R_{\odot}}=1.36\times 10^{9}  \left
(\frac{|a_s|}{\rm fm}\right )^{1/2} \left (\frac{{\rm
eV}/c^2}{m}\right )^{3/2}. 
\end{eqnarray}
Considering standard axions with  $m=10^{-4}\,
{\rm eV}/c^2$ and
$a_s=-5.8\times 10^{-53}\, {\rm m}$ \cite{kc}, corresponding to
$\lambda=-7.4\times
10^{-49}$, we obtain $M_{\rm max}^{\rm exact}=6.5\times 10^{-14}\,
M_{\odot}$ and $(R_{99}^*)^{\rm exact}=3.3\times 10^{-4}\, R_{\odot}$ (the
average density is $\overline{\rho}=2.55\times 10^3\, {\rm g/m^3}$). These
values are ``ridiculously small'' \cite{prd1} as compared to the typical values
of dark matter halos (they rather correspond to the typical size of
asteroids). Obviously, standard
axions cannot form dark matter halos of relevant mass and size.\footnote{The
same conclusion can also
be obtained by
considering the gravitational instability
of an infinite homogeneous distribution of BECs. This quantum Jeans problem has
been studied in detail in \cite{prd1,aacosmo,abrilph} in the general case.}
However, they
could form mini boson stars (axion stars or dark matter stars) of very low mass
which are stable gravitationally bound BECs. They might play a role as dark
matter components (i.e. dark matter halos could be made of mini axion
stars) if they exist in the universe in abundance. These mini
axions stars could collapse into mini black holes of mass $\sim 
10^{-14}\, M_{\odot}$. In that case,  dark matter halos would be made of mini
black holes (their evaporation time $t_e=5120\pi G^2M^3/\hbar c^4\sim
6.6\times 10^{32}\, {\rm s}$ is much larger than the age of the
universe) and behave as CDM. 

This conclusion only applies to the type of axions with attractive
self-interaction that we have considered previously (``standard'' axions).
Axions with a repulsive self-interaction (if they exist)
or axions with an attractive self-interaction, or no self-interaction, and
an extraordinarily small mass could form bigger objects, possibly dark matter
halos. To be specific, let us consider that the smallest dark matter halo
that we know, Willman 1 ($R=33\, {\rm pc}$, $M=0.39\times 10^6\,
M_{\odot}$, $\overline{\rho}=1.75\times 10^{-16}\, {\rm g/m^3}$), is
a pure BEC without atmosphere.\footnote{We assume that dwarf dark matter halos
such as Willman 1
represent the ground state of a self-gravitating BEC (see Appendices D-F of
\cite{kingfermionic}). Larger halos are more complex to study. They may be
formed by hierarchical clustering. They have a core-halo structure in which the
core is a pure BEC and the halo has a Navarro-Frenk-White (NFW) profile
\cite{nfw}. This core-halo structure may also result from a process of
gravitational cooling \cite{seidel94}. In that case, the core is surrounded by
a halo of scalar radiation.} Assuming that the bosons have a
repulsive
self-interaction and using the constraint $(|a_s|/{\rm fm})^2({\rm
eV}/mc^2)\le 1.77\times 10^{-8}$ set by the Bullet Cluster, we obtain
$m=1.69\times 10^{-2}\, {\rm eV}/c^2$, $a_s=1.73\times 10^{-5}\, {\rm fm}$ and
$\lambda=3.72\times 10^{-14}$ (see Appendix D of \cite{kingfermionic}). Assuming
that the bosons have no self-interaction, we obtain $m=2.57\times 10^{-20}\,
{\rm eV}/c^2$ (see Appendix D of \cite{kingfermionic}). Finally, assuming that
the bosons have an attractive self-interaction and using Eqs. (\ref{intro4b})
and (\ref{intro4c}), we obtain $m=1.93\times 10^{-20}\, {\rm eV}/c^2$,
$a_s=-8.29\times 10^{-60}\, {\rm fm}$ and $\lambda=-2.04\times
10^{-86}$. This is a new prediction (especially the scattering length). These
values reproduce the mass and size of
dwarf dark matter halos. On the other hand, the Bullet Cluster constraint is
clearly satisfied.
The values of $m$ and $a_s$ that we have obtained above may be revised by
considering possibly more relevant dark matter halos than
Willman 1 but their orders of magnitude should be correct. Therefore, it is not
impossible that axions cluster into dark matter halos but, for that, their
mass $m$ and scattering length $a_s$ need to have very different values from
their ``standard'' values given previously. We shall call them
``ultralight'' axions. Axionic dark matter halos made of
ultralight axions could  collapse into supermassive black holes of
mass $\sim 
10^{6}\, M_{\odot}$ similar to those reported at the center of
galaxies.

In this paper, we study the collapse of a self-gravitating BEC with 
attractive self-interaction (e.g. an axion star or an axionic dark matter halo)
when its mass is larger than
the maximum mass obtained in \cite{prd1}. To study this complex dynamics, we use
in this
paper a simple analytical model in which the collapse
of the BEC is reduced to the study of the motion of a particle in an effective
potential $V(R)$, where $R$ represents the size of the BEC. For $M>M_{\rm max}$,
the effective potential has no minimum so the
particle descends the potential until the singular point of collapse at $R=0$.
This
mechanical model, based on a Gaussian ansatz for the wave function,  was
introduced in the context of self-gravitating BECs in
\cite{prd1} (see \cite{prep} for
generalizations). This type of
approximation is standard in the study of BECs without self-gravity. It is
known to give a good qualitative description of the evolution of the system but
it is not always
quantitatively accurate. Therefore, it would be important to carry in parallel a
numerical study based on the KGE or GPP equations to compare our
approximate analytical
results to the exact numerical ones. 

The paper is organized as follows. In Sec. \ref{sec_gpp}, we recall the GPP
equations and
their hydrodynamic representation. In Sec. \ref{sec_analytical}, we make a
Gaussian ansatz for
the wave function and recall the dynamical equation, obtained in
\cite{prd1,prep},
satisfied by the radius $R(t)$ of the condensate. The stationary states of this
equation provide an analytical expression of the mass-radius relation of
self-gravitating BECs. We consider in this paper the case of a negative
scattering length ($a_s<0$) for which there is a maximum mass \cite{prd1}.
For
$M<M_{\rm max}$, two equilibrium states exist but only the configuration
corresponding
to the minimum of the effective potential is stable (this is the state with the
largest radius). We calculate the radius, the energy
and the complex pulsation as a function of the mass of the BEC and provide
their asymptotic expressions in the nongravitational limit, the noninteracting
limit, and close to the critical point. For $M\rightarrow M_{\rm max}^{-}$, the
period of the oscillations scales as $(1-M/M_{\rm max})^{-1/4}$.
In Sec. \ref{sec_collapse}, we study
the collapse of the BEC for $M>M_{\rm max}$ and determine the general
expression of the collapse time as a function of the mass of the BEC in the form
of an
integral. In Sec. \ref{sec_tf}, we show that the collapse time scales as
$M^{-1/2}$ in the TF limit $M\gg
M_{\rm max}$. In Sec. \ref{sec_saddle}, we study the
collapse of the BEC close to
the critical point $M\rightarrow M_{\rm max}^+$ corresponding to a saddle node
bifurcation. We find that the collapse time scales as $(M/M_{\rm
max}-1)^{-1/4}$
for $M\rightarrow M_{\rm max}^+$. In Secs. \ref{sec_possible} and \ref{sec_mz},
we consider the possible collapse, explosion, or oscillations of the BEC when
$M<M_{\rm max}$. In Sec. \ref{sec_conclusion}, we conclude our
study by applying our results to the case of axion stars and axionic dark
matter halos. In Appendix \ref{sec_d}, we derive the GP
equation from
the KG equation in the nonrelativistic limit $c\rightarrow +\infty$ and connect
the nonlinearity in the GP equation to the potential in the KG equation. In
Appendix \ref{sec_lh}, we discuss the Lagrangian structure of
the GP equation in the scalar field and hydrodynamic representations and provide
an
alternative derivation of the effective mechanical model. In Appendix
\ref{sec_sscp}, we show that the collapse of the BEC close to the critical
point has a self-similar structure. In Appendix \ref{sec_sum}, we highlight
particular regimes of interest. In Appendix \ref{sec_vna}, we study the validity
of the Newtonian approximation used in our paper. 

\section{The Gross-Pitaevskii-Poisson equations}
\label{sec_gpp}

We consider a self-gravitating BEC at $T=0$ with short-range interactions. The
evolution of the condensate wave
function is governed by the GPP equations (see, e.g., \cite{prd1,prd2,prd3}):
\begin{eqnarray}
\label{gpp1}
i\hbar \frac{\partial\psi}{\partial t}=-\frac{\hbar^2}{2m}\Delta\psi+m\Phi\psi+
\frac{4\pi a_s\hbar^2}{m^2}|\psi|^{2}\psi,
\end{eqnarray}
\begin{equation}
\label{gpp2}
\Delta\Phi=4\pi G |\psi|^2.
\end{equation}
The mass density of the bosons is $\rho=|\psi|^2$. The GP equation (\ref{gpp1})
can
be seen as a nonlinear Schr\"odinger equation with a cubic nonlinearity. The
self-interaction of the bosons is measured by their $s$-scattering length
$a_s$ which can be positive (repulsion) or negative
(attraction).\footnote{The GP
equation (\ref{gpp1}) describes the evolution of the condensate wave function
$\psi({\bf r},t)$ of a gas of bosons in interaction at $T=0$. It is usually
derived from the mean field Schr\"odinger equation \cite{gross,pitaevskii} when
the short-range interaction between the bosons corresponds to binary collisions
that can be modeled by a pair contact potential (see, e.g., Sec. II.A. of
\cite{prd1}). The GP equation can also be derived from the KG equation in the
nonrelativistic limit $c\rightarrow +\infty$ (see Appendix \ref{sec_d}). In
that case,
the cubic nonlinearity in Eq. (\ref{gpp1}) corresponds to a quartic potential of
the form $V(|\phi|)=({\lambda}/{4\hbar c})|\phi|^4$ in the KG equation. The
self-interaction constant $\lambda$ in the scalar field theory is related
to the $s$-scatering length of the bosons by $\lambda=8\pi a_s m c/\hbar$
\cite{prd1}.} 

If we consider a wave function of the form $\psi=\sqrt{\rho({\bf r})}
e^{-iEt/\hbar}$, we obtain the time-independent GP equation
\begin{eqnarray}
\label{gpp3}
-\frac{\hbar^2}{2m}\Delta\psi+m\Phi\psi+
\frac{4\pi a_s\hbar^2}{m^2}|\psi|^{2}\psi=E\psi.
\end{eqnarray}
Equations (\ref{gpp2}) and (\ref{gpp3}) form an eigenvalue problem for the
wavefunction $\psi$, where $E$ is the eigenvalue. We call it the eigenenergy.

By making the Madelung \cite{madelung} transformation
\begin{equation}
\label{gpp4}
\psi({\bf r},t)=\sqrt{{\rho({\bf r},t)}} e^{iS({\bf r},t)/\hbar},
\end{equation}
where
\begin{equation}
\label{gpp6}
S=-i\frac{\hbar}{2}\ln\left (\frac{\psi}{\psi^*}\right )
\end{equation}
is the action and 
\begin{equation}
\label{gpp7}
 {\bf
u}({\bf r},t)=\frac{\nabla S({\bf r},t)}{m}
\end{equation}
is the (irrotational) velocity field, one finds that
the GPP equations are equivalent to the hydrodynamic equations (see, e.g.,
\cite{prd1,prd2,prd3}):
\begin{equation}
\label{gpp8}
\frac{\partial\rho}{\partial t}+\nabla\cdot (\rho {\bf u})=0,
\end{equation}
\begin{equation}
\label{gpp8b}
\frac{\partial S}{\partial t}+\frac{1}{2m}(\nabla S)^2+m\Phi+\frac{4\pi
a_s\hbar^2}{m^2}\rho+Q=0,
\end{equation}
\begin{equation}
\label{gpp9}
\frac{\partial {\bf u}}{\partial t}+({\bf u}\cdot \nabla){\bf
u}=-\frac{1}{\rho}\nabla P-\nabla\Phi-\frac{1}{m}\nabla
Q,
\end{equation}
\begin{equation}
\label{gpp10}
\Delta\Phi=4\pi G\rho,
\end{equation}
called the quantum EP equations. Equation (\ref{gpp8}) is the continuity
equation, Eq. (\ref{gpp8b}) is the quantum Hamilton-Jacobi (or Bernoulli)
equation, and Eq.
(\ref{gpp9}) is the quantum Euler equation. It involves the quantum
potential
\begin{equation}
\label{gpp11}
Q=-\frac{\hbar^2}{2m}\frac{\Delta
\sqrt{\rho}}{\sqrt{\rho}}=-\frac{\hbar^2}{4m}\left\lbrack
\frac{\Delta\rho}{\rho}-\frac{1}{2}\frac{(\nabla\rho)^2}{\rho^2}\right\rbrack
\end{equation}
taking into account the Heisenberg uncertainty principle, and a pressure
\begin{equation}
\label{gpp12}
P=\frac{2\pi a_s\hbar^2}{m^3}\rho^{2}
\end{equation}
taking into account the self-interaction of the bosons
\cite{bogoliubov,revuebec}. The equation of state
(\ref{gpp12}) is that of a polytrope of index $n=1$ \cite{chandra}. The
stationary solution of the quantum Bernoulli equation (\ref{gpp8b}),
corresponding to $S=-Et$,
writes
\begin{equation}
\label{gpp8bc}
E=m\Phi+\frac{4\pi a_s\hbar^2}{m^2}\rho+Q.
\end{equation}
This equation is equivalent to Eq. (\ref{gpp3}). The
stationary solution of the quantum Euler equation (\ref{gpp9}) satisfies the
condition of
hydrostatic equilibrium
\begin{equation}
\label{gpp13}
\nabla P+\rho\nabla\Phi+\frac{\rho}{m}\nabla
Q={\bf 0}.
\end{equation}
This is the gradient of Eq. (\ref{gpp8bc}). 
Combining Eq. (\ref{gpp13}) with the Poisson equation
(\ref{gpp2}), we obtain the differential equation
\begin{equation}
\label{gpp14}
-\frac{4\pi a_s\hbar^2}{m^3}\Delta\rho+\frac{\hbar^2}{2m^2}\Delta
\left (\frac{\Delta\sqrt{\rho}}{\sqrt{\rho}}\right )=4\pi G\rho.
\end{equation}
This equation has been studied analytically and numerically in the
general case (i.e. accounting for the quantum pressure and the self-interaction)
in \cite{prd1,prd2}. The Jeans problem associated with the quantum EP equations
(\ref{gpp8})-(\ref{gpp10}) has been studied in \cite{prd1} in a static
background and in \cite{aacosmo,abrilph} in an  expanding background.

\section{Analytical model}
\label{sec_analytical}

\subsection{The Gaussian ansatz}
\label{sec_ga}

In general, the GPP equations (\ref{gpp1}) and (\ref{gpp2}) must be
solved numerically. However, we can obtain approximate analytical results by
using a Gaussian ansatz for the wave function \cite{prd1,prep}:
\begin{eqnarray}
\label{ga1}
\psi({\bf r},t)=\left \lbrack \frac{M}{\pi^{3/2}R(t)^3}\right
\rbrack^{1/2}e^{-\frac{r^2}{2R(t)^2}}e^{imH(t)r^2/2\hbar},
\end{eqnarray}
where $R(t)$ measures the spatial extension of the system (wave packet). We shall call it  the radius of the BEC. 
In the hydrodynamic representation, the density is given by
\begin{eqnarray}
\label{ga2}
\rho({\bf r},t)=\frac{M}{[\pi R(t)^2]^{3/2}}e^{-\frac{r^2}{R(t)^2}},
\end{eqnarray}
the action by
\begin{eqnarray}
\label{ga2b}
S({\bf r},t)=\frac{1}{2}mH(t)r^2,
\end{eqnarray}
and the velocity field by
\begin{eqnarray}
\label{ga3}
{\bf u}({\bf r},t)=H(t){\bf r}.
\end{eqnarray}
One can easily check  \cite{prd1,prep} that the continuity equation (\ref{gpp8})
is exactly
satisfied by the  ansatz (\ref{ga1}) provided that
\begin{eqnarray}
\label{ga4}
H(t)=\frac{\dot R}{R}.
\end{eqnarray}
This function is similar to the Hubble function in cosmology although the
analogy is purely formal (see Sec. \ref{sec_cosmology} for a
development of this analogy).

\subsection{The gross dynamics}
\label{sec_gd}

With the Gaussian ansatz, one can show (see \cite{prd1,prep} and
Appendix \ref{sec_lh}) that the total energy of the BEC writes
\begin{eqnarray}
E_{\rm tot}=\frac{1}{2}\alpha
M\left (\frac{dR}{dt}\right
)^2+V(R)
\label{gd0}
\end{eqnarray}
with 
\begin{eqnarray}
V(R)=\sigma\frac{\hbar^2M}{m^2R^2}+\zeta\frac{2\pi a_s\hbar^2M^2}{m^3R^3}
-\nu\frac{GM^2}{R},
\label{gd1}
\end{eqnarray}
where $\alpha=3/2$, $\sigma=3/4$, $\zeta=1/(2\pi)^{3/2}$ and
$\nu=1/\sqrt{2\pi}$. The first term in Eq. (\ref{gd0}) is the classical kinetic
energy $\Theta_c$ and the second term in Eq. (\ref{gd0}) is the potential
energy. The potential energy contains the contribution of the quantum kinetic
energy $\Theta_Q$ (or quantum potential), the internal energy $U$ due to the
self-interaction, and the
gravitational energy $W$. It can be shown that the GPP and EP equations conserve
the total energy 
$E_{\rm tot}$ and the mass $M$ \cite{prd1,prep}. Of course, these quantities
must also be conserved by the gross dynamics resulting from the Gaussian ansatz.
Writing $\dot E_{\rm tot}=0$, we find that the dynamical equation satisfied by
the radius of the BEC is
\begin{eqnarray}
\alpha M \frac{d^2R}{dt^2}=-V'(R).
\label{gd2}
\end{eqnarray}
This equation is similar to the equation of motion of a particle of mass $\alpha
M$ moving in an effective potential $V(R)$. This equation can also be obtained
from the virial theorem satisfied by the GPP and EP equations
\cite{prd1,prep}, or from the Euler-Lagrange equations
(see Appendix \ref{sec_lh}), when a trial wave function
parametrized by a function of $R$, such as the one in Eq.
(\ref{ga1}), is introduced.

It can be shown from general arguments \cite{holm} that a stable equilibrium
state of the GPP and EP equations is a minimum of  the total energy $E_{\rm
tot}$ at fixed mass $M$ \cite{prd1}. This is also true for the gross
dynamics: a stable equilibrium state of Eq. (\ref{gd2}) is a minimum of the
total energy $E_{\rm tot}(R,\dot R)$ at fixed mass $M$. A necessary
condition for equilibrium is that $\dot R=0$ meaning that the radius of the
BEC is
stationary. Then,  the equilibrium radius $R$ of the BEC, when an equilibrium
state exists, is a minimum of the effective potential $V(R)$. The condition
$V'(R)=0$ 
leads to the general analytical mass-radius relation
of a self-gravitating BEC with short-range interactions \cite{prd1}:
\begin{eqnarray}
\label{gd5}
M=\frac{2\sigma}{\nu}\frac{\frac{\hbar^2}{Gm^2 R}}{1-\frac{6\pi\zeta
a_s\hbar^2}{\nu Gm^3R^2}}.
\end{eqnarray}
For $M\rightarrow 0$ and $R\rightarrow +\infty$, we
recover the relation
\begin{eqnarray}
\label{gd6}
R \sim \frac{2\sigma}{\nu}\frac{\hbar^2}{GMm^2}
\end{eqnarray}
corresponding to a noninteracting self-gravitating BEC ($a_s=0$) \cite{prd1}.
The radius
$R_{99}$ containing $99\%$ of the mass is $R_{99}=8.955\hbar^2/GMm^2$. It is in
good agreement with the exact result $R_{99}=9.9\hbar^2/GMm^2$
\cite{rb,membrado,prd2}.

\subsection{Analogy with cosmology}
\label{sec_cosmology}

The first integral of motion given by Eqs. (\ref{gd0}) and (\ref{gd1}) can be
rewritten as
\begin{eqnarray}
\label{cosmo0}
\left (\frac{\dot R}{R}\right )^2=\frac{2E_{\rm tot}}{\alpha M
R^2}+\frac{2\nu GM}{\alpha R^3}-\frac{2\sigma\hbar^2}{\alpha
m^2R^4}-\frac{4\pi\zeta a_s\hbar^2M}{\alpha m^3R^5}.\nonumber\\
\end{eqnarray}
In the case where the quantum potential and the self-interaction can be
neglected, it reduces to
\begin{eqnarray}
\label{cosmo1}
\left (\frac{\dot R}{R}\right )^2=\frac{2E_{\rm tot}}{\alpha M
R^2}+\frac{2\nu GM}{\alpha R^3}.
\end{eqnarray}
This is similar to the Friedmann equation in cosmology
\begin{eqnarray}
\label{cosmo3}
H^2=\left (\frac{\dot R}{R}\right )^2=-\frac{kc^2}{R^2}+\frac{8\pi
G}{3c^2}\epsilon
\end{eqnarray}
for a pressureless ($P=0$) universe whose energy density decreases as
$\epsilon\propto R^{-3}$ \cite{weinbergbook} (see the Introduction of
\cite{cosmopoly1} for a short historic of the early development of cosmology).
In
this analogy $R$ plays the role of the scale factor, $H=\dot R/R$ plays the role
of the Hubble parameter, $-2E_{\rm tot}/\alpha M$ plays the role of the
curvature constant $kc^2$, and $2\nu M/\alpha R^3$ plays the role of the mass
density $8\pi \epsilon/3c^2$ with $\epsilon/c^2\propto R^{-3}$. We can therefore
draw certain analogies between
the evolution of a self-gravitating BEC and the evolution of a 
Friedmann-Lema\^itre-Robertson-Walker (FLRW) universe.

{\it Remark:} In a universe filled with a fluid with an equation of state
$P=\alpha\epsilon$, the energy density is related to the scale factor by
$\epsilon\propto R^{-3(1+\alpha)}$. If we take into account all the terms in
Eq. (\ref{cosmo0}), we find that the quantum potential term is analogous to an
energy
density $\epsilon\propto -1/R^4$ in cosmology. This corresponds to $\alpha=1/3$
like for the standard radiation. However, the energy density is negative. On the
other hand, the self-interaction term is analogous  to an energy density
$\epsilon\propto \mp/R^{5}$ in cosmology. This corresponds to
$\alpha=2/3$ (to our knowledge, this coefficient has not been considered in
cosmology). The energy density is negative when $a_s>0$ and a positive 
when $a_s<0$.

\subsection{BECs with negative scattering length}
\label{sec_nsl}

In this paper, we focus on the case of an attractive self-interaction
corresponding to a negative scattering length ($a_s<0$). In that case, there
exists a maximum mass \cite{prd1}:
\begin{eqnarray}
M_{\rm max}=\left (\frac{\sigma^2}{6\pi\zeta\nu}\right
)^{1/2}\frac{\hbar}{\sqrt{Gm|a_s|}}
\label{nsl1}
\end{eqnarray}
corresponding to a radius \cite{prd1}:
\begin{eqnarray}
R_*=\left (\frac{6\pi\zeta|a_s|\hbar^2}{\nu G m^3}\right )^{1/2}.
\label{nsl2}
\end{eqnarray}
Stable equilibrium states exist only for $M<M_{\rm max}$. They have a radius
$R>R_*$. We note that the approximate values of $M_{\rm
max}=1.085 \hbar/\sqrt{Gm|a_s|}$ and $R_{99}^*=4.125
(|a_s|\hbar^2/G m^3)^{1/2}$ obtained within the Gaussian ansatz \cite{prd1} are
relatively
close to the exact values  $M_{\rm
max}^{\rm exact}=1.012 \hbar/\sqrt{Gm|a_s|}$ and $(R_{99}^*)^{\rm exact}=5.5
(|a_s|\hbar^2/G m^3)^{1/2}$ obtained numerically by
solving the GPP equations \cite{prd2}. We note that the maximum mass $M_{\rm
max}$ and the
minimum stable radius $R_*$ are related to each other by
\begin{eqnarray}
M_{\rm max}=\frac{\sigma}{\nu}\frac{\hbar^2}{Gm^2 R_*}.
\label{nsl3}
\end{eqnarray}
It is convenient to introduce the energy scale
\begin{eqnarray}
V_0=\frac{\sigma^2\nu^{1/2}}{(6\pi\zeta)^{3/2}}\frac{\hbar
m^{1/2}G^{1/2}}{|a_s|^{3/2}}.
\label{nsl4}
\end{eqnarray}
Using Eqs. (\ref{nsl1}) and (\ref{nsl2}), we can check that $V_0$ is of the
order of magnitude of $\hbar^2 M_{\rm max}/m^2 R_*$, $|a_s| \hbar^2 M_{\rm
max}^2/m^3 R_*^3$ and $GM_{\rm max}^2/R_*$.  We also
introduce the dynamical time
\begin{eqnarray}
t_D=\left (\frac{\alpha M_{\rm max} R_*^2}{V_0}\right
)^{1/2}=\frac{6\pi\zeta}{\nu}\left (\frac{\alpha}{\sigma}\right
)^{1/2}\frac{|a_s|\hbar}{Gm^2}.
\label{nsl5}
\end{eqnarray}
Using Eqs. (\ref{nsl1}) and (\ref{nsl2}), we can check that $t_D$ is of the
order of magnitude of $(R_{*}^3/GM_{\rm
max})^{1/2}\sim 1/\sqrt{G \rho_{\rm max}}$ where $\rho_{\rm max}=3M_{\rm
max}/4\pi R_*^3$ is a characteristic density equal to the maximum
averaged density of the BEC. For standard axions with $m=10^{-4}\,
{\rm eV}/c^2$ and
$a_s=-5.8\, 10^{-53}\, {\rm m}$, we obtain
$M_{\rm max}=6.9\times 10^{-14}\,
M_{\odot}$, $R_*=1.0\times 10^{-4}\, R_{\odot}$, $V_0=7.1\times 10^{21}\, {\rm
g}\,{\rm m}^2/{\rm s}^2$, $\rho_{\rm max}=9.73\times 10^4\, {\rm g}/{\rm m}^3$,
and $t_D=1.2\times 10^4\, {\rm s}=3.4 \,{\rm hrs}$.
For ultralight axions with $m=1.93\times 10^{-20}\, {\rm eV}/c^2$ and
$a_s=-8.29\times 10^{-60}\, {\rm fm}$, we
obtain $M_{\rm max}=4.18\times 10^{5}\, M_{\odot}$, $R_*=10.4\, {\rm pc}$,
$V_0=5.74\times 10^{46}\, {\rm
g}\,{\rm m}^2/{\rm s}^2$,  $\rho_{\rm max}=6.00\times 10^{-15}\, {\rm g}/{\rm
m}^3$, and $t_D=4.70\times 10^{13}\, {\rm s}=1.49 \,{\rm
Myrs}$.

\subsection{The dynamical equation}
\label{sec_de}

\begin{figure}
\begin{center}
\includegraphics[clip,scale=0.3]{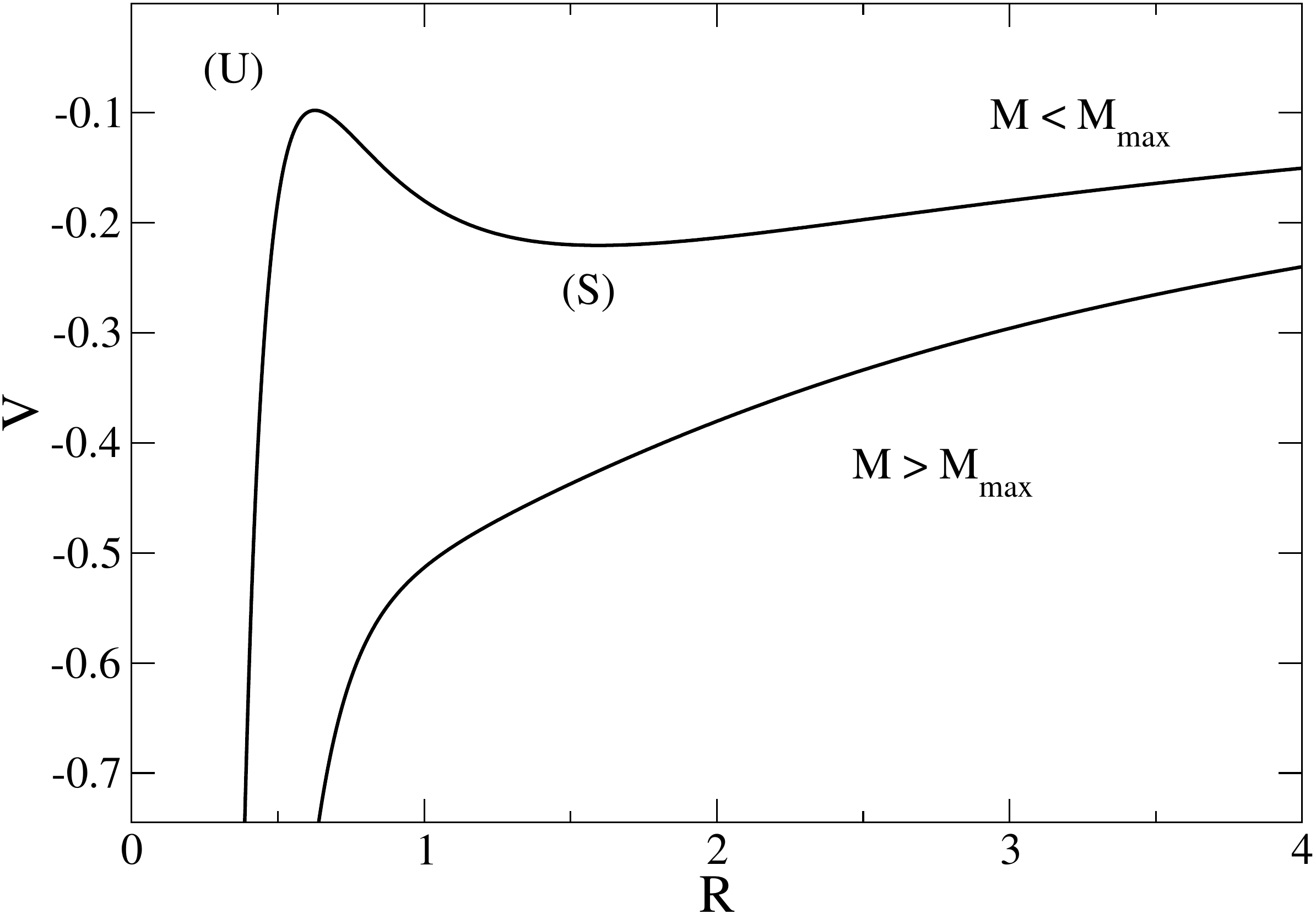}
\caption{Effective potential $V(R)$ as a function of the radius $R$ for
different values of the mass $M$. For illustration, we have taken 
$M=0.9<M_{\rm max}$
and $M=1.1>M_{\rm max}$ where $M_{\rm max}=1$.}
\label{rv}
\end{center}
\end{figure}

We introduce the dimensionless variables
\begin{eqnarray}
\hat M=\frac{M}{M_{\rm max}},\qquad \hat R=\frac{R}{R_*},\qquad \hat
V=\frac{V}{V_0}, 
\label{de1}
\end{eqnarray}
\begin{eqnarray}
\hat t=\frac{t}{t_D},\qquad \hat \omega =\omega t_D.
\label{de2}
\end{eqnarray}
We shall work with these dimensionless variables but, from now on, 
we forget the ``hats'' in order to simplify the notations. 
The
equation of motion (\ref{gd2})  becomes
\begin{eqnarray}
M\frac{d^2R}{dt^2}=-V'(R)
\label{de3}
\end{eqnarray}
with the effective potential
\begin{eqnarray}
V(R)=\frac{M}{R^2}-\frac{M^2}{3R^3}-\frac{M^2}{R}.
\label{de4}
\end{eqnarray}
The
effective potential is plotted in Fig. \ref{rv}.
Equation (\ref{de3})  has the
first
integral
\begin{eqnarray}
E_{\rm tot}=\frac{1}{2}M\left (\frac{dR}{dt}\right )^2+V(R),
\label{de5}
\end{eqnarray}
where $E_{\rm tot}$ is a constant representing the total energy 
of the system. Equation (\ref{de5}) can be rewritten as 
\begin{eqnarray}
\frac{dR}{dt}=\pm\sqrt{\frac{2}{M}[E_{\rm tot}-V(R)]},
\label{de6}
\end{eqnarray}
where the sign $+$ corresponds to an expansion of the BEC  ($\dot R>0$) and the
sign $-$ to
a contraction ($\dot R<0$). Integrating Eq. (\ref{de6}) between
$0$
and $t$, we find that the evolution of the radius $R(t)$ of the BEC is
determined by the equation
\begin{eqnarray}
\int_{R_0}^{R(t)}\frac{dR}{\sqrt{E_{\rm tot}-V(R)}}=\pm\left (\frac{2}{M}\right
)^{1/2} t.
\label{de7}
\end{eqnarray}

\subsection{The mass-radius relation}
\label{sec_mr}

\begin{figure}
\begin{center}
\includegraphics[clip,scale=0.3]{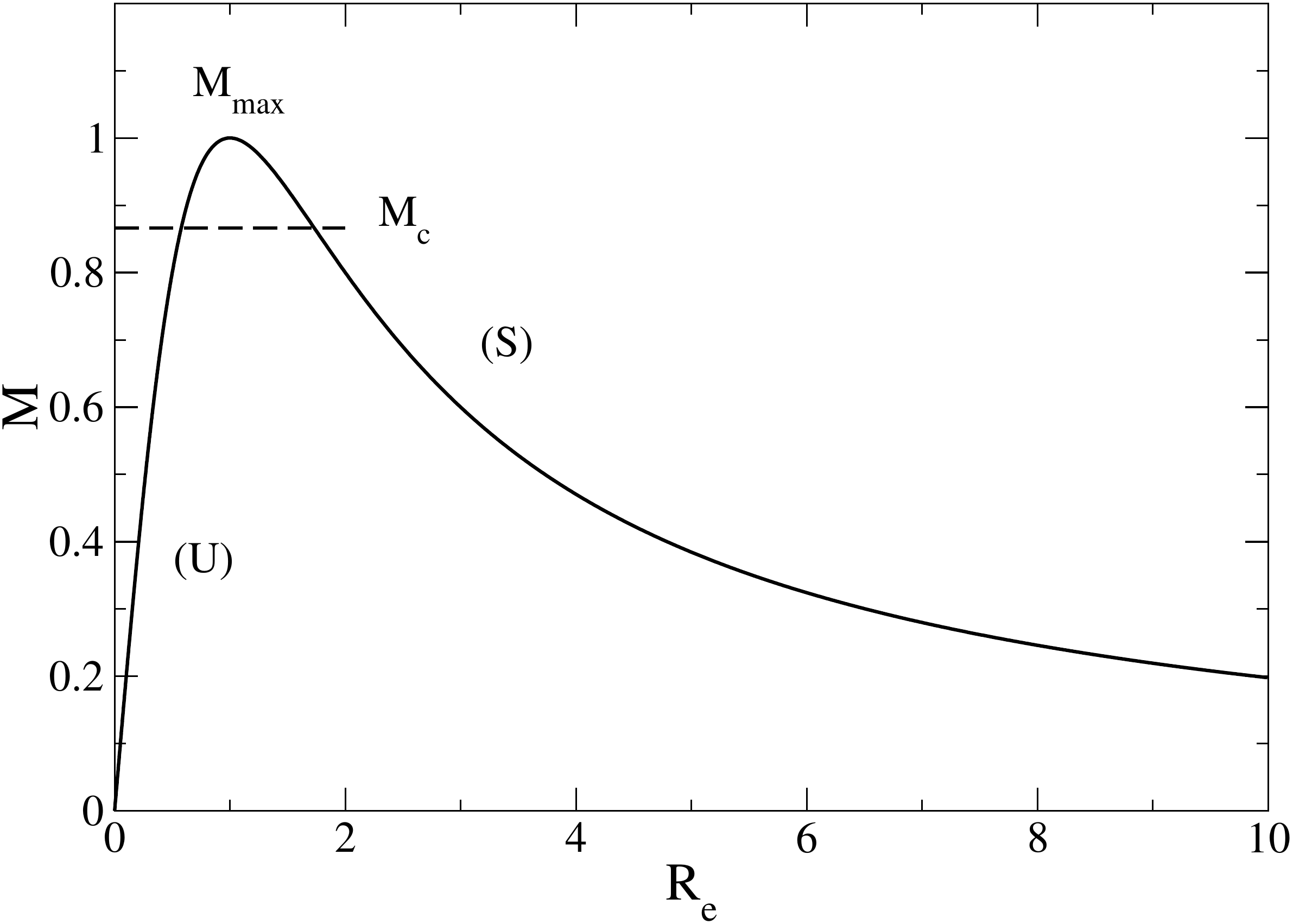}
\caption{Mass-radius relation of a self-gravitating
BEC with attractive self-interaction. The mass
$M_c$ is defined in Sec. \ref{sec_possible}.
}
\label{rm}
\end{center}
\end{figure}

A stable equilibrium state corresponds to a minimum of
the effective potential $V(R)$. The condition $V'(R_e)=0$
leads to the mass-radius relation
\begin{eqnarray}
M=\frac{2R_e}{1+R_e^2},
\label{mr1}
\end{eqnarray}
or
\begin{eqnarray}
R_e=\frac{1\pm\sqrt{1-M^2}}{M}.
\label{mr1b}
\end{eqnarray}
The mass-radius relation is plotted in Fig. \ref{rm}. Equilibrium states exist 
only below a maximum mass which is obtained from the condition $M'(R_e)=0$
yielding $R_*=1$ and $M_{\rm max}=1$. For $M<M_{\rm max}$, there are two
branches determining
two possible radii $R_U(M)<1$ and $R_S(M)>1$ for the same mass $M$. However, a
stable
equilibrium state must be a {\it minimum} of $V(R)$ so it must satisfy
$V''(R_e)>0$. Computing the second derivative of $V(R)$ from Eq. (\ref{de4}) and
using the mass-radius
relation (\ref{mr1}), we get 
\begin{eqnarray}
V''(R_e)=\frac{4(R_e^2-1)}{R_e^3(1+R_e^2)^2}.
\label{mr2}
\end{eqnarray}
From this
expression, we see that the branch (S) corresponding to $R_e>R_*$ is stable
while the branch (U) corresponding to $R_e<R_*$ is unstable. Therefore,
$R_*=1$ represents the minimum possible radius of stable equilibrium states. 
The change of stability in the series of equilibria corresponds to the maximum
mass $M_{\rm max}$ in agreement with the Poincar\'e theorem and with the theory
of catastrophes (see Sec. \ref{sec_pt}).

For $M\rightarrow 0$ and $R_e\rightarrow 0$, we get $M\sim 2R_e$ and
$V''(R_e)\sim -4/R_e^3$. In this limit,
self-gravity is negligible so the equilibrium is due to the balance between the
repulsive quantum pressure and the attractive self-interaction
(nongravitational limit). However, this equilibrium is unstable.

For  $M\rightarrow 0$ and $R_e\rightarrow +\infty$, we get $M\sim 2/R_e$ and
$V''(R_e)\sim 4/R_e^5$. In this limit, the
self-interaction is negligible so the equilibrium is due to the balance between
the repulsive quantum pressure and the attractive self-gravity (noninteracting
limit). This equilibrium is stable. It corresponds to Newtonian noninteracting
BEC stars. 

For  $M\rightarrow M_{\rm max}$ and $R\rightarrow R_*$, we get
\begin{eqnarray}
M\simeq 1-\frac{1}{2}(R_e-1)^2,\qquad R_e\simeq 1\pm\sqrt{2(1-M)},
\label{mr3}
\end{eqnarray}
\begin{eqnarray}
V''(R_e)\sim 2(R_e-1).
\label{mr4}
\end{eqnarray}
These equations describe a saddle-node bifurcation where two equilibria (one
stable
and one unstable) merge and suddenly disappear at the critical point. For future
purposes, it will be necessary to go to next order in the expansion of the
mass-radius relation close to the critical point. We find 
\begin{eqnarray}
M\simeq 1-\frac{1}{2}(R_e-1)^2+\frac{1}{2}(R_e-1)^3,
\label{mr5}
\end{eqnarray}
\begin{eqnarray}
R_e-1=\pm\sqrt{2(1-M)}\left\lbrack 1\pm\frac{1}{2}\sqrt{2(1-M)}\right\rbrack.
\label{mr6}
\end{eqnarray}

\subsection{The total energy}
\label{sec_te}

\begin{figure}
\begin{center}
\includegraphics[clip,scale=0.3]{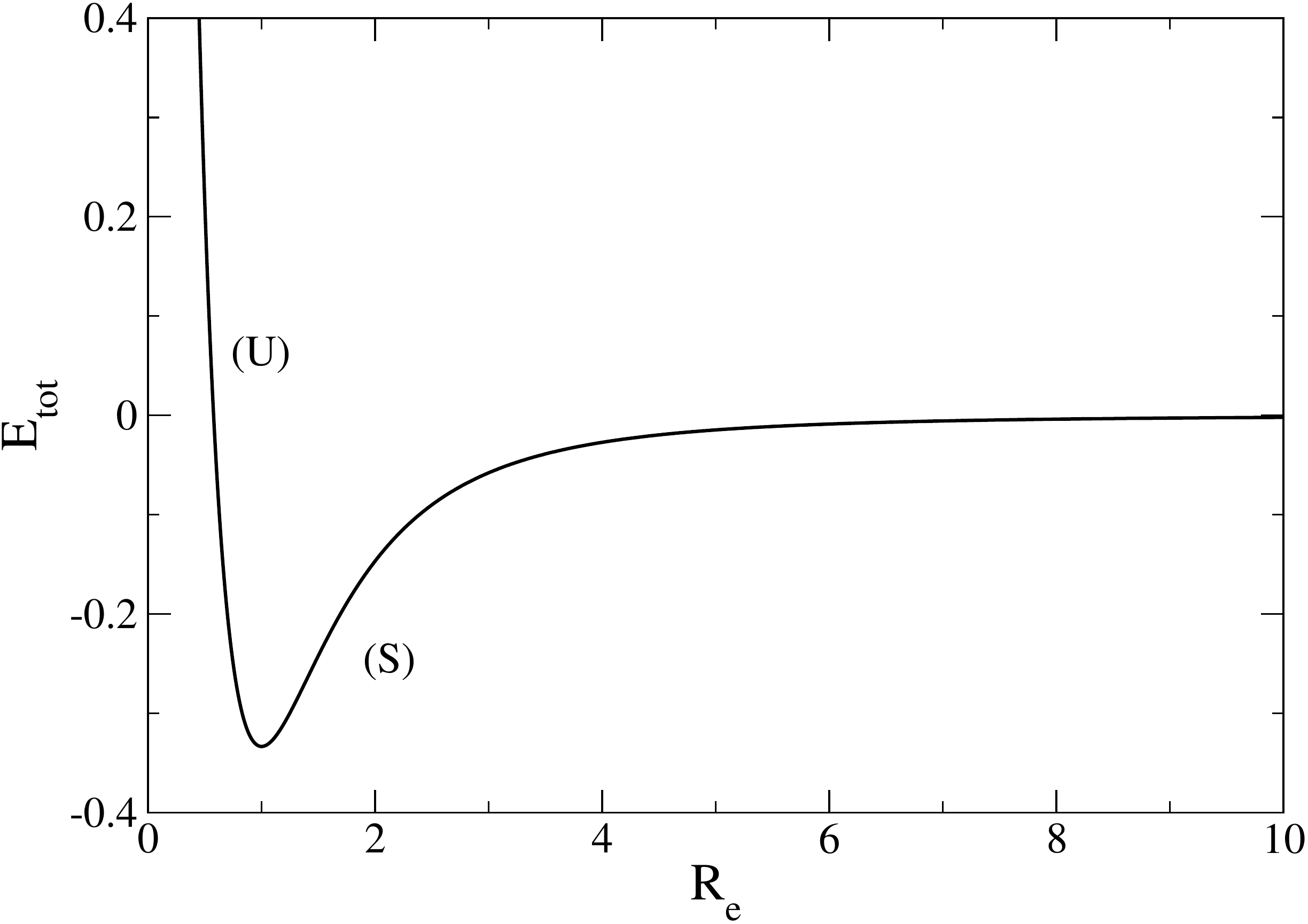}
\caption{Total energy as a function of the radius.}
\label{retot}
\end{center}
\end{figure}

\begin{figure}
\begin{center}
\includegraphics[clip,scale=0.3]{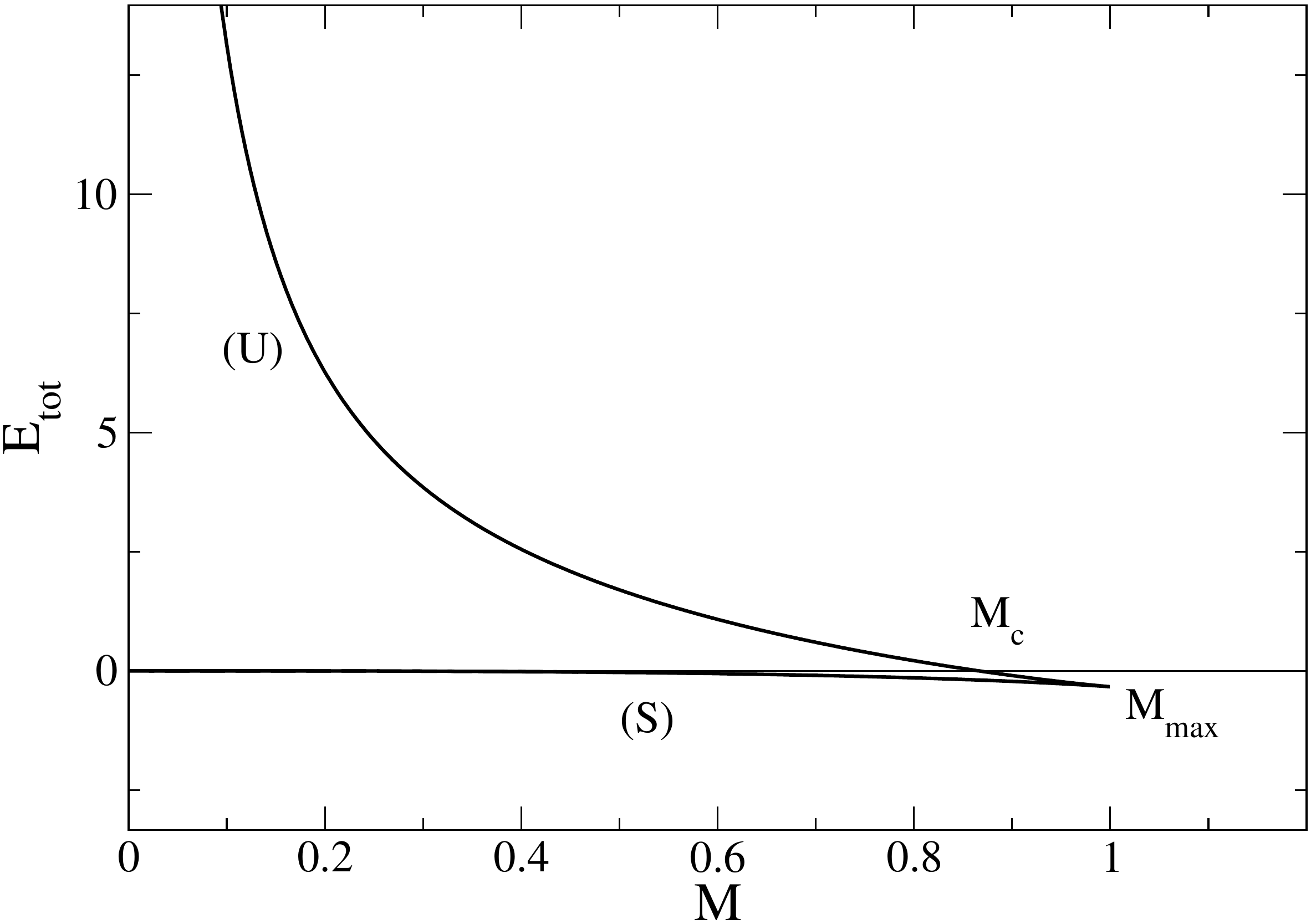}
\caption{Total energy as a function of the mass. The mass
$M_c$ is defined in Sec. \ref{sec_possible}.}
\label{metot}
\end{center}
\end{figure}

At equilibrium, the total energy of a BEC with a mass $M$ is given by $E_{\rm
tot}=V(R_e)$. Therefore
\begin{eqnarray}
E_{\rm tot}=\frac{M}{R_e^2}-\frac{M^2}{3R_e^3}-\frac{M^2}{R_e}.
\label{te1}
\end{eqnarray}
We can obtain  $E_{\rm tot}(R_e)$ by using Eq. (\ref{mr1}) to express $M$  as
a function of $R_e$. This gives
\begin{eqnarray}
E_{\rm tot}=\frac{2(1-3R_e^2)}{3R_e(1+R_e^2)^2}.
\label{te2}
\end{eqnarray}
This function is plotted in Fig. \ref{retot}. We note that the minimum
energy $E_{\rm tot}^{\rm min}=-1/3$ is reached for $R_e=R_*=1$. Therefore, the
change of stability in the series of equilibria corresponds to the minimum
energy $E_{\rm tot}^{\rm min}$. We can obtain $E_{\rm tot}(M)$ by using Eq.
(\ref{mr1b}) to express $R_e$ in terms of $M$. Alternatively, the function
$E_{\rm tot}(M)$ is given in parametric form by Eqs.
(\ref{mr1}) and (\ref{te2}) where the  parameter is $R_e$. This function is
plotted in Fig. \ref{metot}. We can check that the stable branch (S) has a lower
energy that the unstable branch as it should. We also recall that a
necessary (but not sufficient) condition
of nonlinear dynamical stability is that $E_{\rm tot}<0$. We can chek that the
stable branch satisfies this condition.

For $M\rightarrow 0$ and $R_e\rightarrow 0$ (nongravitational limit), we get
$E_{\rm tot}(R_e)\sim 2/3R_e$ and $E_{\rm tot}(M)\sim 4/3M$.

For  $M\rightarrow 0$ and $R_e\rightarrow +\infty$ (noninteracting limit), we
get $E_{\rm tot}(R_e)\sim
-2/R_e^3$ and $E_{\rm tot}(M)\sim -M^3/4$.

For  $M\rightarrow M_{\rm max}$ and $R\rightarrow R_*$ (critical point), we
get
\begin{eqnarray}
E_{\rm tot}(R_e)\simeq -\frac{1}{3}+\frac{5}{6}(R_e-1)^2-\frac{3}{2}(R_e-1)^3,
\label{te3}
\end{eqnarray}
\begin{eqnarray}
E_{\rm tot}(M)\simeq -\frac{1}{3}+\frac{5}{3}(1-M)\mp
\frac{4}{3}\sqrt{2}(1-M)^{3/2},
\label{te4}
\end{eqnarray}
where we used Eq. (\ref{mr6}).

Since the functions $M(R_e)$ and $E_{\rm tot}(R_e)$ achieve their extrema at
the same point $R_e=R_*=1$,\footnote{The intrinsic reason is the
following. A stable steady state of the GPP and EP equations is a
minimum of $E_{\rm tot}$ at fixed mass $M$ \cite{prd1}. Introducing a Lagrange
multiplier $\mu/m$
to take
into account the mass constraint, the cancellation of the first order
variations writes $\delta E_{\rm tot}-(\mu/m)\delta M=0$. This is valid for any
equilibrium state in the series of
equilibria. From that relation, $\delta M=0$ implies $\delta E_{\rm
tot}=0$. Therefore, the mass and the
total energy achieve their extrema at the same point in the series of
equilibria.} the function $E_{\rm
tot}(M)$ presents a {\it spike} at the
critical point $M=M_{\rm max}$ (see Fig. \ref{metot}).

\subsection{The pulsation}
\label{sec_pul}

\begin{figure}
\begin{center}
\includegraphics[clip,scale=0.3]{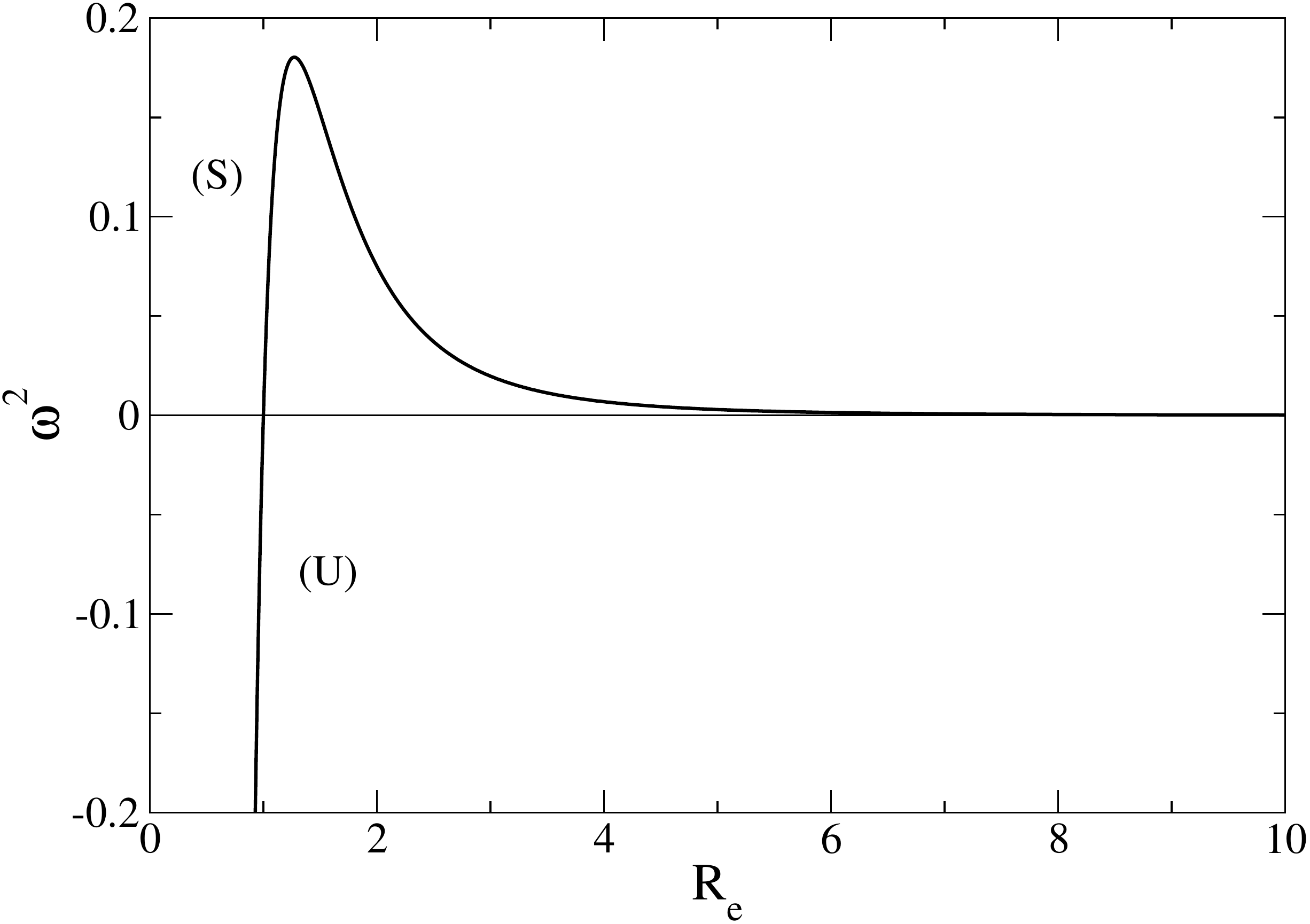}
\caption{Complex pulsation as a function of the radius. 
}
\label{rw2}
\end{center}
\end{figure}

\begin{figure}
\begin{center}
\includegraphics[clip,scale=0.3]{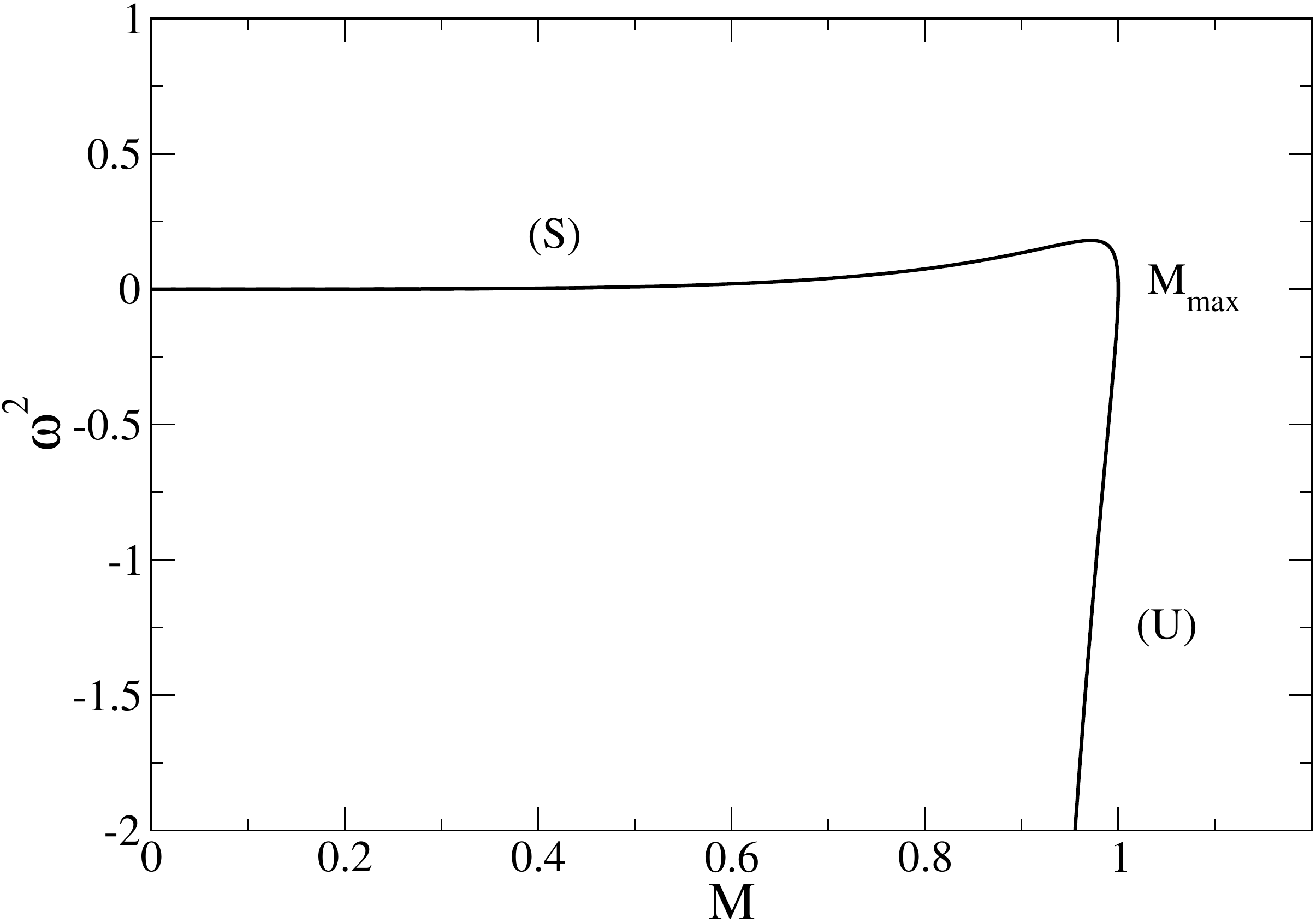}
\caption{Complex pulsation as a function of the mass.
}
\label{mew2}
\end{center}
\end{figure}

We can analyze the stability of an equilibrium state by considering a small
perturbation about equilibrium. We assume $M<M_{\rm max}$ and write
$R(t)=R_e+\epsilon(t)$ with $\epsilon(t)\ll 1$. Substituting this decomposition
into Eq. (\ref{de3}) and keeping only terms that are linear in $\epsilon$, we
obtain
\begin{eqnarray}
\frac{d^2\epsilon}{dt^2}+\omega^2\epsilon=0,
\label{pul1}
\end{eqnarray}
where
\begin{eqnarray}
\omega^2=\frac{V''(R_e)}{M}
\label{pul2}
\end{eqnarray}
is the square of the complex pulsation. Clearly $\omega^2>0$ corresponds to a
stable state which, when displaced from equilibrium, oscillates about
equilibrium with a pulsation $\omega$. By contrast, $\omega^2<0$ corresponds to
an unstable state which, when displaced from equilibrium, evolves away from
equilibrium with a growth rate $\gamma=\sqrt{-\omega^2}$. From these
expressions, we confirm that a minimum of $V(R)$ is stable and a
maximum of $V(R)$ is unstable. Furthermore, using Eqs. (\ref{mr1}) and
(\ref{mr2}), we can express $\omega^2$ as a function of the radius according to
\begin{eqnarray}
\omega^2=\frac{2(R_e^2-1)}{R_e^4(R_e^2+1)}.
\label{pul3}
\end{eqnarray}
This function is plotted in Fig. \ref{rw2}. The pulsation vanishes 
($\omega^2=0$) at the critical point $M=M_{\rm max}=1$ and $R_e=R_*=1$.
The states with $R_e>1$ are stable ($\omega^2>0$) and the states
with $R_e<1$ are unstable ($\omega^2<0$) in agreement with our previous
discussion. We note that
the maximum pulsation $(\omega^2)_{\rm
max}=8(\sqrt{5}-1)/[(\sqrt{5}+3)(1+\sqrt{5})^2]=0.1803...$ is reached for
$R_e=[(1+\sqrt{5})/2]^{1/2}=1.272...$ and
$M=[8(1+\sqrt{5})]^{1/2}/(3+\sqrt{5})=0.9717...$. The function $\omega^2(M)$ is
obtained in
parametric form from Eqs.
(\ref{mr1}) and (\ref{pul3}) where the  parameter is $R_e$. This function is
plotted in Fig. \ref{mew2}.

For $M\rightarrow 0$ and $R_e\rightarrow 0$ (nongravitational limit), we get
$\omega^2\sim -2/R_e^4$ and $\omega^2\sim -32/M^4$.

For  $M\rightarrow 0$ and $R_e\rightarrow +\infty$ (noninteracting limit), we
get $\omega^2\sim 2/R_e^4$ and $\omega^2\sim M^4/8$.

For  $M\rightarrow M_{\rm max}$ and $R_e\rightarrow R_*$ (critical point), we
get 
\begin{eqnarray}
\omega^2\sim 2(R_e-1)\sim \pm 2\sqrt{2(1-M)}.
\label{pul4}
\end{eqnarray}

\subsection{The Poincar\'e theorem}
\label{sec_pt}

According to Eqs. (\ref{mr1}), (\ref{mr2}) and
(\ref{pul3}), we have the relation
\begin{eqnarray}
\frac{dM}{dR_e}=-\frac{1}{2}R_e^3 V''(R_e)=-\frac{1}{2}MR_e^3\omega^2.
\label{pt2}
\end{eqnarray}
Close to the critical point, Eq. (\ref{pt2}) reduces to
\begin{eqnarray}
\frac{dM}{dR_e}=-\frac{1}{2} V''(R_e)=-\frac{\omega^2}{2}
\label{pt1}
\end{eqnarray}
which can be directly obtained from Eqs. (\ref{mr3}), (\ref{mr4}) and
(\ref{pul4}). These relations first show that the change of stability
($\omega^2=0$) corresponds to the turning point of mass ($M'(R_e)=0$).
Furthermore, they relate the stability of the system to the slope of the
mass-radius relation
$M(R_e)$.  The branch where the mass increases with the radius ($M'(R_e)>0$) is
unstable ($\omega^2<0$) and the branch where the mass decreases with the radius
($M'(R_e)<0$) is stable ($\omega^2>0$). These results are particular cases of
the Poincar\'e theorem on the series of equilibria
\cite{poincare} (see \cite{katzpoincare,ijmpb} for some applications of the
Poincar\'e theorem to self-gravitating systems). They have been formalized in
the theory of catastrophes.

\section{Collapse of the BEC when $M\ge M_{\rm max}$}
\label{sec_collapse}

We consider a BEC with a mass  $M\ge M_{\rm max}=1$ so that no
equilibrium state exists. In that case, the BEC is expected to collapse and
form a black hole. Here, we study the collapse analytically by using the gross
dynamics defined by Eqs. (\ref{de3}) and (\ref{de4}).

\subsection{The collapse time}
\label{sec_ct}

To be specific, we consider an initial condition such that $\dot R_0=0$ (no
initial velocity) although more general initial conditions could be considered
as well. For this initial condition the total energy is $E_{\rm tot}=V(R_0)$,
where $R_0$ is the initial radius of the BEC. In that case, according to Eq.
(\ref{de7}), the evolution of
the radius $R(t)$ of the BEC is given by
\begin{eqnarray}
\int_{R(t)}^{R_0}\frac{dR}{\sqrt{V(R_0)-V(R)}}=\left (\frac{2}{M}\right
)^{1/2} t,
\label{ct2}
\end{eqnarray}
where the sign has been chosen so that $R(t)$ decreases with time since 
we are considering a collapse solution. The collapse time, corresponding to
$R(t_{\rm coll})=0$, is given by
\begin{eqnarray}
t_{\rm coll}=\left(\frac{M}{2}\right
)^{1/2}\int_{0}^{R_0}\frac{dR}{\sqrt{V(R_0)-V(R) }}.
\label{ct3}
\end{eqnarray}
Using Eq. (\ref{ct3}), we can rewrite Eq. (\ref{ct2}) in the form
\begin{eqnarray}
\int_0^{R(t)}\frac{dR}{\sqrt{V(R_0)-V(R)}}=\left (\frac{2}{M}\right
)^{1/2}(t_{\rm coll}-t).
\label{ct4}
\end{eqnarray}
For the potential of Eq.  (\ref{de4})  the
collapse time is
given by
\begin{eqnarray}
t_{\rm coll}=\sqrt{\frac{M}{2}}\int_{0}^{R_0}\frac{dR}{\sqrt{\frac{M}{R_0^2}
-\frac{M^2}{3R_0^3}
-\frac {M^2}{R_0}-\frac{M}{R^2}+\frac{M^2}{3R^3}+\frac{M^2}{R}
}}.\nonumber\\
\label{ct5}
\end{eqnarray}
This is a function $t_{\rm coll}(M,R_0)$ of the
mass $M$ of the BEC and of its initial radius $R_0$.

\begin{figure}
\begin{center}
\includegraphics[clip,scale=0.3]{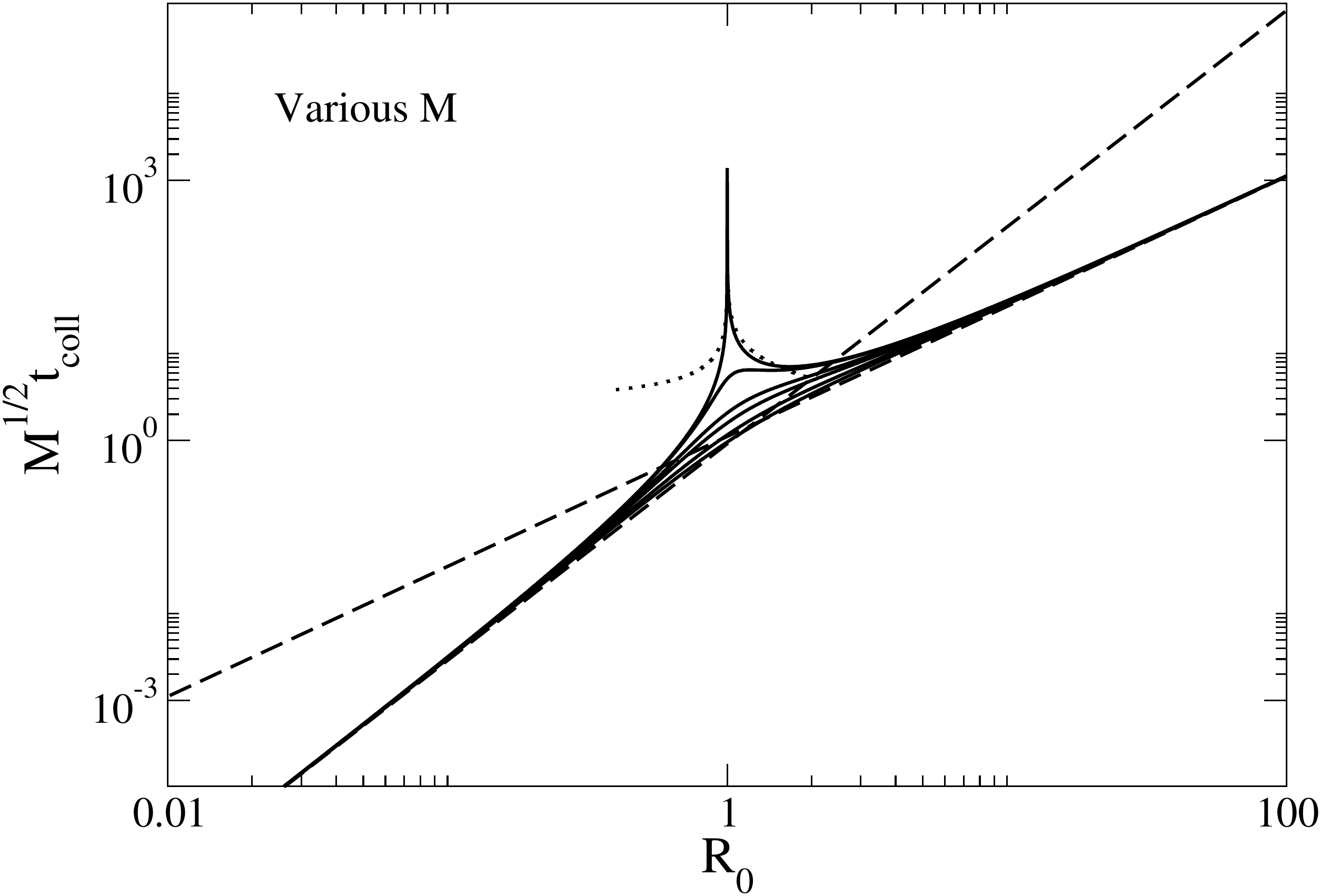}
\caption{Collapse time of the self-gravitating BEC as a function of its initial
radius $R_0$ for different masses $M=1, 1.01, 1.1, 1.2, 1.5, 2$ (top to
bottom). For
$M=1$, the collapse time diverges when $R_0\rightarrow 1$ (dotted lines) as
detailed in Figs.
\ref{r0tcollm1singplus} and \ref{r0tcollm1singmoins}. We note that the
asymptotic behaviors
of $\sqrt{M}t_{\rm coll}(M,R_0)$ for $R_0\rightarrow 0$ and $R_0\rightarrow
+\infty$ are independent of $M$ (dashed lines). Furthermore, for $M\rightarrow
+\infty$, the
function $R_0 \!\mapsto\! M^{1/2}t_{\rm coll}(M,R_0)$ tends to the
function $A(R_0)$ defined by Eq. (\ref{la2}) and plotted in Fig. \ref{r0A}.
}
\label{rotcoll}
\end{center}
\end{figure}

\begin{figure}
\begin{center}
\includegraphics[clip,scale=0.3]{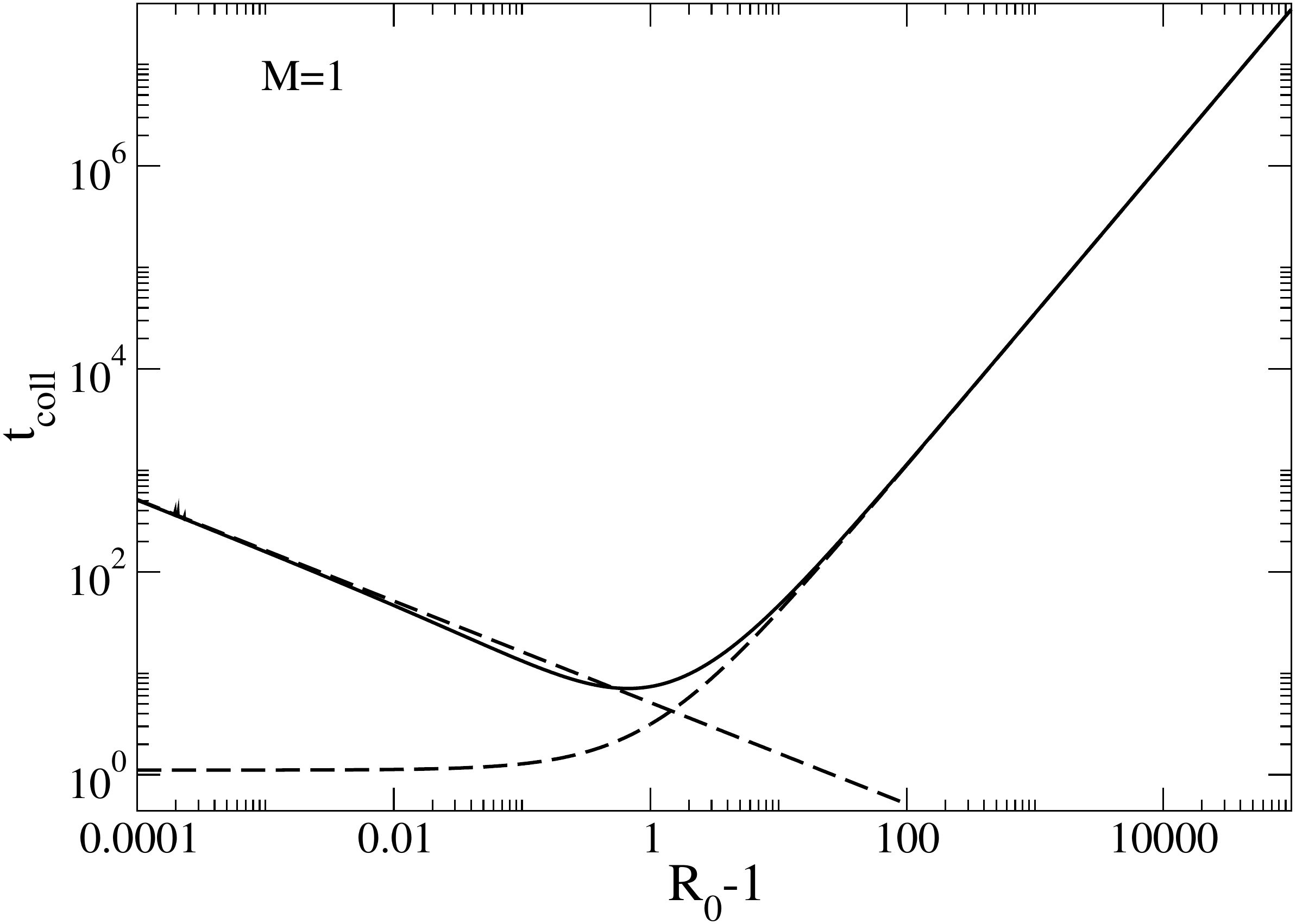}
\caption{Collapse time of the self-gravitating BEC as a function
of its initial
radius $R_0\ge 1$ for $M=1$. 
}
\label{r0tcollm1singplus}
\end{center}
\end{figure}

\begin{figure}
\begin{center}
\includegraphics[clip,scale=0.3]{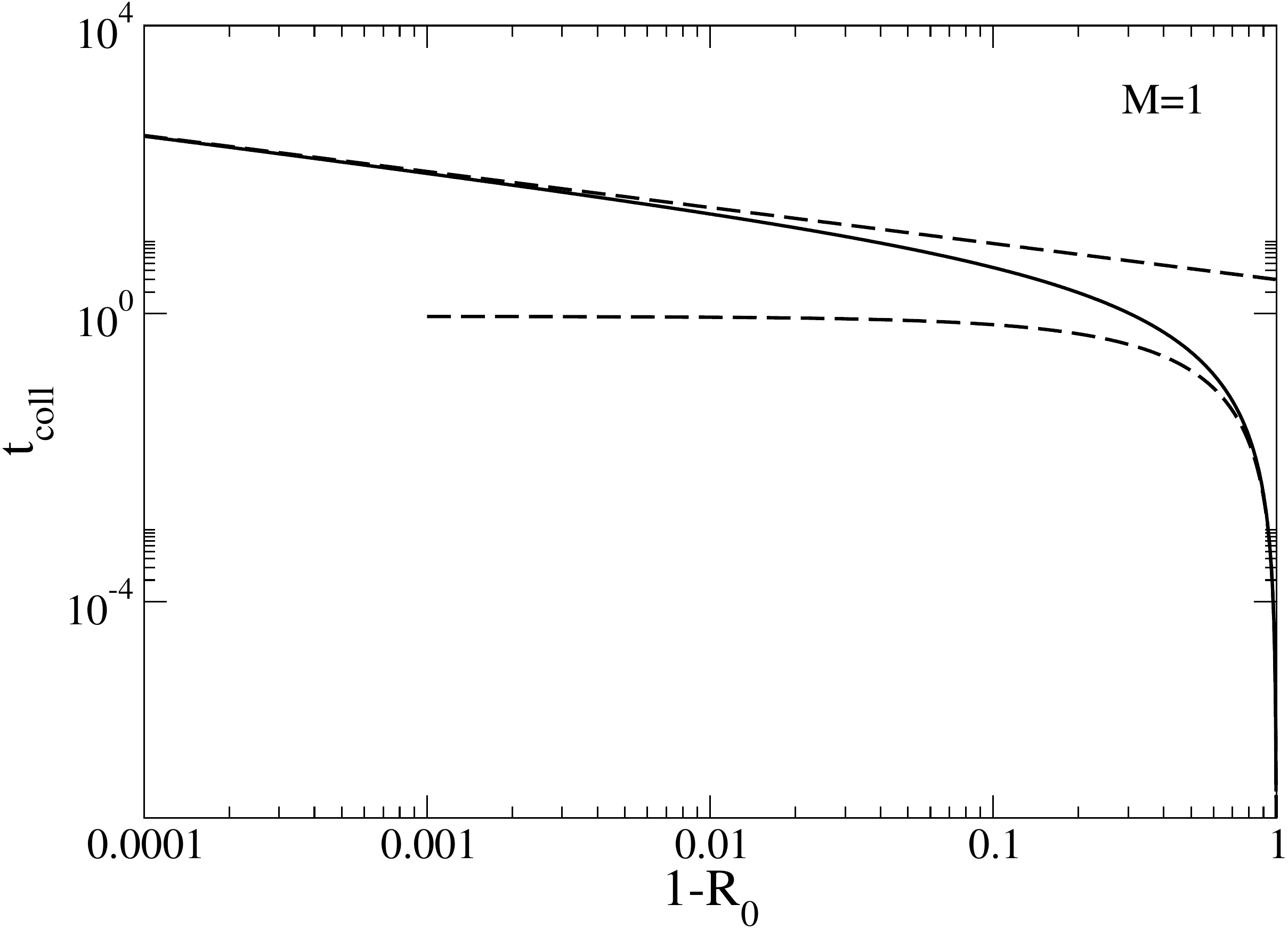}
\caption{Collapse time of the self-gravitating BEC as a function
of its initial
radius $R_0\le 1$ for $M=1$.
}
\label{r0tcollm1singmoins}
\end{center}
\end{figure}

\begin{figure}
\begin{center}
\includegraphics[clip,scale=0.3]{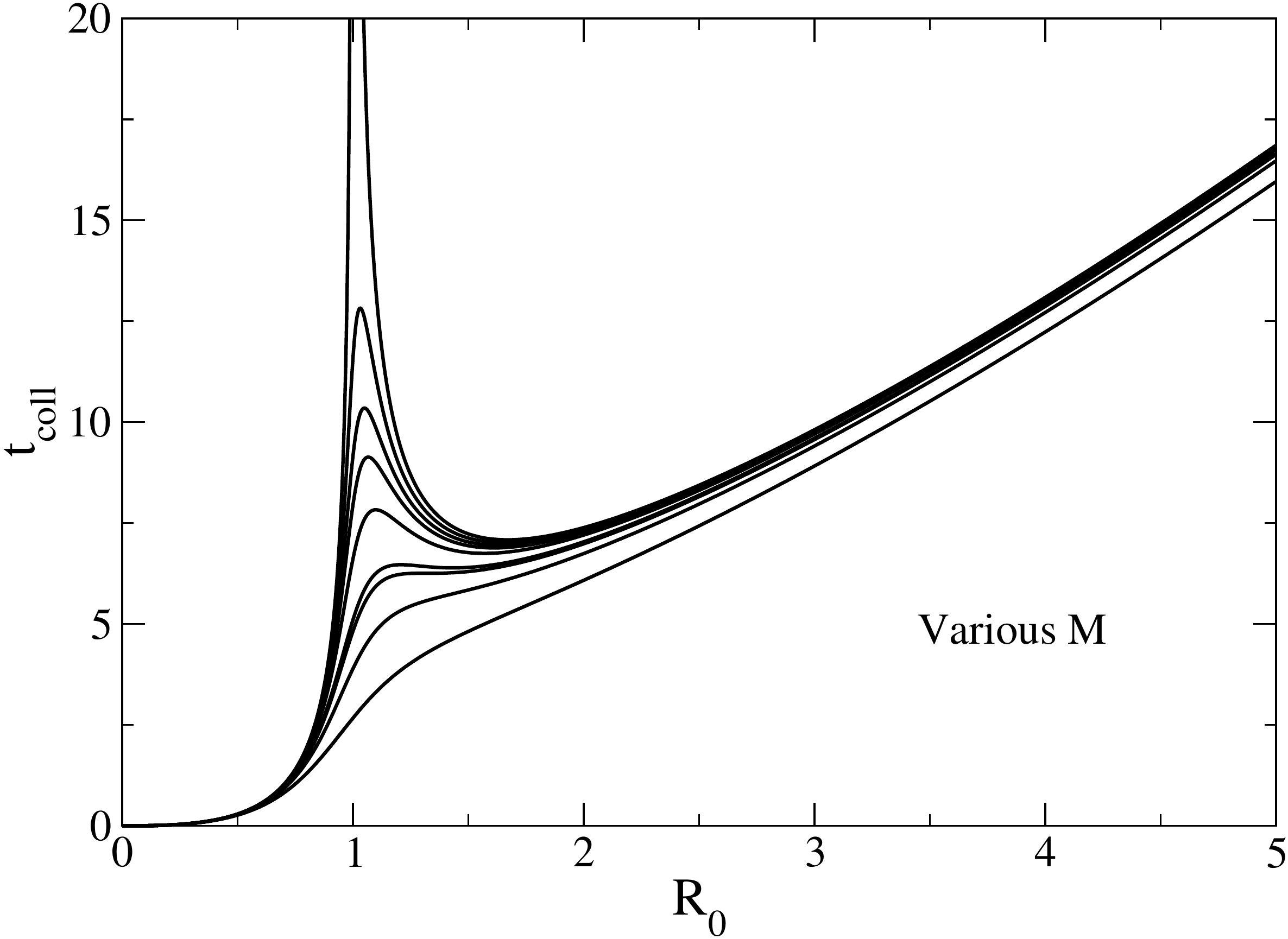}
\caption{Collapse time of the self-gravitating BEC as a function of its initial
radius $R_0$ for different masses $M=1, 1.001, 1.002, 1.003, 1.005, 1.01,
1.0116, 1.02, 1.05$ (top to
bottom).
For $1\le M\le M_*=1.0116$, the collapse time presents a local maximum at
$((R_0)_M,(t_{\rm coll})_M)$ and a 
local minimum at $((R_0)_m,(t_{\rm coll})_m)$.}
\label{r0tcollminmax}
\end{center}
\end{figure}

\begin{figure}
\begin{center}
\includegraphics[clip,scale=0.3]{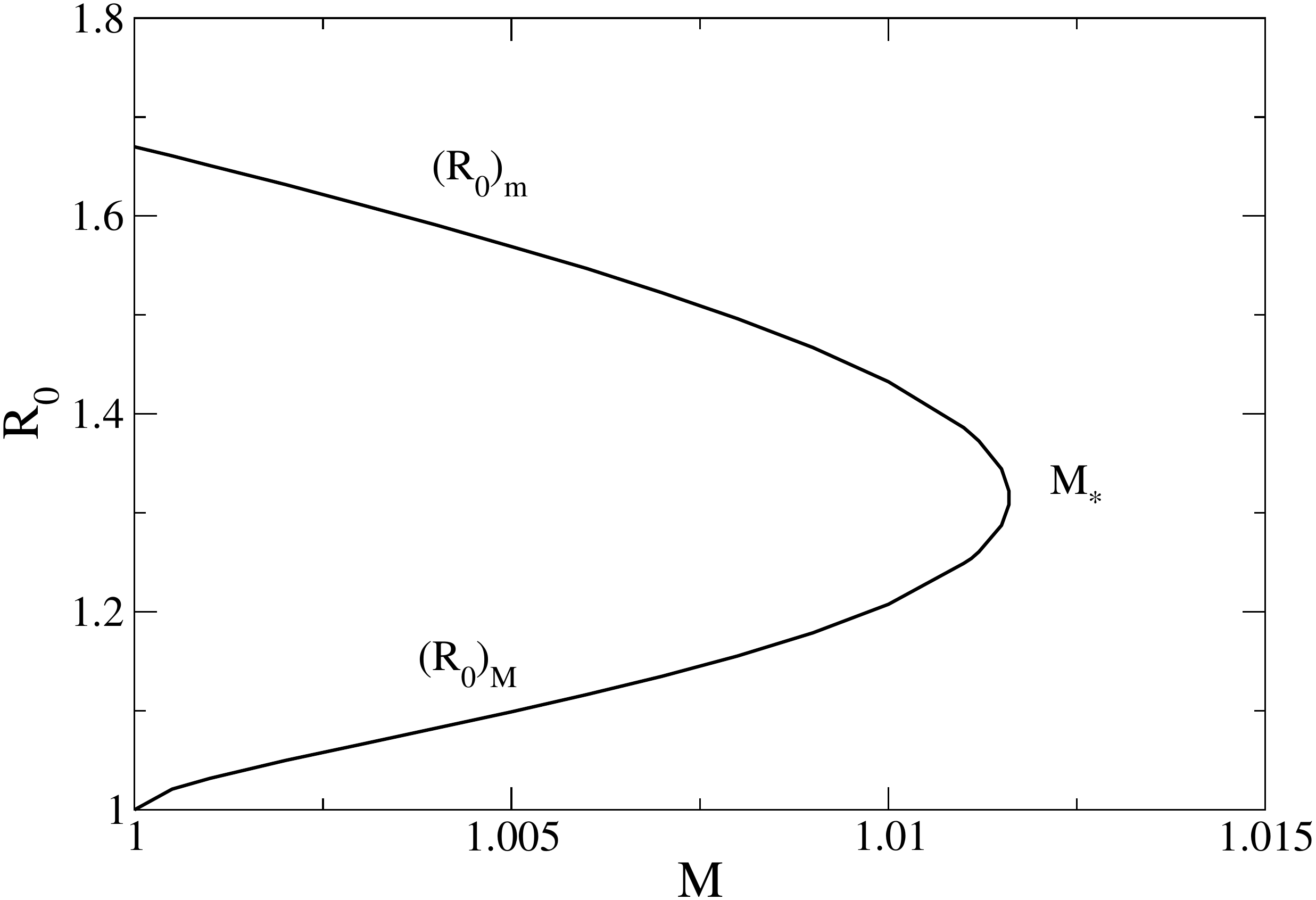}
\caption{Evolution of $(R_0)_M$ and $(R_0)_m$ as a function of $M$.}
\label{mrminmax}
\end{center}
\end{figure}

\begin{figure}
\begin{center}
\includegraphics[clip,scale=0.3]{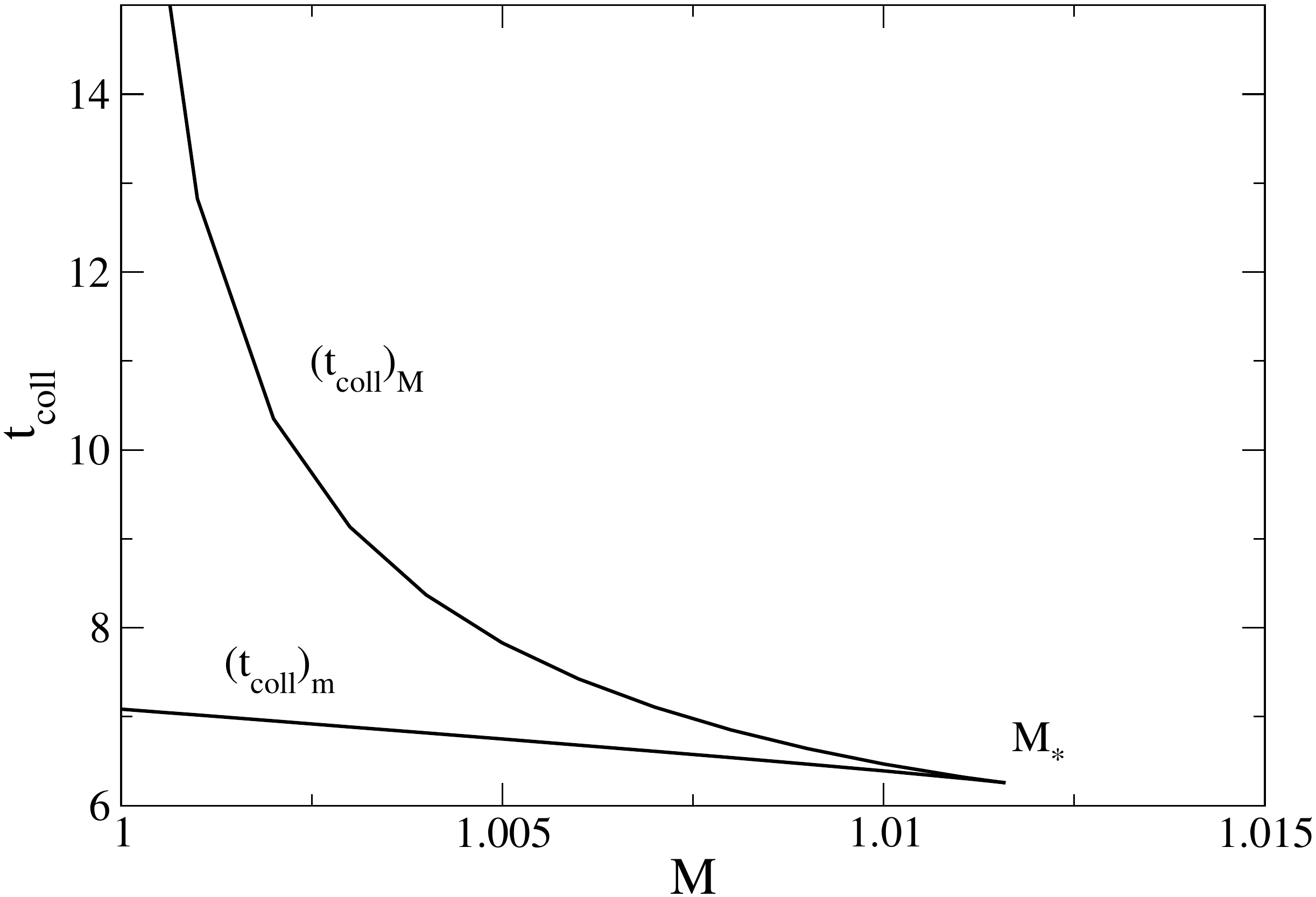}
\caption{Evolution of $(t_{\rm coll})_M$ and  $(t_{\rm coll})_m$ as a function
of $M$.}
\label{mtminmax}
\end{center}
\end{figure}

The function 
$R_0 \!\mapsto\! t_{\rm coll}(M,R_0)$ is plotted in Figs.
\ref{rotcoll}, \ref{r0tcollm1singplus}, \ref{r0tcollm1singmoins} and
\ref{r0tcollminmax} for different values of the mass $M$. The asymptotic
behaviors of 
$R_0  \!\mapsto\!  t_{\rm coll}(M,R_0)$ for $R_0\rightarrow 0$ and
$R_0\rightarrow +\infty$ are given by Eqs. (\ref{rs4}) and (\ref{rl4}).
When $M=1$, the collapse time  $t_{\rm coll}(M=1,R_0)$ diverges when
$R_0\rightarrow 1$ according to Eqs. (\ref{ru3}) and (\ref{ru6}). For
$M\rightarrow 1^+$ and
$R_0\rightarrow 1$, the function  $R_0 \!\mapsto\! t_{\rm
coll}(M,R_0)$ has a self-similar structure discussed in Appendix \ref{sec_sst}
and
illustrated in Figs. \ref{ssg} and \ref{ssgR0neg}. For $1\le M\le M_*\simeq
1.0116$, the
function $R_0 \!\mapsto\! t_{\rm coll}(R_0,M)$ presents a local maximum
at $((R_0)_M,(t_{\rm coll})_M)$ and a 
local minimum at $((R_0)_m,(t_{\rm coll})_m)$ as illustrated in Fig.
\ref{r0tcollminmax}. These characteristic values are plotted as a function of
$M$ in Figs. \ref{mrminmax} and \ref{mtminmax}. For $M=1$, we find
$((R_0)_M,(t_{\rm coll})_M)=(1,+\infty)$ and $((R_0)_m,(t_{\rm
coll})_m)=(1.67...,7.0857...)$. For $M=M_*$, we find
$((R_0)_M,(t_{\rm coll})_M)=((R_0)_m,(t_{\rm
coll})_m)=(1.31...,6.257...)$.

\begin{figure}
\begin{center}
\includegraphics[clip,scale=0.3]{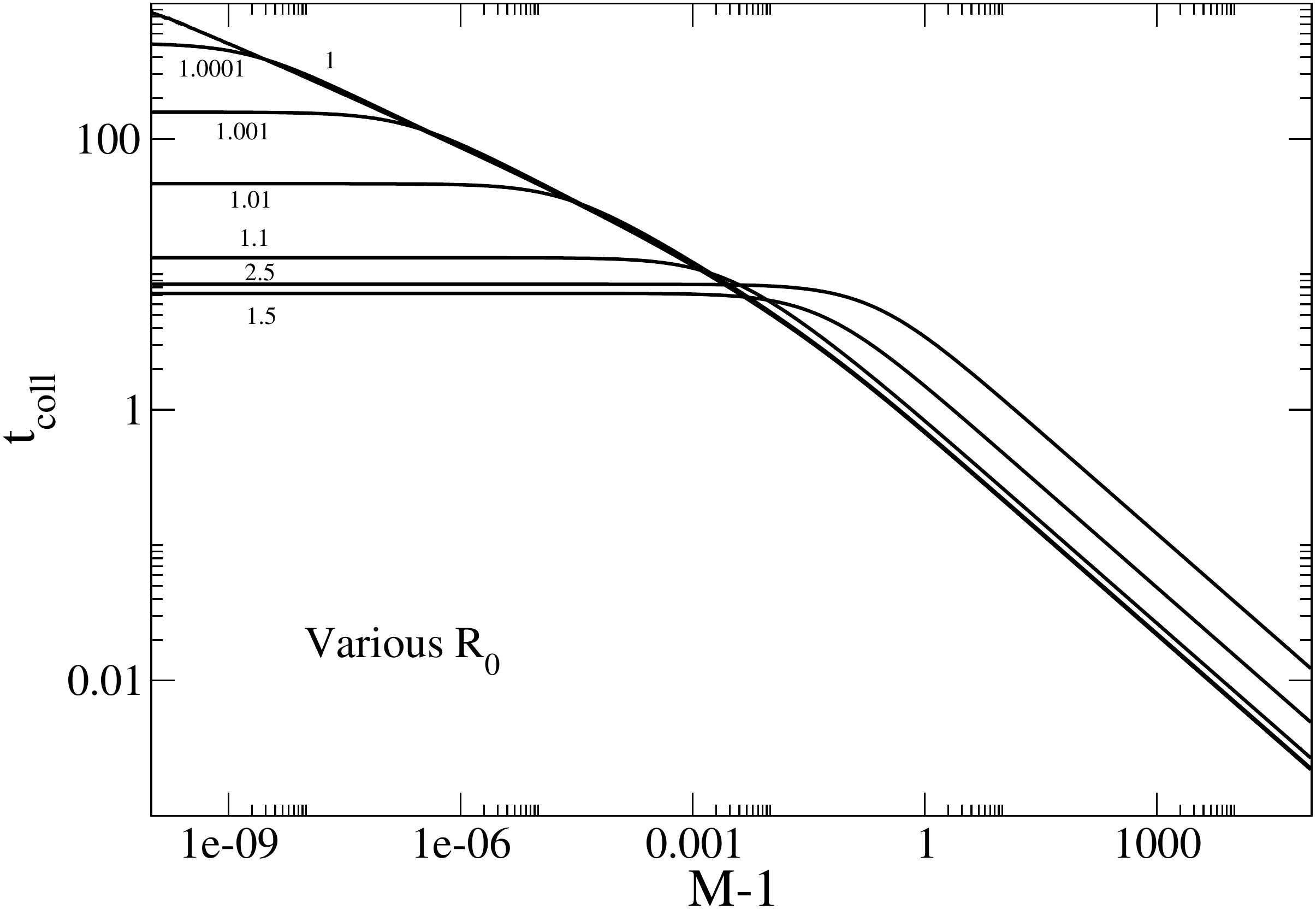}
\caption{Collapse time of the self-gravitating BEC as a function of its mass $M$
for different values of the initial radius $R_0=1, 1.0001, 1.001, 1.01, 1.1,
1.5, 2.5$ (for clarity, we have restricted ourselves to $R_0>1$). For
$R_0=1$, the collapse time diverges when $M\rightarrow 1^+$ as
detailed in Fig. \ref{Mtcoll}.  We note that the evolution of
$t_{\rm coll}(M=1,R_0)$ is non monotonic for $R_0\ge 1$ in agreement
with the results of Fig. \ref{r0tcollm1singplus}. The plateau corresponding to
$M\rightarrow 1$
first decreases as $R_0$ increases from $R_0=1$ to $(R_0)_m=1.67...$, then
increases as $R_0$
passes above $(R_0)_m=1.67...$.}
\label{mtcollmultirofinale}
\end{center}
\end{figure}

\begin{figure}
\begin{center}
\includegraphics[clip,scale=0.3]{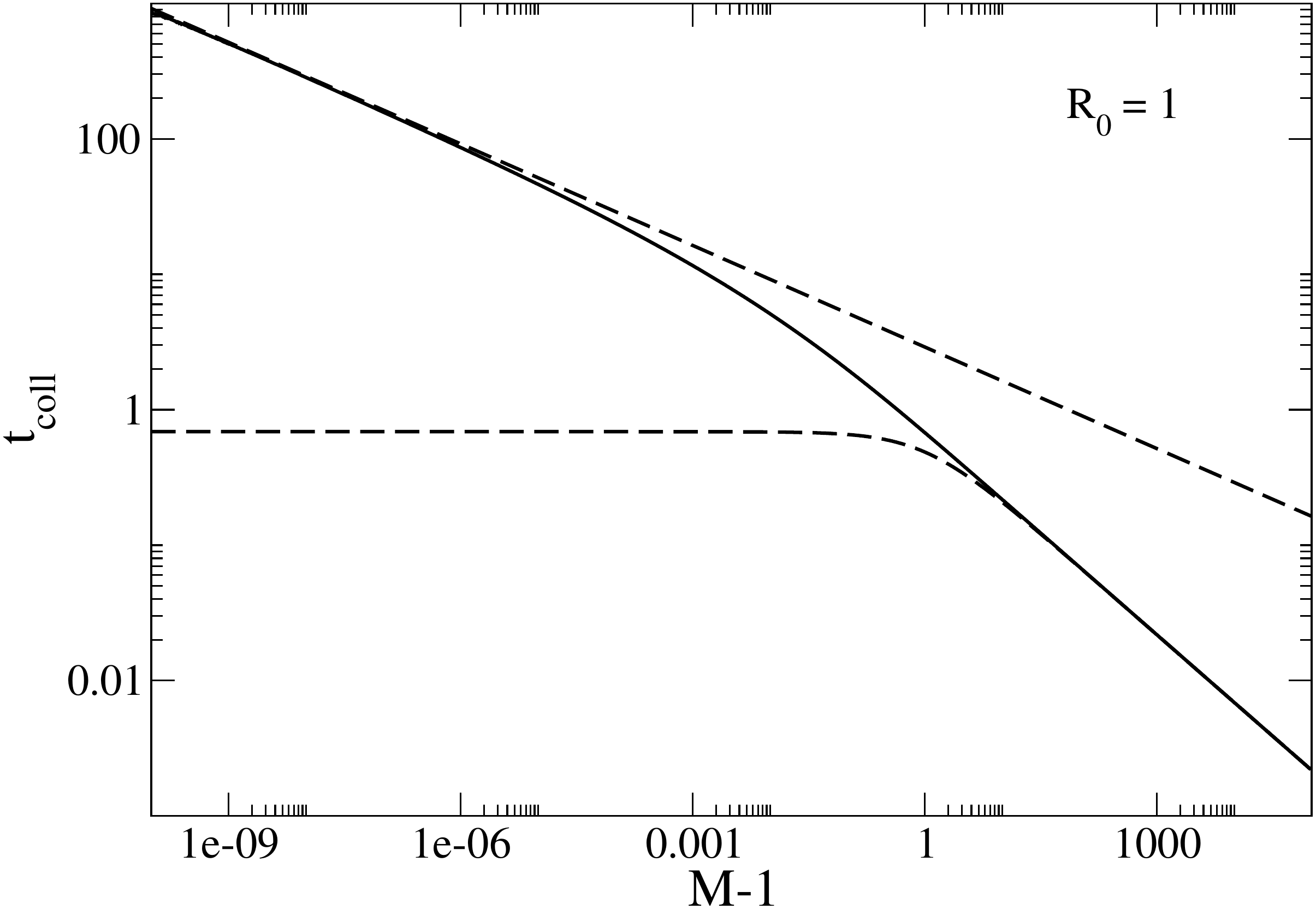}
\caption{Collapse time of the self-gravitating BEC as a function of its mass
when started from $R_0=1$.
}
\label{Mtcoll}
\end{center}
\end{figure}

The function 
$M \!\mapsto\!  t_{\rm coll}(M,R_0)$ is plotted in
Figs. \ref{mtcollmultirofinale} and \ref{Mtcoll}  for different values of the
initial radius $R_0$. The asymptotic behavior of 
$M \!\mapsto\!  t_{\rm coll}(M,R_0)$ for $M\rightarrow +\infty$ is given by
Eq. (\ref{la1}). The value of $t_{\rm coll}(M,R_0)$ at $M=1$ is
plotted as a function of $R_0$ in Figs. \ref{r0tcollm1singplus},
\ref{r0tcollm1singmoins} and \ref{r0tcollminmax} (as we have already mentioned,
the function $R_0 \!\mapsto\! t_{\rm coll}(R_0,M=1)$ is non monotonic).
When
$R_0=1$, the collapse time  $t_{\rm coll}(M,R_0=1)$ diverges when
$M\rightarrow 1^+$ according to Eqs. (\ref{saddle10}). For
$M\rightarrow 1^+$ and
$R_0\rightarrow 1$, the function  $M \!\mapsto\! t_{\rm
coll}(M,R_0)$ has a self-similar structure as shown in Appendixes \ref{sec_ssu}
and \ref{sec_ssd} and illustrated in  Fig. \ref{ssfplus}.

\begin{figure}
\begin{center}
\includegraphics[clip,scale=0.3]{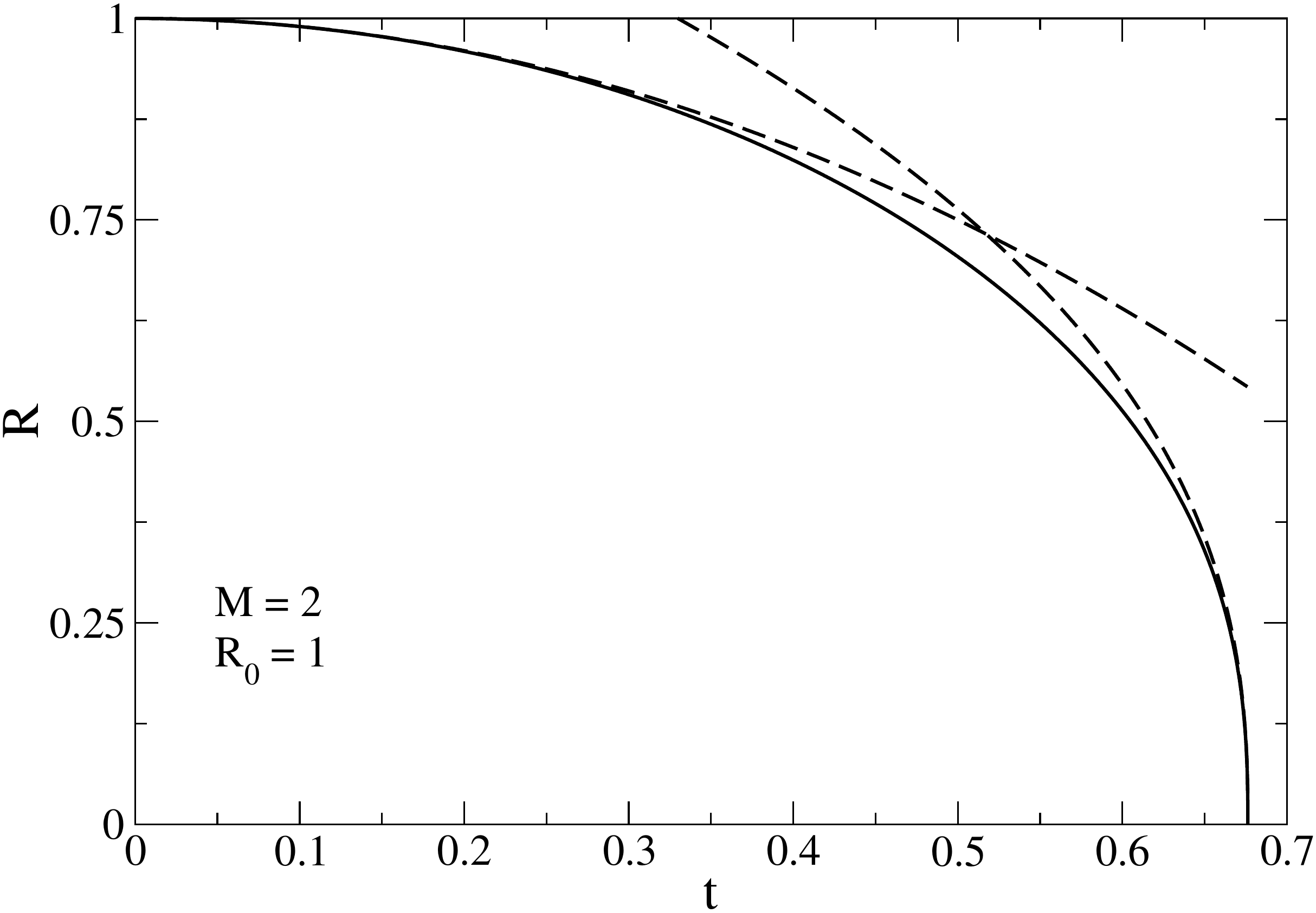}
\caption{Collapse of the self-gravitating BEC starting from $R_0=1$ for $M=2$.
The
collapse time is $t_{\rm coll}=0.676301...$.
}
\label{tr}
\end{center}
\end{figure}

\begin{figure}
\begin{center}
\includegraphics[clip,scale=0.3]{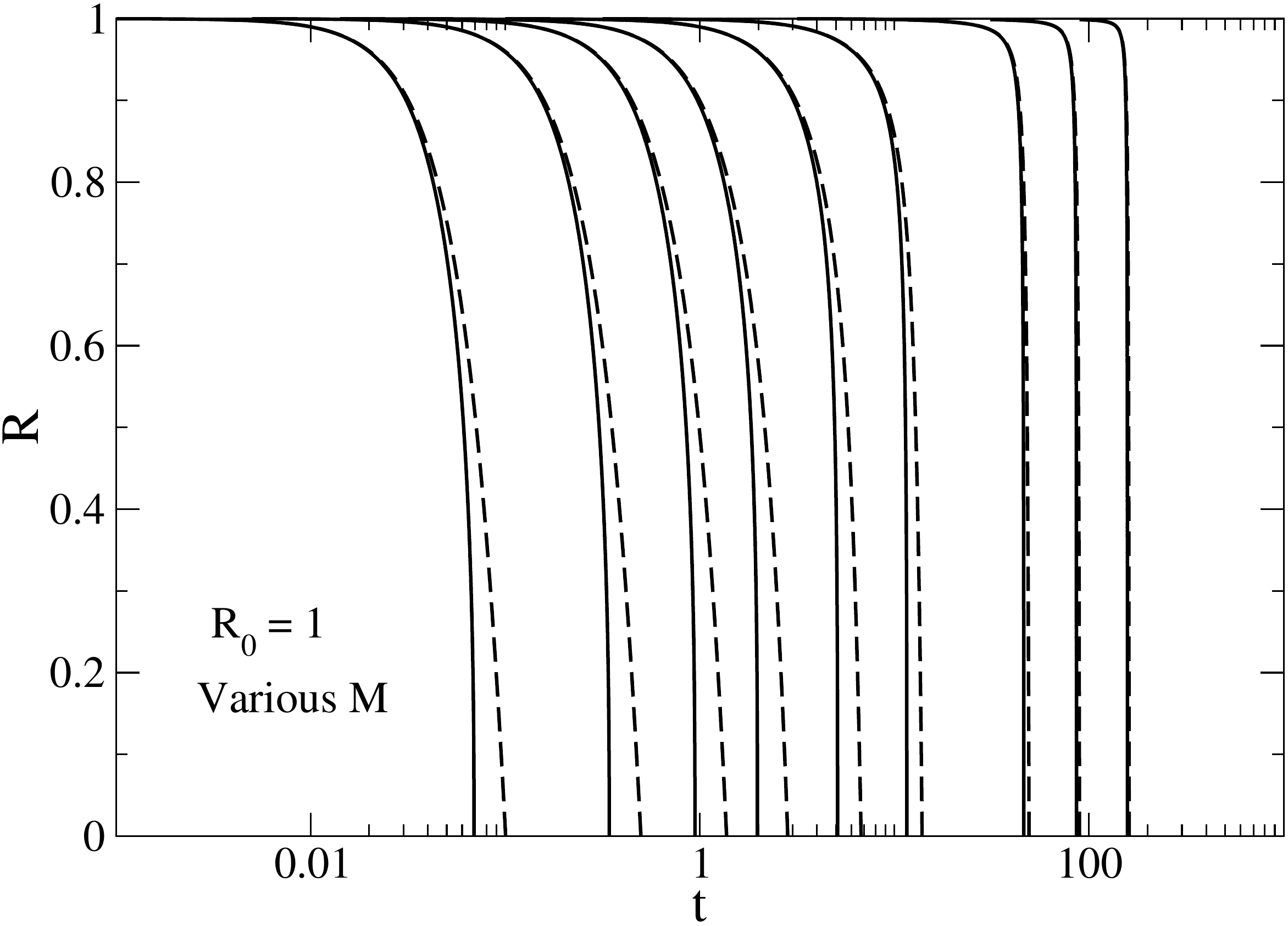}
\caption{Collapse of the self-gravitating BEC starting from $R_0=1$ for
different values of the mass $M=100, 5, 1.5, 1.1, 1.01, 1.001, 1.00001,
1.000001, 1.0000001$ (left to right). The dashed lines correspond to the
invariant profile of Appendix \ref{sec_sst} [see Eq. (\ref{sst5})] valid close
to the critical point $M=1$.}
\label{trmultiM}
\end{center}
\end{figure}

The evolution
of the radius $R(t)$ of the 
BEC is represented in Fig. \ref{tr} for $M=2$ and $R_0=1$ and in Fig.
\ref{trmultiM} for different values of $M$. The asymptotic
behaviors of $R(t)$ for
$t\rightarrow 0$ and $t\rightarrow t_{\rm coll}$ are given by Eqs. (\ref{bz4})
and (\ref{bco3}). The universal self-similar evolution of the scaled radius
close to the
critical point $(M,R)=(1,1)$ is given by Eqs. (\ref{ssu2}), (\ref{ssd2}) and
(\ref{sst3}) and
illustrated in Figs. \ref{txmu}, \ref{txnegmu}, and \ref{txmore}.

\subsection{Behavior of $R(t)$ for $t\rightarrow 0$}
\label{sec_rtz}

For $t\rightarrow 0$, corresponding to $R\rightarrow R_0$, we can make the
approximation $V(R)\simeq V(R_0)+V'(R_0)(R-R_0)$ with
$V'(R_0)>0$. Equation (\ref{ct2}) becomes
\begin{eqnarray}
\int_{R(t)}^{R_0}\frac{dR}{\sqrt{V'(R_0)(R_0-R)}}=\left
(\frac{2}{M}\right
)^{1/2}t,
\label{bz1}
\end{eqnarray}
leading to 
\begin{eqnarray}
R(t)\simeq R_0-\frac{1}{2M}V'(R_0)t^2.
\label{bz2}
\end{eqnarray}
For the potential of Eq.  (\ref{de4}), Eq. (\ref{bz2}) takes the form 
\begin{eqnarray}
R(t)\simeq R_0-\frac{1}{2}\left\lbrack
-\frac{2}{R_0^3}+\frac{M}{R_0^4}+\frac{M}{R_0^2}\right\rbrack t^2.
\label{bz4}
\end{eqnarray}
In particular, for $R_0=1$, we get
\begin{eqnarray}
R(t)\simeq 1-(M-1) t^2.
\label{bz5}
\end{eqnarray}

\subsection{Behavior of $R(t)$ for $t\rightarrow t_{\rm coll}$}
\label{sec_rti}

For $t\rightarrow t_{\rm coll}$, corresponding to $R\rightarrow 0$, using the
expression  (\ref{de4}) of the potential,  we can make the
approximation
\begin{eqnarray}
V(R_0)-V(R)\simeq \frac{M^2}{3R^3}
\label{bco2}
\end{eqnarray}
in Eq.  (\ref{ct4}).
In this limit, the self-gravity and the quantum force are negligible with
respect
to the attractive self-interaction. Performing the integral in Eq. (\ref{ct4})
with the
approximation of Eq. (\ref{bco2}), we obtain
\begin{eqnarray}
R(t)\sim \left (\frac{25}{6}\right )^{1/5}M^{1/5} (t_{\rm coll}-t)^{2/5}.
\label{bco3}
\end{eqnarray}
We note that this scaling is different from the scaling
$R(t)\propto (t_{\rm coll}-t)^{1/2}$ obtained by exactly
solving the nongravitational GP equation with an attractive
self-interaction \cite{sulem}. This shows that the Gaussian ansatz is not
always accurate.

\subsection{Behavior of $t_{\rm coll}(M,R_0)$ for $R_0\rightarrow 0$}
\label{sec_asz}

For fixed $M$ and $R_0\rightarrow 0$, we can neglect the quantum potential
($\propto 1/R^2$) and the self-gravity ($\propto 1/R$) in front of the
self-interaction ($\propto 1/R^3$). In that case, Eq. (\ref{ct5}) reduces to  
\begin{eqnarray}
t_{\rm
coll}\sim \sqrt{\frac{M}{2}}\int_{0}^{R_0}\frac{dR}{\sqrt{-\frac{M^2}
{3R_0^3}+\frac{M^2}{3R^3}}}.
\label{rs1}
\end{eqnarray}
This can be rewritten as
\begin{eqnarray}
t_{\rm
coll}\sim
\sqrt{\frac{3}{2M}}R_0^{5/2}\int_{0}^{1}\frac{dx}{\sqrt{\frac{1}{x^3}-1}}.
\label{rs2}
\end{eqnarray}
Using
\begin{eqnarray}
\int_{0}^{1}\frac{dx}{\sqrt{\frac{1}{x^3}-1}}=\frac{\sqrt{\pi}\,\Gamma(5/6)}{
\Gamma(1/3)}=0.746834...,
\label{rs3}
\end{eqnarray}
we obtain
\begin{eqnarray}
t_{\rm coll}\sim \sqrt{\frac{3\pi}{2}}\frac{\Gamma(5/6)}{
\Gamma(1/3)}\frac{R_0^{5/2}}{\sqrt{M}}\sim
0.914681...\frac{R_0^{5/2}}{\sqrt{M}}.
\label{rs4}
\end{eqnarray}

\subsection{Behavior of $t_{\rm coll}(M,R_0)$ for $R_0\rightarrow +\infty$}
\label{sec_asi}

For fixed $M$ and $R_0\rightarrow +\infty$, we can neglect the quantum potential
($\propto 1/R^2$) and the self-interaction ($\propto 1/R^3$) in
front of the self-gravity ($\propto 1/R$). In that case, Eq.
(\ref{ct5}) reduces to  
\begin{eqnarray}
t_{\rm
coll}\sim \sqrt{\frac{M}{2}}\int_{0}^{R_0}\frac{dR}{\sqrt{-\frac{M^2}
{R_0}+\frac{M^2}{R}}}.
\label{rl1}
\end{eqnarray}
This can be rewritten as
\begin{eqnarray}
t_{\rm
coll}\sim
\frac{1}{\sqrt{2M}}R_0^{3/2}\int_{0}^{1}\frac{dx}{\sqrt{\frac{1}{x}-1}}.
\label{rl2}
\end{eqnarray}
Using
\begin{eqnarray}
\int_{0}^{1}\frac{dx}{\sqrt{\frac{1}{x}-1}}=\frac{\pi}{2},
\label{rl3}
\end{eqnarray}
we obtain
\begin{eqnarray}
t_{\rm coll}\sim \frac{\pi}{2\sqrt{2}}\frac{R_0^{3/2}}{\sqrt{M}}\sim 
{1.11072...}\frac{R_0^{3/2}}{\sqrt{M}}.
\label{rl4}
\end{eqnarray}

\section{The Thomas-Fermi limit $M\rightarrow +\infty$}
\label{sec_tf}

The limit $M\rightarrow +\infty$ corresponds to the  TF
approximation in which  the quantum potential ($\propto M$) can be
neglected
in front of the self-gravity  ($\propto M^2$) and the self-interaction  
($\propto M^2$).

\subsection{The evolution of the radius of the BEC}
\label{sec_tfa}

In the TF limit, the
evolution of the radius of the BEC is given by
\begin{eqnarray}
\int_{R(t)}^{R_0}\frac{dR}{\sqrt{-\frac{1}{
3R_0^3 }
-\frac {1}{R_0}+\frac{1}{3R^3}+\frac{1}{R}
}}=\sqrt{2M}t.
\label{la5}
\end{eqnarray}
This equation describes the collapse of the BEC under the action of the
self-interaction and the self-gravity when $E_{\rm tot}<0$. If $R_0\ll 1$, we
can neglect the
self-gravity in front of the self-interaction and we obtain
\begin{eqnarray}
\int_{\frac{R(t)}{R_0}}^{1}\frac{dx}{\sqrt{-1+\frac{1}{x^3}}}=\sqrt{\frac{2M}{
3R_0^5}} t.
\label{la6}
\end{eqnarray}
This equation describes the collapse of the BEC under the action of the
self-interaction alone when $E_{\rm tot}<0$ (this asymptotic
limit $R_0\ll 1$
is actually valid for all $M$). The integral has the analytical
expression
\begin{equation}
\int_a^1\frac{dx}{\sqrt{-1+\frac{1}{x^3}
}}=\frac{\sqrt{\pi}\Gamma(5/6)}{\Gamma(1/3)}-\frac{2}{5}a^{5/2}\,
_2F_1\left (\frac{1}{2},\frac{5}{6},\frac{11}{6},a^3\right ).
\label{la6b}
\end{equation}
Combining Eqs. (\ref{la6}) and (\ref{la6b}), we obtain the evolution of the
radius of the BEC under the form $t=t(R)$. If $R_0\gg 1$, we can neglect the
self-interaction in
front of the self-gravity for sufficiently short times and we obtain 
\begin{eqnarray}
\int_{\frac{R(t)}{R_0}}^{1}\frac{dx}{\sqrt{-1+\frac{1}{x}
}}=\sqrt{\frac{2M}{R_0^3}}t.
\label{la7}
\end{eqnarray}
This equation describes the collapse of the BEC under the action of the
self-gravity alone when $E_{\rm tot}<0$ (this asymptotic limit
$R_0\gg 1$
is actually valid for all $M$). The
integral has the analytical
expression
\begin{eqnarray}
\int_a^1 \frac{dx}{\sqrt{-1+\frac{1}{x}
}}=\sqrt{a(1-a)}+\cos^{-1}(\sqrt{a}).
\label{la8}
\end{eqnarray}
Combining Eqs. (\ref{la7}) and (\ref{la8}), we obtain the evolution of the
radius of the BEC under the form $t=t(R)$. This solution, which describes the
gravitational collapse of a homogeneous sphere with a
negative energy  $E_{\rm tot}<0$ was first found by
Mestel \cite{mestel}. It is usually written in parametric form
as
\begin{eqnarray}
R=R_0\cos^2\theta,\qquad
\sqrt{\frac{2M}{R_0^3}}t=\theta+\frac{1}{2}\sin(2\theta),
\label{penston1}
\end{eqnarray}
where $\theta$ is a parameter going from $\theta=0$ to $\theta=\pi/2$
(collapse). This solution also occurs in cosmology. It describes the expansion,
then the collapse, of a pressureless FLRW universe with curvature $k=+1$
\cite{weinbergbook}.
It is usually written in parametric form as
\begin{eqnarray}
R=\frac{1}{2}R_0(1-\cos\Theta),\qquad
\sqrt{\frac{2M}{R_0^3}}t=\frac{1}{2}(\Theta-\sin\Theta-\pi),\nonumber\\
\label{penston2}
\end{eqnarray}
where $\Theta$ is a parameter going from $\Theta=0$ (Big Bang) to $\Theta=2\pi$
(Big Crunch) passing by $\Theta=\pi$ (maximum expansion). This is the equation
of a cycloid. The solutions of Eqs.
(\ref{penston1}) and (\ref{penston2}) are
related to each other by the change of variables $\Theta=2\theta+\pi$.

\subsection{Behavior of $t_{\rm coll}(M,R_0)$ for $M\rightarrow +\infty$}
\label{sec_mi}

\begin{figure}
\begin{center}
\includegraphics[clip,scale=0.3]{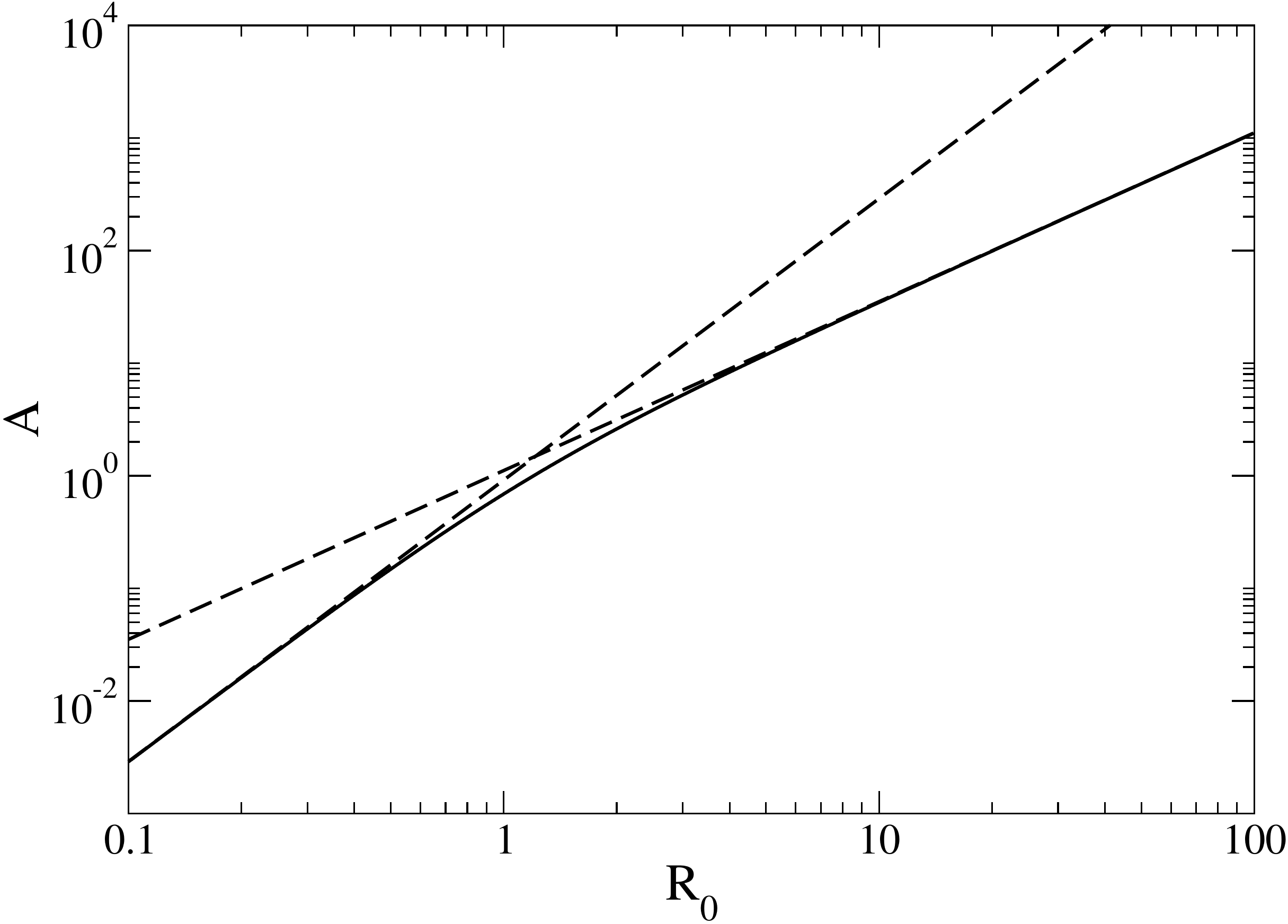}
\caption{The function $A(R_0)$.
}
\label{r0A}
\end{center}
\end{figure}

In the TF limit, the expression of the collapse time given by Eq. (\ref{ct5})
 reduces to 
\begin{eqnarray}
t_{\rm coll}\sim
\frac{A(R_0)}{M^{1/2}}
\label{la1}
\end{eqnarray}
with
\begin{eqnarray}
A(R_0)=\frac{1}{\sqrt{2}}\int_{0}^{R_0}\frac{dR}{\sqrt{-\frac{1}{
3R_0^3 }
-\frac {1}{R_0}+\frac{1}{3R^3}+\frac{1}{R}
}}.
\label{la2}
\end{eqnarray}
This function is plotted in Fig.
\ref{r0A}.
For  $R_0=1$, we find $A(1)=0.688033...$. For $R_0\rightarrow 0$, we can neglect
the self-gravity with
respect to the self-interaction
and obtain
\begin{eqnarray}
A(R_0)
\sim  \sqrt{\frac{3\pi}{2}}\frac{\Gamma(5/6)}{\Gamma(1/3)}R_0^{5/2}\sim
0.914681... R_0^{5/2},
\label{la4}
\end{eqnarray}
where we have used Eq. (\ref{rs3}). For $R_0\rightarrow +\infty$, we
can neglect the self-interaction with respect to the self-gravity and  obtain
\begin{eqnarray}
A(R_0)\sim 
\frac{\pi}{2\sqrt{2}}R_0^{3/2}\sim 1.11072... R_0^{3/2},
\label{la3}
\end{eqnarray}
where we have used Eq. (\ref{rl3}). This result can also be
directly obtained from Eq. (\ref{penston1}) with $\theta=\pi/2$. The asymptotic
limits $R_0\rightarrow 0$ and $R_0\rightarrow +\infty$ are actually valid for
all $M$ as shown in Secs. \ref{sec_asz} and \ref{sec_asi}.

\section{Collapse of the BEC when started close to the critical point
$M\rightarrow M_{\rm max}^+$}
\label{sec_saddle}

\subsection{The normal form of the potential}

\begin{figure}
\begin{center}
\includegraphics[clip,scale=0.3]{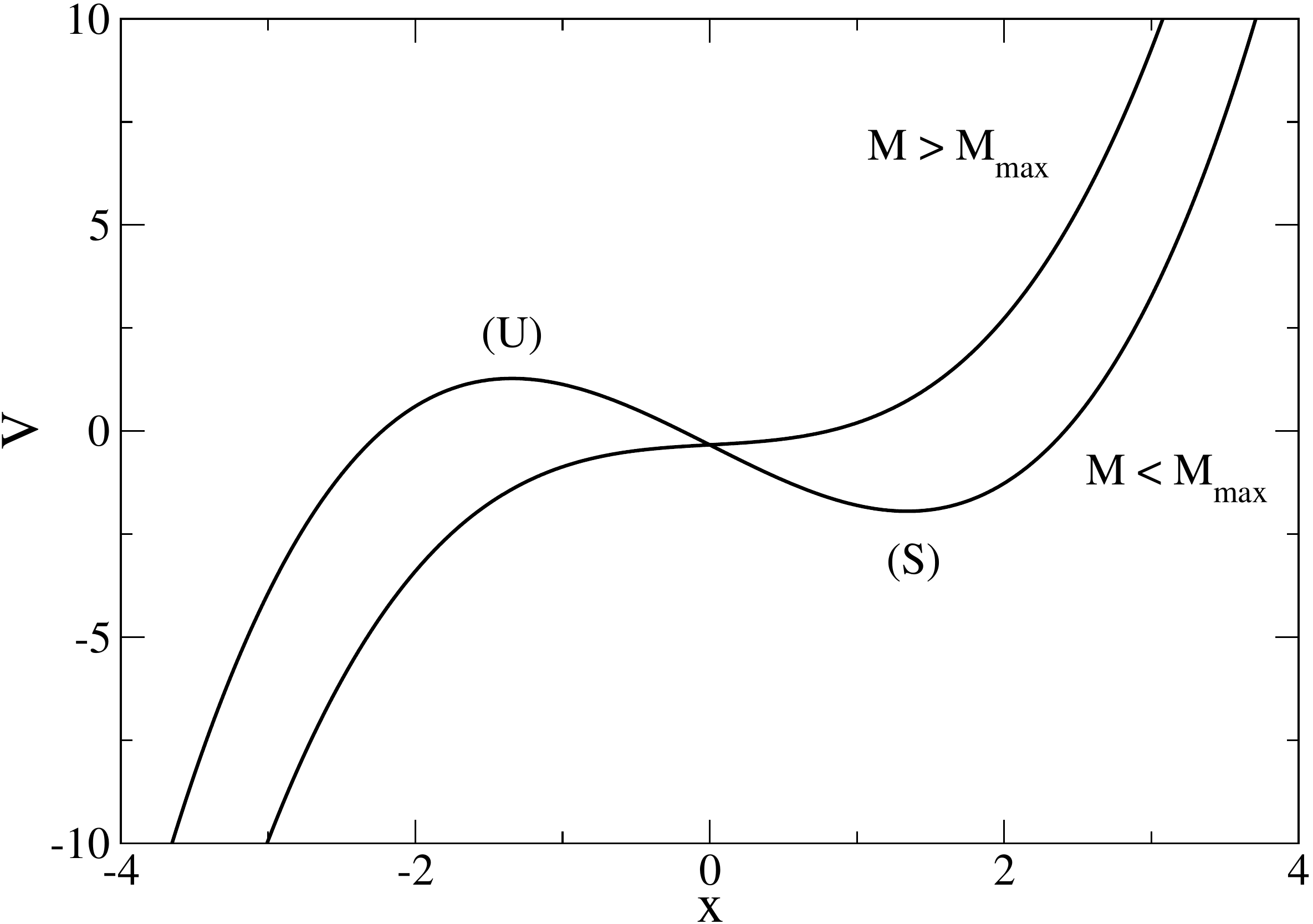}
\caption{Normal form of the effective potential close to the critical point
corresponding to a saddle-node bifurcation.}
\label{saddlenode}
\end{center}
\end{figure}

We consider the collapse of the BEC when $R_0\rightarrow R_*=1$
and $M\rightarrow M_{\rm max}^+=1^+$.
Since we are close to the critical point, corresponding to a saddle-node
bifurcation,  we can approximate the potential $V(R)$ by its normal
form\footnote{The terms $-(4/3)(M-1)^2$, $2(M-1)^2x$ and $-3(M-1)x^2$ in the
potential turn out to be always subdominant in our applications so, for
simplicity, we do not consider them.} 
\begin{eqnarray}
V(x)=-\frac{1}{3}-\frac{5}{3}(M-1)-2(1-M)x+\frac{1}{3}x^3,
\label{saddle1}
\end{eqnarray}
where we have
written $x=R-1$. This
approximation is valid when $M\rightarrow 1$ and $x\ll
1$. The effective potential of Eq. (\ref{saddle1}) is plotted in Fig.
\ref{saddlenode} for
illustration. In this approximation, for $M<M_{\rm max}=1$, the equilibrium
states, corresponding to $V'(x_e)=0$, are
given by  $x_e=\pm \sqrt{2(1-M)}$. This returns the expression (\ref{mr3}) of
the mass-radius relation close to the critical point. We also recover the
correct expression (\ref{te4})
of the total energy. On the other hand,
$V''(x_e)=\pm 2\sqrt{2(1-M)}$. This immediately shows that the solution
$x_e=+\sqrt{2(1-M)}$ is a
minimum (stable) and
the solution $x_e=-\sqrt{2(1-M)}$ is a maximum (unstable). Finally, the
square of the complex
pulsation is given by $\omega^2(x_e)=V''(x_e)=\pm
2\sqrt{2(1-M)}$ which returns the result of Eq.
(\ref{pul4}).

The equation of motion corresponding to the normal form (\ref{saddle1}) of the
potential is
\begin{eqnarray}
\frac{d^2x}{dt^2}=-V'(x)=2(1-M)-x^2.
\label{saddle2}
\end{eqnarray}
Its first integral is 
\begin{eqnarray}
E_{\rm tot}=\frac{1}{2}\left ( \frac{dx}{dt}\right )^2+V(x).
\label{saddle2b}
\end{eqnarray}
We consider $M\ge 1$. Assuming  $\dot x=0$ and $x=x_0$ at $t=0$, we obtain
$E_{\rm
tot}=V(x_0)$ so that Eq. (\ref{saddle2b}) yields
\begin{eqnarray}
\frac{dx}{dt}=-\sqrt{2[V(x_0)-V(x)]},
\label{saddle3}
\end{eqnarray}
where the sign has been chosen so that $x(t)$ decreases with time since 
we are considering a collapse solution. Integrating this relation between $0$
and $t$, we obtain
\begin{eqnarray}
\int_{x(t)}^{x_0} \frac{dx}{\sqrt{2[V(x_0)-V(x)]}}=t.
\label{saddle4}
\end{eqnarray}
Substituting the expression of $V(x)$ from Eq. (\ref{saddle1}) into Eq.
(\ref{saddle4}), we get
\begin{eqnarray}
\int_{x(t)}^{x_0} \frac{dx}{\sqrt{4(M-1)(x_0-x)+\frac{2}{3}(x_0^3-x^3)}}=t.
\label{saddle5}
\end{eqnarray}
The effective particle released from $x_0$ without initial velocity ($\dot
x=0$) runs down the potential in the direction of increasingly negative values
of $x$. The collapse time, corresponding to $x\rightarrow -\infty$,  is given by
\begin{eqnarray}
t_{\rm coll}=\int_{-\infty}^{x_0}
\frac{dx}{\sqrt{4(M-1)(x_0-x)+\frac{2}{3}(x_0^3-x^3)}}.
\label{saddle5b}
\end{eqnarray}
Using Eq. (\ref{saddle5b}), we can rewrite Eq. (\ref{saddle5}) in the form
\begin{eqnarray}
\int_{-\infty}^{x(t)}
\frac{dx}{\sqrt{4(M-1)(x_0-x)+\frac{2}{3}(x_0^3-x^3)}}=t_{\rm
coll}-t.\nonumber\\
\label{saddle9}
\end{eqnarray}
For $t\rightarrow 0$,
corresponding to
$x\rightarrow x_0$, we can make the approximation $V(x)\simeq
V(x_0)+V'(x_0)(x-x_0)$ so
that Eq.  (\ref{saddle4}) can be integrated into
\begin{eqnarray}
x\simeq x_0 -\frac{1}{2}V'(x_0) t^2.
\label{saddle15}
\end{eqnarray}
For the potential of Eq. (\ref{saddle1}), we get
\begin{eqnarray}
x\simeq x_0 -\frac{1}{2} [2(M-1)+x_0^2] t^2.
\label{saddle15b}
\end{eqnarray}
This expression is consistent with the general result of
Eq. (\ref{bz4}) valid for {\it any} $M>1$ and any $R_0$. For
$t\rightarrow t_{\rm coll}$, corresponding to $x\rightarrow -\infty$, we
can make the approximation $V(x_0)-V(x)\simeq -x^3/3$ so that Eq. 
(\ref{saddle4}) can be
integrated into
\begin{eqnarray}
x\sim -\frac{6}{(t_{\rm coll}-t)^2}.
\label{saddle16}
\end{eqnarray}
The evolution of the radius of the BEC  $x(t;M,x_0)$ given by Eq.
(\ref{saddle5}) and the collapse time
$t_{\rm coll}(M,x_0)$ given by Eq. (\ref{saddle5b}) can be
put in a self-similar form as shown in Appendix \ref{sec_sscp} and
illustrated in Figs. \ref{txmu}, \ref{txnegmu},
\ref{txmore} and in Figs.  \ref{ssfplus}, \ref{ssg}, \ref{ssgR0neg}.

\subsection{The case $M=1$ and $R_0\rightarrow 1$}
\label{sec_uu}

For $M=1$, the collapse time diverges when $R_0\rightarrow 1$
because the effective potential given by Eq. (\ref{de4}) presents an inflexion
point
at $R=1$ when $M=1$. From Eq. (\ref{saddle5b}), we obtain
\begin{eqnarray}
t_{\rm coll}(M=1,R_0)\sim
\sqrt{\frac{3}{2}}\int_{-\infty}^{x_0}\frac{dx}{\sqrt{x_0^3-x^3}}.
\label{ru1}
\end{eqnarray}
Assuming $x_0>0$ and making the change of variables $y=x/x_0$, we
get
\begin{eqnarray}
t_{\rm coll}(M=1,R_0)\sim
\sqrt{\frac{3}{2}}\frac{1}{x_0^{1/2}}\int_{-\infty}^{1}\frac{dy}{\sqrt{
1-y^3}}.
\label{ru2}
\end{eqnarray}
Therefore, when $R_0\rightarrow 1^+$:
\begin{eqnarray}
t_{\rm coll}(M=1,R_0)\sim K_+ (R_0-1)^{-1/2}
\label{ru3}
\end{eqnarray}
with 
\begin{eqnarray}
K_+=\sqrt{\frac{3}{2}}\int_{-\infty}^{1}\frac{dy}{\sqrt{
1-y^3}}=\sqrt{\frac{3\pi}{2}}\frac{\Gamma(1/3)}{\Gamma(5/6)}
=5.15195...\nonumber\\
\label{ru4}
\end{eqnarray}
Assuming $x_0<0$ and making the change of
variables $y=x/|x_0|$, we
get
\begin{eqnarray}
t_{\rm coll}(M=1,R_0)\sim
\sqrt{\frac{3}{2}}\frac{1}{|x_0|^{1/2}}\int_{-\infty}^{-1}\frac{dy}{\sqrt{
-1-y^3}}.
\label{ru5}
\end{eqnarray}
Therefore, when $R_0\rightarrow 1^-$:
\begin{eqnarray}
t_{\rm coll}(M=1,R_0)\sim K_- (1-R_0)^{-1/2}
\label{ru6}
\end{eqnarray}
with
\begin{eqnarray}
K_-=\sqrt{\frac{3}{2}}\int_{-\infty}^{-1}\frac{dy}{\sqrt{
-1-y^3}}=\sqrt{6\pi}\frac{\Gamma(7/6)}{\Gamma(2/3)}=2.97448...\nonumber\\
\label{ru7}
\end{eqnarray}

\subsection{The case $R_0=1$ and $M\rightarrow 1^+$}
\label{sec_dd}

For $R_0=1$, the collapse time diverges when $M\rightarrow 1^+$ because the
effective potential given by Eq. (\ref{de4}) presents an inflexion point
at $R=1$ when $M=1$. From Eq. (\ref{saddle5b}), we obtain
\begin{eqnarray}
t_{\rm coll}(M,R_0=1)=\int_{-\infty}^{0}
\frac{dx}{\sqrt{-4(M-1)x-\frac{2}{3}x^3}}.
\label{saddle5c}
\end{eqnarray}
Making the change of variables $y=x/\sqrt{6(M-1)}$, we find that the collapse
time is given,  when  $M\rightarrow 1^+$, by
\begin{eqnarray}
t_{\rm coll}(M,R_0=1)\sim B (M-1)^{-1/4}
\label{saddle10}
\end{eqnarray}
with
\begin{eqnarray}
B=\left
(\frac{3}{8}\right )^{1/4}\int_{-\infty}^0
\frac{dy}{\sqrt{-y-y^3}}\nonumber\\
=\left
(\frac{3}{8}\right )^{1/4}\int_{-\infty}^0
\frac{du}{\sqrt{-\sinh(u)}}=2.90178...
\label{saddle11}
\end{eqnarray}
To get the second line, we have made the change of variables $y=\sinh(u)$.
For $M>M_{\rm max}$, the collapse time diverges at the
critical
point
as $t_{\rm coll}\sim 2.90178... (M-1)^{-1/4}$. Interestingly, this is the same
scaling as for the divergence of the period of the oscillations $2\pi/\omega\sim
3.73600... (1-M)^{-1/4}$
close
to the critical
point for $M<M_{\rm max}$ [see Eq. (\ref{pul4})]. The prefactor is, however,
different being
$2.90178...$ for $t_{\rm coll}$ ($M>M_{\rm max}$) and $3.73600...$
for $2\pi/\omega$ ($M<M_{\rm max}$). 

\subsection{The Painlev\'e equation}
\label{sec_pain}

If the mass $M$ of the system increases with time as $M(t)=1+at/2$ (for
example by accreting matter around it or because of a continuous inflow of
particles), the equation of motion (\ref{saddle2}) can be rewritten as
\begin{eqnarray}
\frac{d^2x}{dt^2}=-at-x^2,
\label{pain1}
\end{eqnarray}
where the origin of times has been chosen such that $M\le M_{\rm max}$ for
$t\le 0$ and $M\ge M_{\rm max}$ for
$t\ge 0$. In this way, an initially stable system loses its stability at $t=0$
and collapses. Equation (\ref{pain1}) is the celebrated
Painlev\'e I equation. It has been studied in detail in \cite{pomeau} in a model
of supernovae that presents a saddle-center bifurcation similar to our system.

\section{Possible collapse, explosion and oscillations of the BEC when $M<M_{\rm
max}$}
\label{sec_possible}

\subsection{The critical mass $M_c$}

\begin{figure}
\begin{center}
\includegraphics[clip,scale=0.3]{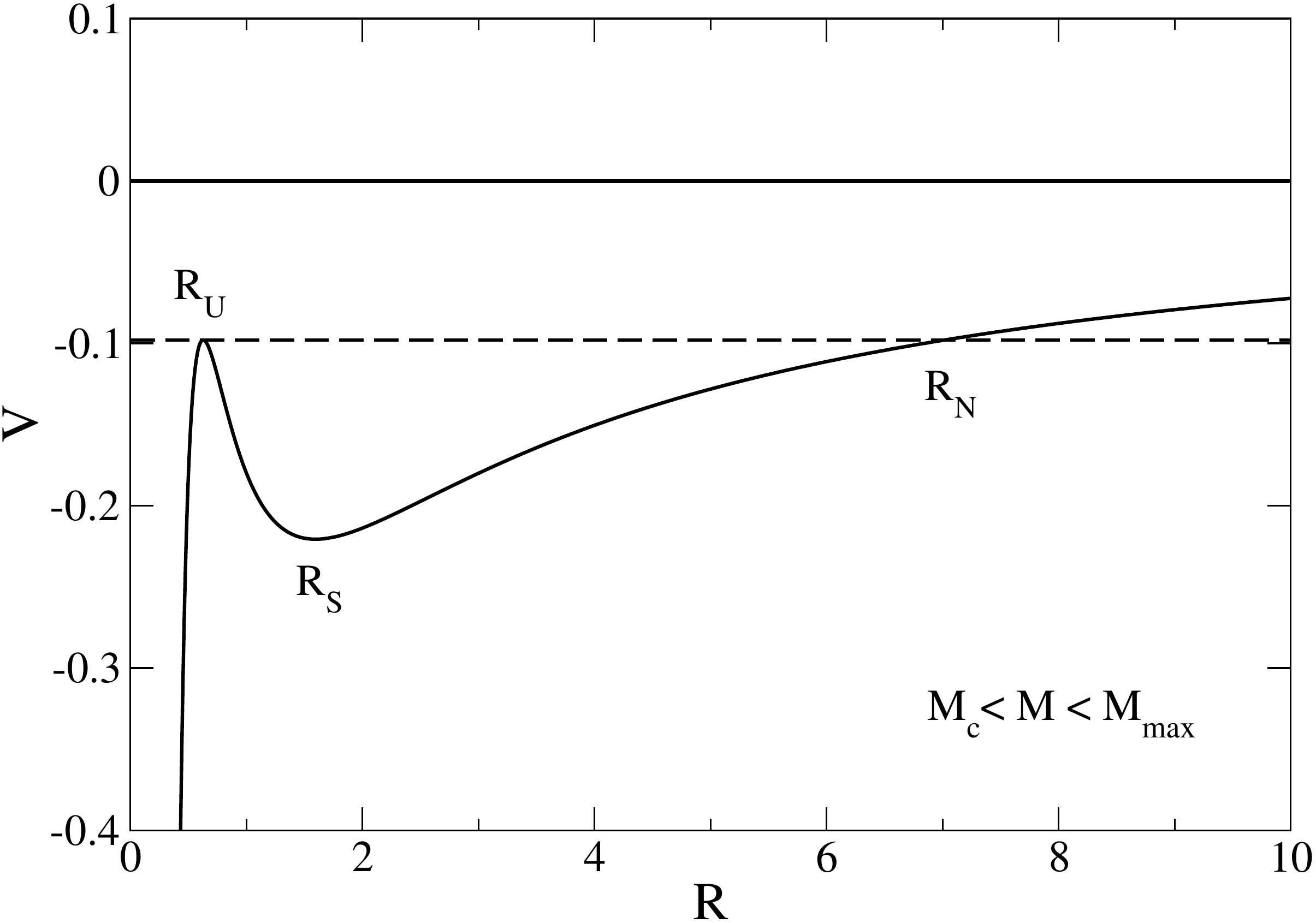}
\caption{Effective potential $V(R)$ as a function of the radius $R$ for
$M_c< M=0.9<M_{\rm max}$. In that case $V_{\rm max}<0$.}
\label{rvM0p9}
\end{center}
\end{figure}

\begin{figure}
\begin{center}
\includegraphics[clip,scale=0.3]{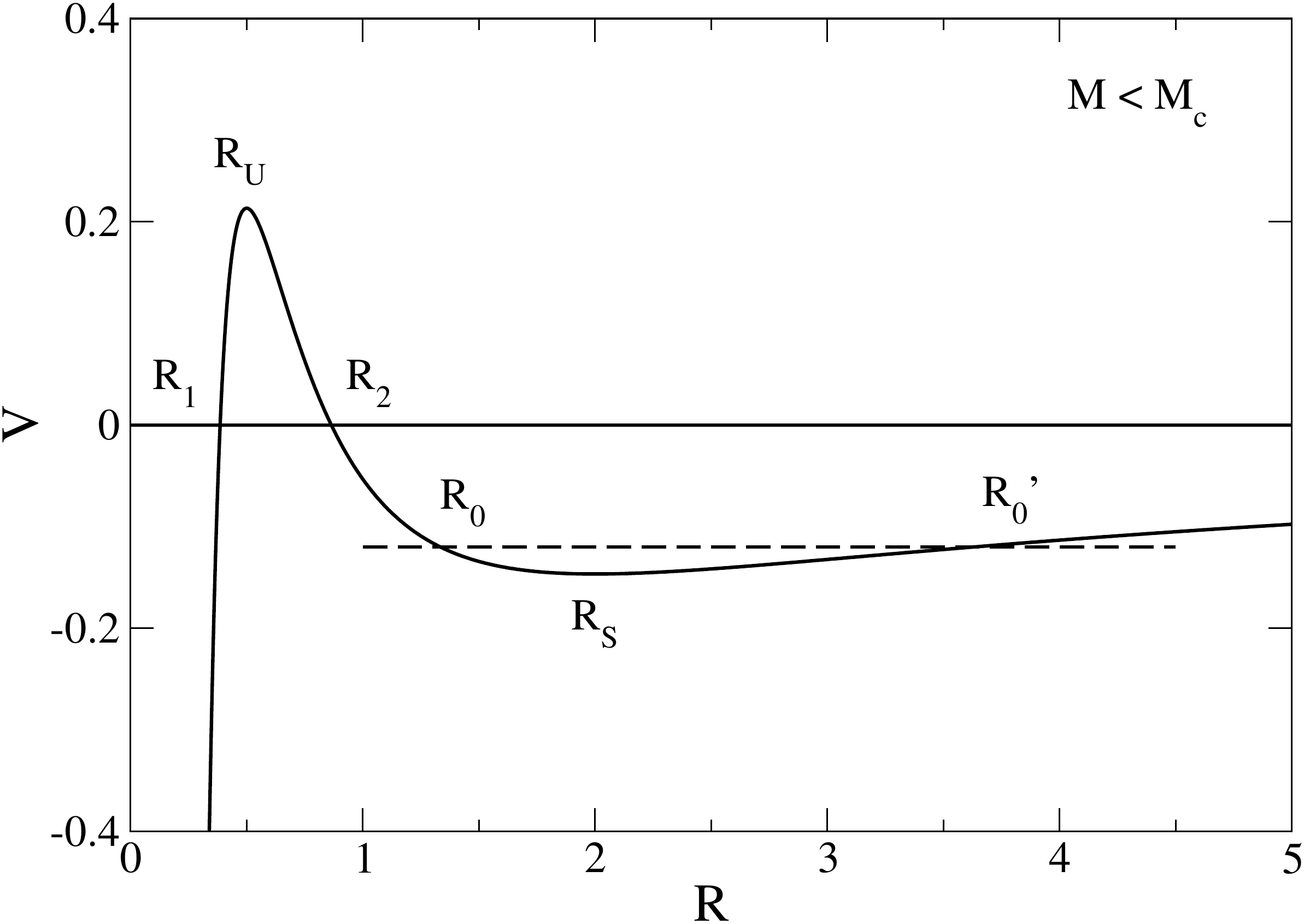}
\caption{Effective potential $V(R)$ as a function of the radius $R$ for
$M=0.8<M_{c}$. In that case $V_{\rm max}>0$.}
\label{rvM0p8}
\end{center}
\end{figure}

When $M<M_{\rm max}$, the effective potential is plotted in Figs. \ref{rvM0p9}
and \ref{rvM0p8}. The (local)
maximum of the potential $V(R)$ corresponds to the energy of the unstable
equilibrium state: $V_{\rm max}=V(R_U)=E_{\rm tot}(R_U)$. According to Eq.
(\ref{te2}),
$V_{\rm
max}$ is positive when $R_U<R_c$ with
\begin{eqnarray}
R_c=\frac{1}{\sqrt{3}}=0.57735....
\label{cm1}
\end{eqnarray}
From Eq. (\ref{mr1}), this corresponds to a mass $M<M_c$ with
\begin{eqnarray}
M_c=\frac{\sqrt{3}}{2}=0.866025....
\label{cm2}
\end{eqnarray}
This is a new critical mass that has not been introduced before. Restoring the
dimensional variables, we have  $M_c=0.9396 \hbar/\sqrt{Gm|a_s|}$ and
$R_{99}^c=2.382 (|a_s|\hbar^2/G m^3)^{1/2}$. For standard
axions with $m=10^{-4}\,
{\rm eV}/c^2$ and
$a_s=-5.8\, 10^{-53}\, {\rm m}$, we obtain
$M_{c}=5.98\times 10^{-14}\,
M_{\odot}$ and $R_c=5.8\times 10^{-5}\, R_{\odot}$. For ultralight axions with
$m=1.93\times 10^{-20}\, {\rm eV}/c^2$ and
$a_s=-8.29\times 10^{-60}\, {\rm fm}$, we
obtain $M_{c}=3.62\times 10^{5}\, M_{\odot}$ and $R_c=6\, {\rm pc}$.

When $M_c<M<M_{\rm max}$, the local maximum of the effective potential is
negative: $V_{\rm max}<0$ (see Fig. \ref{rvM0p9}). If $R_0<R_U$, the BEC
collapses. If
$R_U<R<R_N$ where $R_N(M)$ is such that $V(R_N)=V(R_U)$ the
BEC oscillates about its stable equilibrium
state corresponding to $R=R_S$. If $R>R_N$, the BEC collapses. When slightly
perturbed, the unstable
equilibrium state at $R_U$ (unstable branch) with $E_{\rm max}<0$ can either
collapse to a black hole
or oscillate about the stable equilibrium at $R_S$.

When $M<M_c$, the local maximum of the effective potential is
positive:
$V_{\rm
max}>0$ (see Fig. \ref{rvM0p8}). In that case, we introduce the radii $R_1$ and
$R_2>R_1$ at which
the effective potential vanishes: $V(R_1)=V(R_2)=0$. They are given by 
\begin{eqnarray}
R_1=\frac{1-\sqrt{1-\left (\frac{M}{M_c}\right
)^2}}{2M},
\label{cm3}
\end{eqnarray}
\begin{eqnarray}
R_2=\frac{1+\sqrt{1-\left (\frac{M}{M_c}\right )^2}}{2M}.
\label{cm4}
\end{eqnarray}
If $R_0<R_U$, the BEC collapses. If $R_U<R\le R_2$, the BEC
explodes. If $R>R_2$, the BEC oscillates about its stable equilibrium state 
corresponding to $R=R_S$. When slightly perturbed, the unstable
equilibrium state at $R_U$ (unstable branch) with $E_{\rm max}>0$ can either
collapse to a black hole or explode and disperse away.

\subsection{Collapse of the BEC}

\subsubsection{Collapse time when $R_0\rightarrow R_U^-$}
\label{sec_qw}

When $R_0<R_U$, the BEC collapses. The results of Sec. \ref{sec_collapse} still
apply. The
collapse time diverges when $R_0\rightarrow R_U^-$
because the effective potential given by Eq. (\ref{de4}) presents a maximum
at $R_U$. Writing $R_0=R_U-x_0$ and $R=R_U-x$, expanding Eq.
(\ref{ct3}) for
$x_0\ll 1$ and $x\ll 1$, and using $V'(R_U)=0$ and $V''(R_U)<0$, we obtain
\begin{eqnarray}
t_{\rm coll}\sim
\sqrt{\frac{M}{-V''(R_U)}}\int_{x_0}^{R_U}\frac{dx}{\sqrt{x^2-x_0^2}}.
\label{ctu1}
\end{eqnarray}
\begin{figure}
\begin{center}
\includegraphics[clip,scale=0.3]{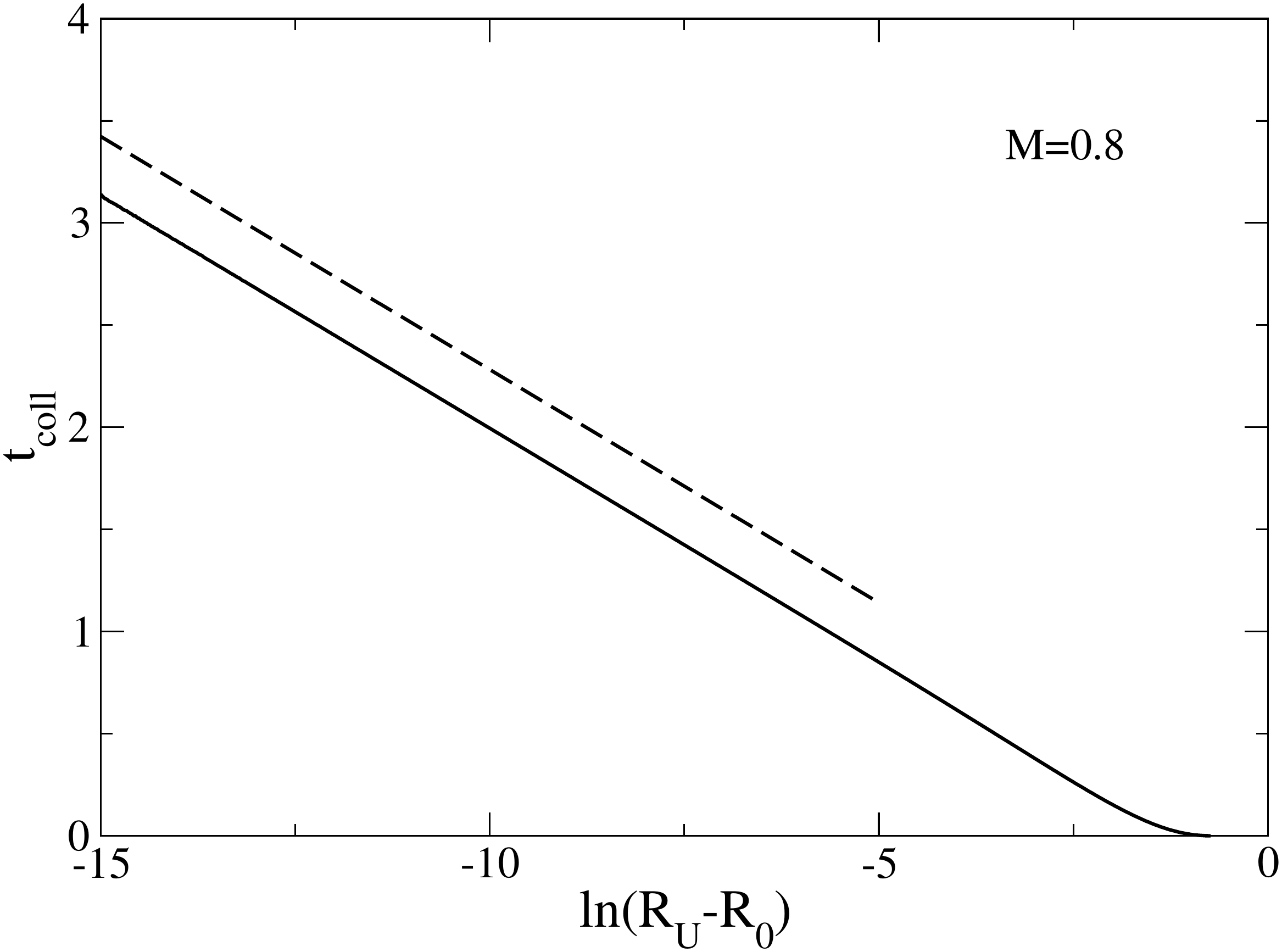}
\caption{Collapse time of the BEC when $R_0\rightarrow R_U^{-}$ (we have taken
$M=0.8$). The dashed line corresponds to the asymptotic expression of Eq.
(\ref{ctu4}).}
\label{tcollsousM0p8}
\end{center}
\end{figure}
Making the change of variables $y=x/x_0$, we
get
\begin{eqnarray}
t_{\rm coll}\sim
\sqrt{\frac{M}{-V''(R_U)}}\int_{1}^{R_U/x_0}\frac{dy}{\sqrt{
y^2-1}}.
\label{ctu2}
\end{eqnarray}
Using
\begin{eqnarray}
\int_{1}^{z}\frac{dy}{\sqrt{
y^2-1}}=\ln\left (z+\sqrt{z^2-1}\right ),
\label{ctu3}
\end{eqnarray}
we obtain for $R_0\rightarrow R_U^-$:
\begin{eqnarray}
t_{\rm coll}\sim
-\sqrt{\frac{M}{-V''(R_U)}}\ln\left
(\frac{R_U-R_0}{2R_U}\right ).
\label{ctu4}
\end{eqnarray}
The divergence is logarithmic.  The collapse time is plotted as a function of
$\ln(R_U-R_0)$ in Fig. \ref{tcollsousM0p8}, and compared with the asymptotic
result of Eq. (\ref{ctu4}). This formula is only valid at leading order which
explains the offset with the exact solution on this figure.

{\it Remark:} For $M=M_c$, we have $R_U=1/\sqrt{3}$ and $V''(R_U)=-9\sqrt{3}/2$.

\subsubsection{Collapse time when $M_c<M<M_{\rm max}$ and $R_0\rightarrow
R_N^+$}
\label{sec_fio}

We consider the case $M_c<M<M_{\rm max}$. When
$R_0>R_N$, the BEC collapses. The results of Sec. \ref{sec_collapse}
still apply. The collapse
time diverges when $R_0\rightarrow R_N^+$
because the effective potential given by Eq. (\ref{de4}) presents a maximum
at $R_U$. Therefore, the effective particle takes an
infinitly long time to reach that maximum before collapsing. Writing
$R_0=R_N+\epsilon$ and $R=R_U+x$, expanding Eq.
(\ref{ct3}) for
$\epsilon\ll 1$ and $x\ll 1$, and using $V'(R_U)=0$,  $V''(R_U)<0$
 and $V'(R_N)>0$, we obtain  
\begin{eqnarray}
t_{\rm coll}\sim
\sqrt{\frac{M}{2}}\int_{-R_U}^{R_N-R_U+\epsilon}\frac{dx}{\sqrt{
V'(R_N)\epsilon-V''(R_U)\frac{x^2}{2}} }.\nonumber\\
\label{ctn1}
\end{eqnarray}
Making the change of variables 
\begin{eqnarray}
y=\sqrt{\frac{-V''(R_U)}{2V'(R_N)\epsilon}}x,
\label{ctn2}
\end{eqnarray}
we get
\begin{eqnarray}
t_{\rm coll}\sim
\sqrt{\frac{M}{-V''(R_U)}}\int_{-R_U\sqrt{\frac{-V''(R_U)}{2V'(R_N)\epsilon}}}^{
(R_N-R_U)\sqrt{\frac{-V''(R_U)}{
2V'(R_N)\epsilon}}}\frac{dy}{\sqrt{1+y^2}}.\nonumber\\
\label{ctn3}
\end{eqnarray}
Using
\begin{eqnarray}
\int\frac{dy}{\sqrt{1+y^2}}=\sinh^{-1}(y),
\label{ctn4}
\end{eqnarray}
the forgoing equation can be rewritten as
\begin{eqnarray}
t_{\rm coll}\sim
\sqrt{\frac{M}{-V''(R_U)}}\Biggl\lbrace \sinh^{-1}\left
\lbrack(R_N-R_U)\sqrt{\frac{-V''(R_U)}{ 2V'(R_N)\epsilon}}\right
\rbrack\nonumber\\
+\sinh^{-1}\left
(R_U\sqrt{\frac{-V''(R_U)}{ 2V'(R_N)\epsilon}}\right )\Biggr\rbrace.\qquad
\label{ctn5}
\end{eqnarray}
Using the equivalent $\sinh^{-1}(y)\sim \ln(2y)$
valid for $y\rightarrow +\infty$, we finally obtain
\begin{eqnarray}
t_{\rm coll}\sim
\sqrt{\frac{M}{-V''(R_U)}}\ln\left\lbrack
-\frac{4(R_N-R_U)R_UV''(R_U)}{2V'(R_N)(R_0-R_N)}\right\rbrack.\nonumber\\
\label{ctn6}
\end{eqnarray}
The divergence is logarithmic. This formula is only
valid at leading order.

{\it Remark:} For $M\rightarrow M_c$, we have $R_U\rightarrow 1/\sqrt{3}$,
$V''(R_U)\rightarrow -9\sqrt{3}/2$, $R_N\sim M_c^2/[3(M-M_c)]$ and $V'(R_N)\sim
9(M-M_c)^2/M_c^2$. To obtain these asymptotic results, we have used $R_U\simeq
1/\sqrt{3}+(4/3)(M-M_c)$, $V(R_U)\sim -3(M-M_c)$  for $M\rightarrow M_c$ and
$V(R)\sim -M^2/R$ for $R\rightarrow +\infty$.

\subsection{Explosion of the BEC}

We consider the case $M<M_c$. We assume  $R_U<R_0\le R_2$ and $\dot
R_0=0$ (implying $E_{\rm tot}=V(R_0)$) leading to the explosion of the
BEC. In that case, according to Eq. (\ref{de7}), the evolution of the radius
$R(t)$ of the BEC is given by 
\begin{eqnarray}
\int_{R_0}^{R(t)}\frac{dR}{\sqrt{V(R_0)-V(R)}}=\left (\frac{2}{M}\right
)^{1/2} t,
\label{e1}
\end{eqnarray}
where the sign has been chosen so that $R(t)$ increases with time since we are
considering an explosive solution.

\subsubsection{Behavior of $R(t)$ for $t\rightarrow 0$}

For $t\rightarrow 0$, corresponding to $R\rightarrow R_0$, we can make the
approximation $V(R)\simeq V(R_0)+V'(R_0)(R-R_0)$ with
$V'(R_0)<0$ so that Eq. (\ref{e1}) becomes
\begin{eqnarray}
\int_{R_0}^{R(t)}\frac{dR}{\sqrt{-V'(R_0)(R-R_0)}}=\left
(\frac{2}{M}\right
)^{1/2}t
\label{e2}
\end{eqnarray}
leading to 
\begin{eqnarray}
R(t)\simeq R_0-\frac{1}{2M}V'(R_0)t^2.
\label{e3}
\end{eqnarray}
For the potential of Eq.  (\ref{de4}),
Eq. (\ref{e3})
takes the form 
\begin{eqnarray}
R(t)\simeq R_0-\frac{1}{2}\left\lbrack
-\frac{2}{R_0^3}+\frac{M}{R_0^4}+\frac{M}{R_0^2}\right\rbrack t^2.
\label{bz4b}
\end{eqnarray}

\subsubsection{Behavior of $R(t)$ for $t\rightarrow +\infty$ when $R_U<R_0<R_2$}
\label{sec_gppos}

\begin{figure}
\begin{center}
\includegraphics[clip,scale=0.3]{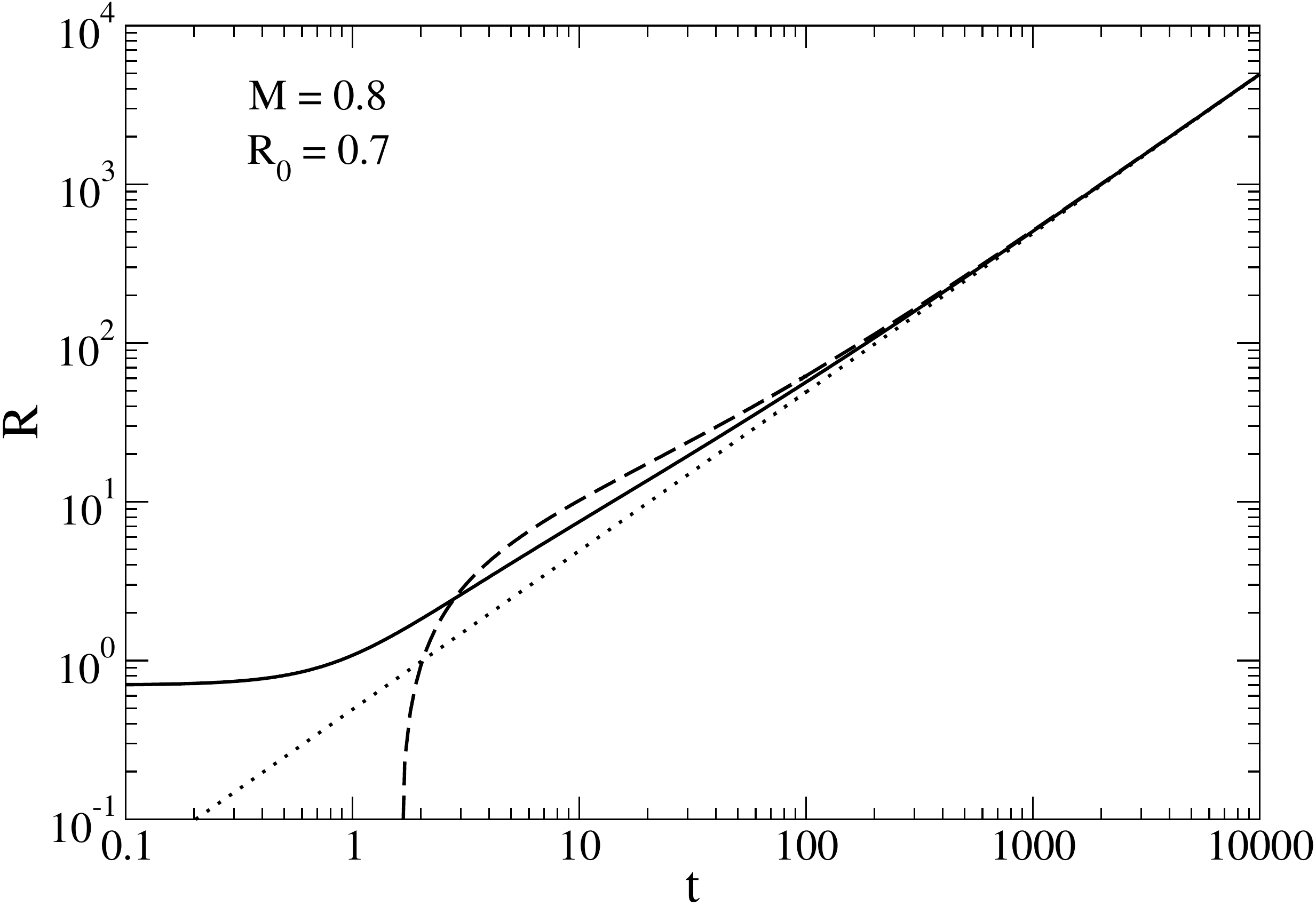}
\caption{Explosion of the BEC for $M=0.8<M_c$ and $R_U<R_0=0.7<R_2$
(for this initial radius $V(R_0)=0.0964025$). The dotted line corresponds to the
asymptotic expression given by Eq. (\ref{e4}) and the dashed line corresponds to
the asymptotic expression given by Eq. (\ref{e8}).}
\label{explosionM0p8a0p7}
\end{center}
\end{figure}

We assume $R_U<R_0<R_2$. For $t\rightarrow +\infty$, corresponding to
$R\rightarrow +\infty$, the potential $V(R)\rightarrow 0$. In this limit, the
effective particle has a free motion so that, at leading order, Eq. (\ref{e1})
leads to 
\begin{eqnarray}
R(t)\sim \sqrt{\frac{2V(R_0)}{M}}t.
\label{e4}
\end{eqnarray}
It possible to take into account the other terms in the potential $V(R)$
perturbatively. For late times, the effective potential is dominated by the
gravity term:
$V(R)\sim -M^2/R$. Therefore, we can rewrite Eq. (\ref{e1}) as
\begin{eqnarray}
\int_{R_0}^{R(t)}\frac{dR}{\sqrt{V(R_0)+\frac{M^2}{R}}}\simeq \left
(\frac{2}{M}\right
)^{1/2} t
\label{e5}
\end{eqnarray}
and treat the gravitational term as a small correction. We
get
\begin{eqnarray}
R(t)- \frac{M^2}{2V(R_0)}\ln R(t)\simeq \sqrt{\frac{2V(R_0)}{M}}t,
\label{e7}
\end{eqnarray}
leading to
\begin{eqnarray}
R(t)\simeq \sqrt{\frac{2V(R_0)}{M}}t+\frac{M^2}{2V(R_0)}\ln \left\lbrack
\sqrt{\frac{2V(R_0)}{M}}t \right\rbrack.
\label{e8}
\end{eqnarray}
This result can also be obtained by evaluating the integral in Eq. (\ref{e5})
with the identity
\begin{eqnarray}
\int\frac{dx}{\sqrt{1+\frac{a}{x}}}=\sqrt{x(a+x)}-a\ln(\sqrt{x}+\sqrt{a+x})
\label{e6}
\end{eqnarray}
and taking the limit $R\rightarrow +\infty$. The temporal evolution of the
radius of the BEC is shown in Fig. \ref{explosionM0p8a0p7}. Equation (\ref{e5})
describes the explosion of a
homogeneous self-gravitating sphere with positive energy  $E_{\rm tot}>0$.
Combining Eqs. (\ref{e5}) and (\ref{e6}), we obtain the evolution of the radius
of the system under the form $t=t(R)$. This
solution also occurs in cosmology. It
describes the expansion of a pressureless FLRW universe with curvature
$k=-1$ \cite{weinbergbook}. It is usually written in parametric form as
\begin{eqnarray}
\frac{V(R_0)}{M^2}R(t)=\frac{1}{2}(\cosh\Psi-1),
\label{penston5}
\end{eqnarray}
\begin{eqnarray}
\frac{V(R_0)}{M^2}\sqrt{ \frac { 2V(R_0) } { M } } t=\frac { 1 } { 2 }
(\sinh\Psi-\Psi-\sinh\Psi_0+\Psi_0),\nonumber\\
\label{penston6}
\end{eqnarray}
where $\Psi$ is a parameter going from $\Psi_0$ to $+\infty$.

{\it Remark:} One can check that the correction due to the quantum potential is
not divergent when $t\rightarrow +\infty$ so the asymptotic result of Eq.
(\ref{e8}) cannot
be improved.

\subsubsection{Behavior of $R(t)$ for $t\rightarrow
+\infty$ when $R_0=R_2$}
\label{sec_gpnul}

We assume that $R_0=R_2$. In that case $E_{\rm tot}=V(R_0)=0$ so the equation of
motion
reduces
to
\begin{eqnarray}
\int_{R_2}^{R(t)}\frac{dR}{\sqrt{-\frac{M}{R^2}+\frac{M^2}{3R^3}+\frac{M^2}{R}}}
=\left (\frac{2}{M}\right
)^{1/2} t.
\label{e9}
\end{eqnarray}

For $t\rightarrow +\infty$, corresponding to $R\rightarrow +\infty$, the
effective potential is dominated by the gravity term. We obtain at leading
order
\begin{eqnarray}
R(t)\sim \left (\frac{9}{2}M\right )^{1/3}t^{2/3}.
\label{e12}
\end{eqnarray}
This solution describes the explosion of a
homogeneous self-gravitating sphere with vanishing energy  $E_{\rm tot}=0$. This
solution also occurs in cosmology. It
describes the expansion of a pressureless FLRW universe without curvature
($k=0$) \cite{weinbergbook}. This
is the famous Einstein-de Sitter (EdS) solution. The $t^{2/3}$ behavior of
$R(t)$ in Eq. (\ref{e12}) when $E_{\rm tot}=0$  can be contrasted from
its linear behavior in Eq.
(\ref{e4}) when $E_{\rm tot}>0$.

\begin{figure}
\begin{center}
\includegraphics[clip,scale=0.3]{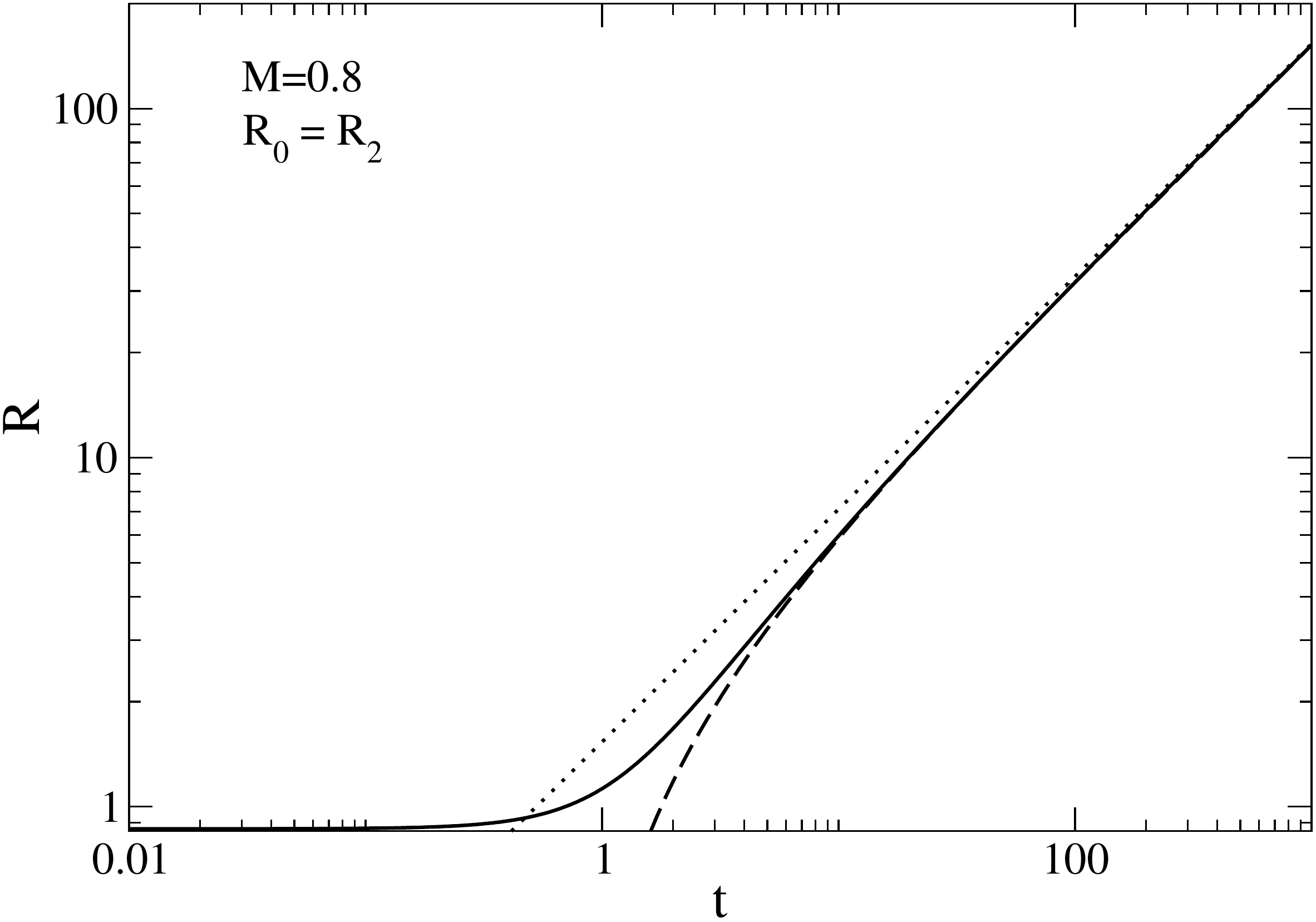}
\caption{Explosion of the BEC for $M=0.8<M_c$ and $R_0=R_2=0.86436...$
(for this initial radius $V(R_0)=0$). The dotted line corresponds to the
asymptotic expression given by Eq. (\ref{e12}) and the dashed line corresponds
to the asymptotic expression given by Eq. (\ref{e13}).}
\label{explosionM0p8a0p86436}
\end{center}
\end{figure}

We can take into account the correction coming from
the quantum potential perturbatively. We start from the equation
\begin{eqnarray}
\int_{R_2}^{R(t)}\frac{dR}{\sqrt{-\frac{M}{R^2}+\frac{M^2}{R}}}
\simeq \left (\frac{2}{M}\right
)^{1/2} t
\label{e10q}
\end{eqnarray}
and treat the quantum potential as a small correction. We obtain 
\begin{eqnarray}
R(t)\simeq \left (\frac{9}{2}M\right )^{1/3}t^{2/3}\left\lbrack
1-\left
(\frac{2}{9}\right )^{1/3}\frac{1}{M^{4/3}}\frac{1}{t^{2/3}}\right\rbrack.
\label{e13}
\end{eqnarray}
This result can also be obtained by evaluating the integral in
Eq. (\ref{e10q})
with the identity
\begin{eqnarray}
\int\frac{dx}{\sqrt{-\frac{1}{x^2}+\frac{a}{x}}}=\frac{2}{3a^2}(2+ax)\sqrt{ax-1}
\label{e6w}
\end{eqnarray}
and taking the limit $R\rightarrow +\infty$. The temporal evolution of the
radius of the BEC is shown in Fig. \ref{explosionM0p8a0p86436}.
We note that Eq. (\ref{e10q})
describes the explosion of the BEC under the action of the quantum potential
and the self-gravity when $E_{\rm tot}=0$.
Combining Eqs. (\ref{e10q}) and (\ref{e6w}), we obtain the evolution of the
radius of the BEC in that case under the form $t=t(R)$.

{\it Remark:} One can check that the correction due to the self-interaction is
not divergent when $t\rightarrow +\infty$ so the asymptotic result of Eq.
(\ref{e13})
cannot
be improved.

\subsection{Oscillations of the BEC}

\begin{figure}
\begin{center}
\includegraphics[clip,scale=0.3]{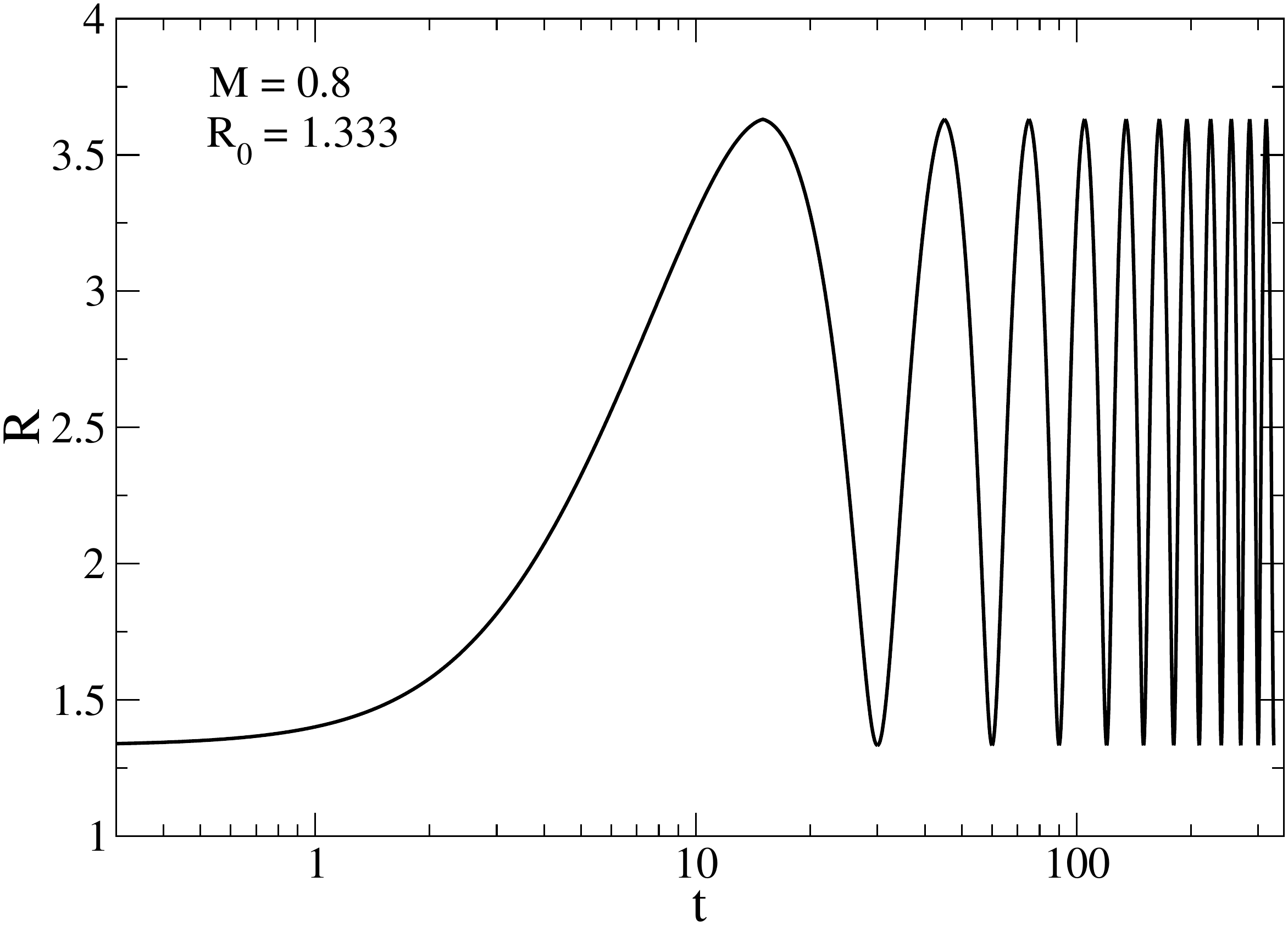}
\caption{Oscillations of the BEC for $M=0.8<M_c$ and $R_0=1.333$
(for this initial radius $V(R_0)=-0.12$).}
\label{oscill}
\end{center}
\end{figure}

In certain cases, the BEC  oscillates. When $M_c<M<M_{\rm max}$, this happens
when $R_U<R_0<R_N$. When $M<M_{c}$, this happens when $R_0>R_2$. The period of
the oscillations is given by
\begin{eqnarray}
T=\pm\sqrt{2M}\int_{R_0}^{R_0'}\frac{dR}{\sqrt{V(R_0)-V(R)}},
\label{osc1}
\end{eqnarray}
where $R_0'$ is determined by the condition
$V(R_0')=V(R_0)$. The sign $+$ corresponds to $R_0<R_0'$ and the sign $-$
corresponds to $R_0>R_0'$.  The BEC oscillates about the equilibrium radius
$R_S$, between the extremal radii $R_0$ and $R_0'$. The pulsation is
$\omega=2\pi/T$. The oscillations of the BEC are illustrated in Fig.
\ref{oscill}.

\subsubsection{The oscillations close to equilibrium}

Close to the minimum $R_S$, the effective potential can be approximated by the
quadratic form $V(R)=V(R_S)+(1/2)V''(R_S)(R-R_S)^2$. In that
case $R_0'=2R_s-R_0$. Making the change of variables $x=(R-R_S)/(R_S-R_0)$,
we find that the period of the oscillations is given by
\begin{eqnarray}
T=\sqrt{\frac{4M}{V''(R_S)}}\int_{-1}^{+1}\frac{dx}{\sqrt{1-x^2}}.
\label{oe1}
\end{eqnarray}
Using
\begin{eqnarray}
\int_{-1}^{+1}\frac{dx}{\sqrt{1-x^2}}=\pi,
\label{oe2}
\end{eqnarray}
we obtain
\begin{eqnarray}
T=2\pi\sqrt{\frac{M}{V''(R_S)}}.
\label{oe3}
\end{eqnarray}
This returns the result of Eq. (\ref{pul2}).

\subsubsection{The case  $M_c<M<M_{\rm max}$ and $R_0\rightarrow R_U^+$}

We consider the case  $M_c<M<M_{\rm max}$. The period of the oscillations
diverges when $R_0\rightarrow R_U^+$ because the effective
potential given by Eq. (\ref{de4}) presents a maximum
at $R_U$. Writing $R_0=R_U+x_0$ and $R=R_U+x$, expanding Eq.
(\ref{osc1}) for
$x_0\ll 1$ and $x\ll 1$, and using $V'(R_U)=0$ and $V''(R_U)<0$, we obtain
\begin{eqnarray}
T\simeq 
\sqrt{\frac{4M}{-V''(R_U)}}\int_{x_0}^{R_N-R_U}\frac{dx}{\sqrt{x^2-x_0^2}}.
\label{osc2}
\end{eqnarray}
Making the change of variables $y=x/x_0$, we
get
\begin{eqnarray}
T\simeq
\sqrt{\frac{4M}{-V''(R_U)}}\int_{1}^{(R_N-R_U)/x_0}\frac{dy}{\sqrt{
y^2-1}}.
\label{osc3}
\end{eqnarray}
Using Eq. (\ref{ctu3}), we obtain for $R_0\rightarrow R_U^+$:
\begin{eqnarray}
T\sim \sqrt{\frac{4M}{-V''(R_U)}}\ln\left (2\frac{R_N-R_U}{R_0-R_U}\right ).
\label{osc4}
\end{eqnarray}
The divergence is logarithmic. This expression can be compared with the
expression (\ref{ctu4}) of the collapse time of the BEC when $R_0\rightarrow
R_U^-$.

\begin{figure}
\begin{center}
\includegraphics[clip,scale=0.3]{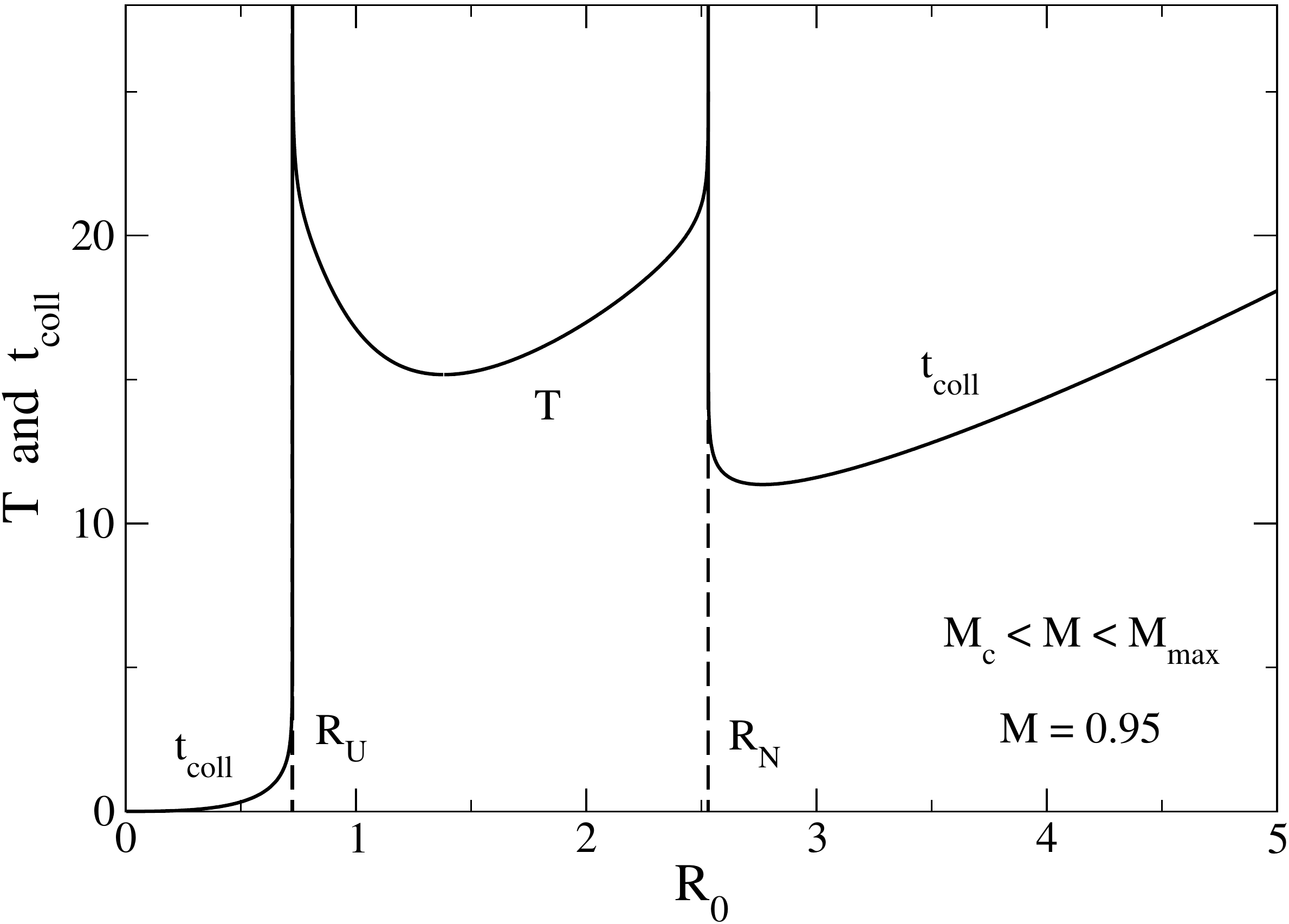}
\caption{Period of the oscillations of the BEC as a function of $R_0\in
[R_U,R_N]$ when $M_c<M<M_{\rm max}$ (we have taken $M=0.95$). The period of the
oscillations diverges
according to Eq. (\ref{osc4}) for
$R_0\rightarrow R_U^+$ and according to Eq.  (\ref{osc6w}) for 
$R_0\rightarrow
R_N^-$. Its minimum is given by Eq. (\ref{oe3}). We have also represented
the collapse time for $R_0<R_U$ and $R_0>R_N$. It diverges
according to Eq. (\ref{ctu4}) for
$R_0\rightarrow R_U^-$ and according to Eq. (\ref{ctn6}) for  $R_0\rightarrow
R_N^+$. It tends to $0$ for
$R_0\rightarrow 0$ according to Eq. (\ref{rs4})  and to $+\infty$ for
$R_0\rightarrow +\infty$ according to Eq. (\ref{rl4}).}
\label{malT}
\end{center}
\end{figure}

When  $R_0\rightarrow R_U^+$, $R_0'$ can be deduced from the
relation
\begin{eqnarray}
R_0-R_U=\sqrt{\frac{2V'(R_N)(R_0'-R_N)}{V''(R_U)}}
\label{osc5}
\end{eqnarray}
obtained by equating $V(R_0)\simeq V(R_U)+(1/2)V''(R_U)(R_0-R_U)^2$ and 
$V(R_0')\simeq V(R_N)+V'(R_N)(R_0'-R_N)$, and using $V(R_U)=V(R_N)$. If we
start from $R_0\rightarrow R_N^{-}$, the period of the oscillations is given by
Eq. (\ref{osc4}) with the substitution of Eq. (\ref{osc5}) and the
final replacement of $R_0'$ by $R_0$. This gives
\begin{eqnarray}
T\sim \sqrt{\frac{M}{-V''(R_U)}}\ln\left \lbrack
\frac{2V''(R_U)(R_N-R_U)^2}{V'(R_N)(R_0-R_N)}\right \rbrack.
\label{osc6w}
\end{eqnarray}
The formulae (\ref{osc4}) and (\ref{osc6w}) are only
valid at leading order. These results are illustrated in Fig. \ref{malT}.

{\it Remark:} the expressions of the different quantities appearing in
Eqs. (\ref{osc4}) and (\ref{osc6w}) when  $M\rightarrow M_c$ have been given at
the end of Sec. \ref{sec_fio}.

\subsubsection{The case  $M<M_{c}$ and $R_0\rightarrow R_2^+$}

We consider the case $M<M_{c}$. When $R_0\rightarrow R_2^+$, $R_0'$ is
rejected to infinity. For large values of $R$,
the effective potential can be approximated by $V(R)\sim -M^2/R$.
Writing $V(R_0)\simeq V'(R_2)(R_0-R_2)$ and  $V(R_0')\simeq -M^2/R_0'$,
the condition $V(R_0')=V(R_0)$ gives
\begin{eqnarray}
R_0'\sim \frac{M^2}{|V'(R_2)|(R_0-R_2)}.
\label{osc6}
\end{eqnarray}
When  $R_0\rightarrow R_2^+$, Eq. (\ref{osc1}) can be approximated by
\begin{eqnarray}
T\simeq \sqrt{2M}\int_{R_0}^{R_0'}\frac{dR}{\sqrt{V(R_0)+\frac{M^2}{R}}}.
\label{osc7}
\end{eqnarray}
Writing $V(R_0)\simeq V'(R_2)(R_0-R_2)$, we obtain
\begin{eqnarray}
T\simeq
\sqrt{\frac{2M}{|V'(R_2)|(R_0-R_2)}}\nonumber\\
\times\int_{R_0}^{R_0'}\frac{dR}{\sqrt{
-1
+\frac{M^2}{R|V'(R_2)|(R_0-R_2)}}}.
\label{osc8}
\end{eqnarray}
Making the change of variables $x=R|V'(R_2)|(R_0-R_2)/M^2$, we get
\begin{eqnarray}
T\sim
\frac{\sqrt{2}M^{5/2}}{|V'(R_2)|^{3/2}(R_0-R_2)^{3/2}}\int_{0}^{1}\frac{dx}
{\sqrt{\frac{1}{x}-1}}.
\label{osc9}
\end{eqnarray}
Using the identity (\ref{rl3}), we find
that, when $R_0\rightarrow R_2^+$, the period of the oscillations
is given by
\begin{eqnarray}
T\sim
\frac{\pi M^{5/2}}{\sqrt{2}|V'(R_2)|^{3/2}(R_0-R_2)^{3/2}}.
\label{osc10}
\end{eqnarray}

\begin{figure}
\begin{center}
\includegraphics[clip,scale=0.3]{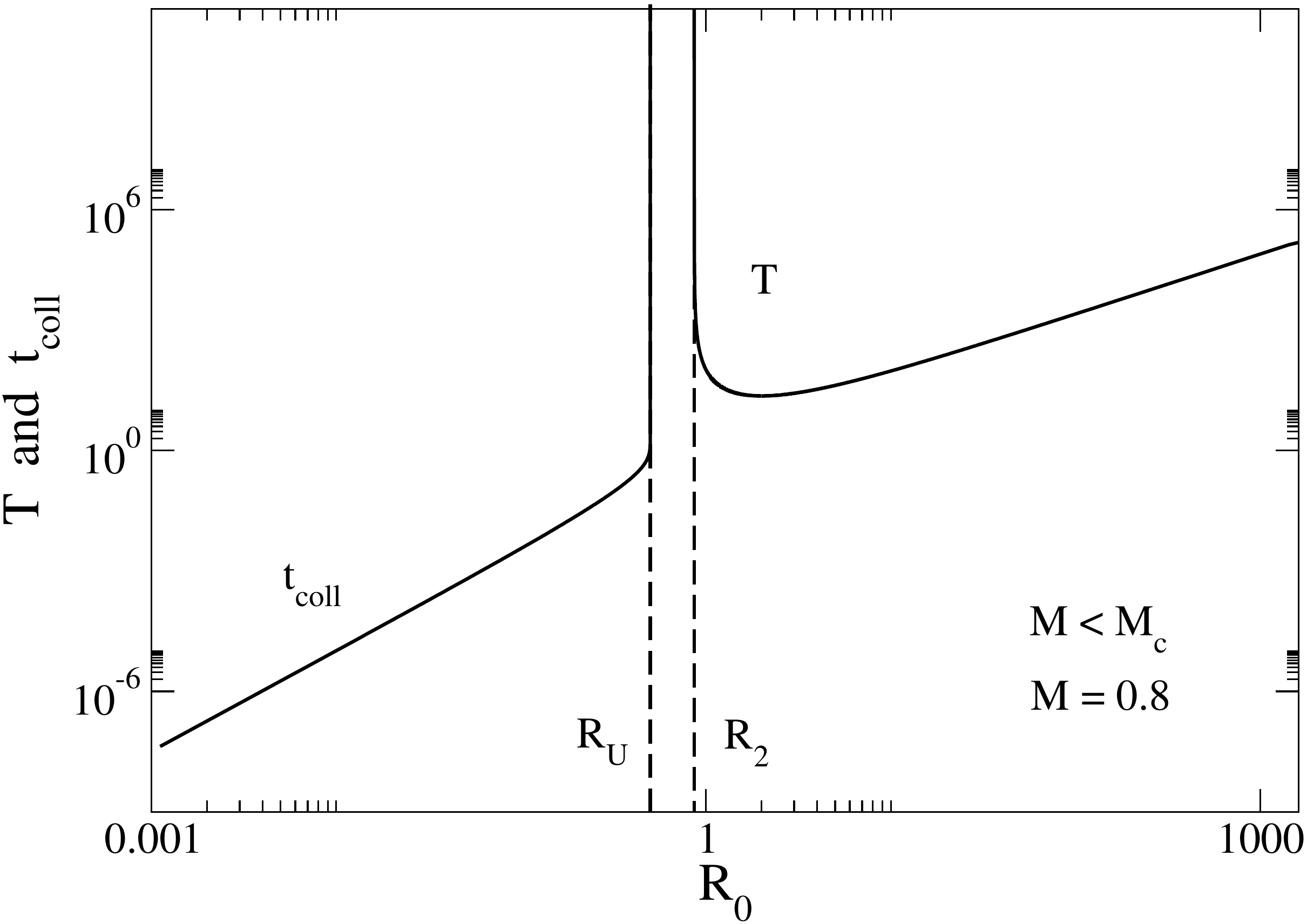}
\caption{Period of the oscillations of the BEC as a function of $R_0\ge R_2$
when $M<M_c$ (we have taken $M=0.8$). The period of the
oscillations
diverges according to Eq. (\ref{osc10}) for
$R_0\rightarrow R_2^+$ and according to Eq.  (\ref{osc11}) for
$R_0\rightarrow +\infty$. Its minimum
is given by Eq. (\ref{oe3}). We have also represented the collapse time of the
BEC as a function of $R_0\le R_U$.  It diverges according to Eq. (\ref{ctu4})
for $R_0\rightarrow R_U^-$. It tends to $0$ for
$R_0\rightarrow 0$ according to Eq. (\ref{rs4}).}
\label{malTbis}
\end{center}
\end{figure}

If we start from  $R_0\rightarrow +\infty$, the period of the oscillations is
given by Eq. (\ref{osc10}) with
the substitution of Eq. (\ref{osc6}) and the final replacement of $R_0'$ by
$R_0$.
This gives
\begin{eqnarray}
T\sim
\frac{\pi}{\sqrt{2M}}R_0^{3/2}.
\label{osc11}
\end{eqnarray} 
These results are illustrated in Fig. \ref{malTbis}.

{\it Remark:} For $M\rightarrow M_c$, we have $R_2\simeq
1/\sqrt{3}+(2/3^{3/4})\sqrt{M_c-M}$, and 
$V'(R_2)\sim -3^{7/4}\sqrt{M_c-M}$.

\section{The limit $M\rightarrow 0$}
\label{sec_mz}

In this section, we investigate the limit $M\rightarrow 0$. In this limit, 
\begin{eqnarray}
R_U\sim \frac{M}{2},\qquad R_S\sim \frac{2}{M},
\label{e14}
\end{eqnarray}
\begin{eqnarray}
V(R_U)\sim \frac{4}{3M},\qquad V(R_S)\sim -\frac{M^3}{4},
\label{e15}
\end{eqnarray}
\begin{eqnarray}
R_1\sim \frac{M}{3},\qquad R_2\sim \frac{1}{M}.
\label{e16}
\end{eqnarray}
The expressions of $R_U$, $V(R_U)$ and $R_1$ for $M\rightarrow 0$ can be
obtained by neglecting the gravitational term in the effective potential
(\ref{de4}). This corresponds to the nongravitational limit. The expressions of
$R_S$, $V(R_S)$ and $R_2$ for $M\rightarrow 0$ can be
obtained by neglecting the self-interaction term in the effective potential
(\ref{de4}). This corresponds to the noninteracting
limit.

\subsection{The nongravitational limit $M\rightarrow 0$ and $R\sim M\ll 1$}
\label{sec_mzng}

For $M\rightarrow 0$ and $R\sim M\ll 1$, we can neglect the
gravitational term in the effective potential
(\ref{de4}). According to Eq. (\ref{de7}), the
evolution of the radius of the BEC is given by
\begin{eqnarray}
\int_{R(t)}^{R_0}\frac{dR}{\sqrt{\frac{M}{R_0^2}-\frac{M^2}{3R_0^3}-\frac{M}{R^2
} +\frac{M^2}{3R^3}}}=\pm\left
(\frac{2}{M}\right
)^{1/2} t,
\label{e17}
\end{eqnarray}
where the sign $+$ corresponds to a contraction of the BEC and
the sign $-$ corresponds to an expansion of the BEC. Since we assume $R\sim M$,
we define $R_0=qM$. With the
change of variables $x=R/M$, the foregoing equation can be rewritten as
\begin{eqnarray}
\int_{\frac{R(t)}{M}}^{q}\frac{dx}{\sqrt{\frac{1}{q^2}-\frac{1}{3q^3}-\frac{1}{
x^2
} +\frac{1}{3x^3}}}=\pm\frac{\sqrt{2}}{M^2}t.
\label{e18}
\end{eqnarray}
This equation describes the evolution of the BEC under the
action of the self-interaction and the quantum potential for any $E_{\rm
tot}$.

\begin{figure}
\begin{center}
\includegraphics[clip,scale=0.3]{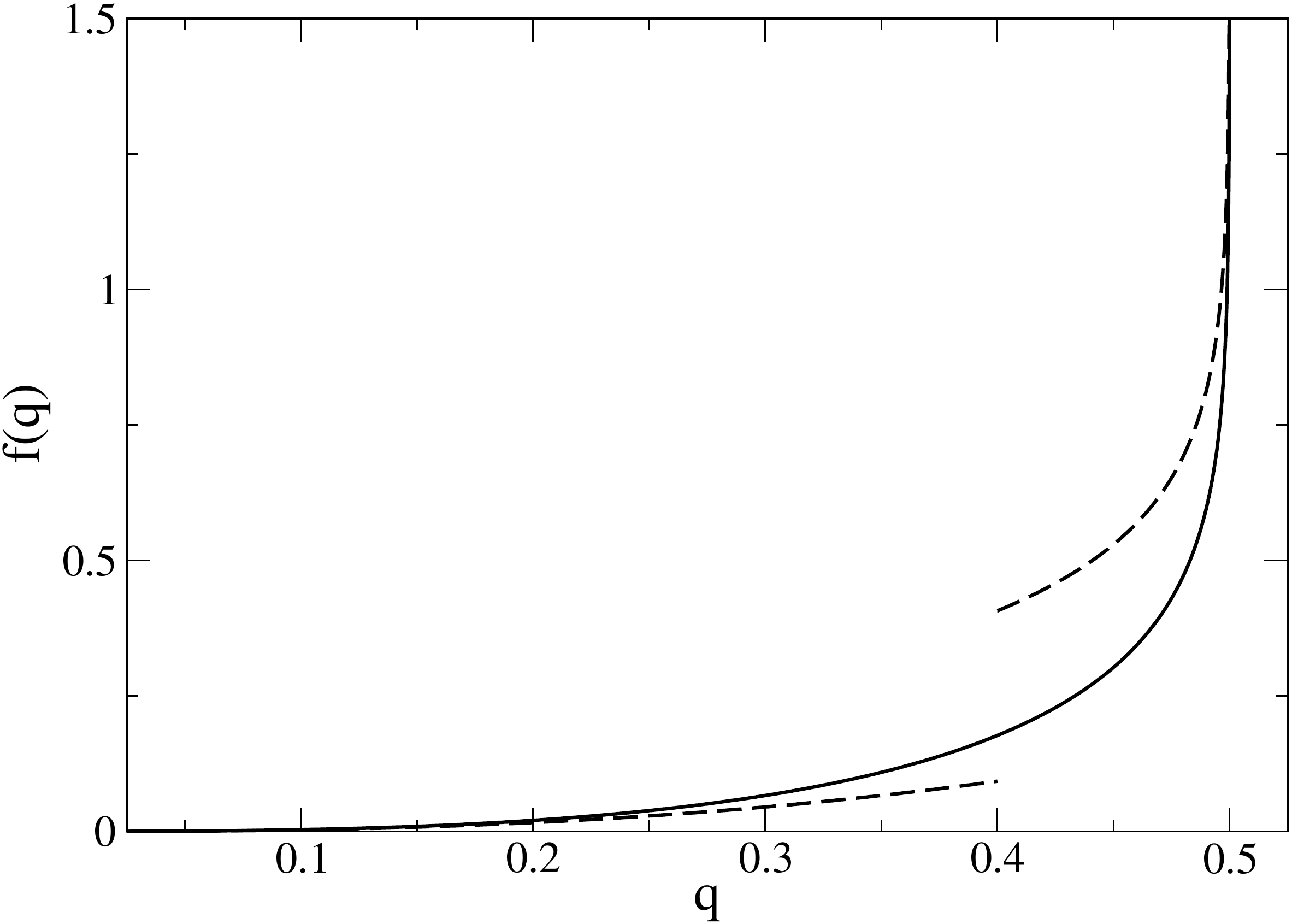}
\caption{The function  $f(q)$. The dashed lines correspond to
the asymptotic behaviors given by Eqs. (\ref{e20a}) and (\ref{e20b}).}
\label{qf}
\end{center}
\end{figure}

We consider the collapse of the BEC starting from $R_0\le R_U$. This 
corresponds to $q\le 1/2$. The collapse time, obtained from the condition
$R(t_{\rm
coll})=0$, is given by
\begin{eqnarray}
t_{\rm coll}(M,R_0)=f\left (\frac{R_0}{M}\right )M^2
\label{e19}
\end{eqnarray}
with
\begin{eqnarray}
f(q)=\frac{1}{\sqrt{2}}\int_{0}^{q}\frac{dx}{\sqrt{\frac{1}{q^2}
-\frac{1}{3q^3}-\frac{1}
{x^2}+\frac{1}{3x^3}}}.
\label{e20}
\end{eqnarray}
This function is plotted in Fig. \ref{qf}. For $q\rightarrow 0$, using Eq.
(\ref{rs3}), we get
\begin{eqnarray}
f(q)\sim
\sqrt{\frac{3\pi}{2}}\frac{\Gamma(5/6)}{\Gamma(1/3)}q^{5/2}.
\label{e20a}
\end{eqnarray}
Substituting this asymptotic result in Eq. (\ref{e19}), we recover Eq.
(\ref{rs4}). For $q\rightarrow 1/2$, using Eq. (\ref{ctu3}), we get
\begin{eqnarray}
f(q)\sim -\frac{1}{4\sqrt{2}}\ln\left
(\frac{1}{2}-q\right ).
\label{e20b}
\end{eqnarray}
Substituting this asymptotic result in Eq. (\ref{e19}), we recover Eq.
(\ref{ctu4}) with $R_U\sim M/2$ and $V''(R_U)\sim -32/M^3$.
For $R_0=R_1$, corresponding to $q=1/3$, we have $E_{\rm tot}=V(R_0)=0$. Using
the identity
\begin{eqnarray}
\int_{s}^{1/3}\frac{dx}{\sqrt{-\frac{1}{x^2}
+\frac{1}{3x^3}}}=\qquad\qquad\qquad\qquad\qquad\nonumber\\
\frac{1}{12}\left\lbrack
(1+2s)\sqrt{3s(1-3s)}+\cos^{-1}(\sqrt{3s})\right\rbrack,
\label{e21}
\end{eqnarray}
we find that Eq. (\ref{e18}) with the sign $+$ (collapse) becomes
\begin{eqnarray}
\frac{1}{12}\Biggl\lbrack
\left (1+\frac{2R}{M}\right
)\sqrt{\frac{3R}{M}\left
(1-\frac{3R}{M}\right )}\nonumber\\
+\cos^{-1}\left (\sqrt{\frac{3R}{M}}\right
)\Biggr\rbrack=\frac{\sqrt{2}}{M^2}t.
\end{eqnarray}
This equation describes the collapse of the BEC under the
action of the self-interaction and the quantum potential when $E_{\rm tot}=0$.
On the other hand, using $f(1/3)=\pi/24\sqrt{2}$, the
collapse time is exactly given by
\begin{eqnarray}
t_{\rm coll}(M,R_0=R_1)=\frac{\pi}{24\sqrt{2}}M^2.
\label{e22}
\end{eqnarray}
This expression is valid for sufficiently small values of $M$ so that
the self-gravity of the BEC can be neglected. 

{\it Remark:} if we assume $E_{\rm tot}=0$ and neglect the quantum potential in
Eq. (\ref{e17}), we obtain
\begin{eqnarray}
R(t)=R_0\left (1-\frac{5}{\sqrt{6}}\frac{M^{1/2}}{R_0^{5/2}}t\right )^{2/5},
\label{e23}
\end{eqnarray}
leading to
\begin{eqnarray}
t_{\rm coll}=\frac{\sqrt{6}}{5}\frac{R_0^{5/2}}{M^{1/2}}.
\label{e24}
\end{eqnarray}
Equation (\ref{e23})  describes the collapse of the BEC with $E_{\rm tot}=0$
under the action of the self-interaction alone (the explosive solution is
obtained from Eq. (\ref{e23}) by changing the sign $-$ into $+$).
Taking $R_0=M/3$, we get $t_{\rm coll}=\sqrt{2}M^2/45$ which is different by
a factor $\sim 3$ from Eq. (\ref{e22}) obtained by taking the quantum
potential into account.

\subsection{The noninteracting limit $M\rightarrow 0$ and $R\sim 1/M\gg 1$}
\label{sec_niq}

For $M\rightarrow 0$ and $R\sim 1/M\gg 1$, we can neglect the
self-interaction in the effective potential (\ref{de4}).  According to Eq.
(\ref{de7}), the
evolution of the radius of the BEC is given by
\begin{eqnarray}
\int_{R_0}^{R(t)}\frac{dR}{\sqrt{\frac{M}{R_0^2}-\frac{M^2}{R_0}-\frac{
M}{R^2
} +\frac{M^2}{R}}}=\pm\left
(\frac{2}{M}\right
)^{1/2} t,
\label{ni1}
\end{eqnarray}
where the sign $+$ corresponds to an expansion of the BEC and
the sign $-$ corresponds to a contraction of the BEC. Since we assume $R\sim
1/M$, we define $R_0=q/M$. With the change of variables
$x=RM$, the foregoing equation can be rewritten as
\begin{eqnarray}
\int_{q}^{MR(t)}\frac{dx}{\sqrt{\frac{1}{q^2}-\frac{1}{q}-\frac{
1 } {x^2} +\frac{1}{x}}}=\pm\sqrt{2}M^2 t.
\label{ni2}
\end{eqnarray}
This equation describes the evolution of the BEC under the
action of the self-gravity and the quantum potential for any $E_{\rm
tot}$. The integral can be calculated analytically using the identity
\begin{eqnarray}
\int \frac{dx}{\sqrt{a-\frac{
1 } {x^2}
+\frac{1}{x}}}=\frac{1}{a}\sqrt{ax^2+x-1}\nonumber\\
-\frac{1}{2a^{3/2}}
\ln\left\lbrack 1+2ax+2\sqrt{a}\sqrt{ax^2+x-1}\right\rbrack.
\label{ni2b}
\end{eqnarray}
Combining Eqs. (\ref{ni2}) and (\ref{ni2b}), we can obtain an
analytical expression of the evolution of the radius of the BEC under the form
$t=t(R)$.
For $R_0=R_2$,
corresponding to $q=1$, we have $E_{\rm tot}=V(R_0)=0$. In that
case, using the identity
\begin{eqnarray}
\int\frac{dx}{\sqrt{-\frac{
1 } {x^2} +\frac{1}{x}}}=\frac{2}{3}(2+x)\sqrt{x-1},
\label{ni3}
\end{eqnarray}
we find that Eq. (\ref{ni1}) with the sign $+$
becomes
\begin{eqnarray}
\frac{2}{3}(2+MR)\sqrt{MR-1}=\sqrt{2}M^2 t.
\label{ni4}
\end{eqnarray}
This equation describes the explosion of the BEC under
the action of the self-gravity and the quantum potential when $E_{\rm
tot}=0$.

Equation (\ref{ni1}) describe the evolution of the BEC under the action of
the quantum potential and the self-gravity. For sufficiently small values
of $R_0$ and sufficiently short times, self-gravity can be neglected and Eq.
(\ref{ni1}) reduces to
\begin{eqnarray}
\int_{R_0}^{R(t)}\frac{dR}{\sqrt{\frac{M}{R_0^2}-\frac{
M}{R^2
}}}=\left
(\frac{2}{M}\right
)^{1/2} t.
\label{ni5}
\end{eqnarray}
This equation describes the explosion of the BEC under the action of the
quantum potential alone.  It can be integrated into
\begin{eqnarray}
R=R_0\sqrt{1+\frac{2t^2}{R_0^4}}.
\label{ni6}
\end{eqnarray}

\section{Conclusion}
\label{sec_conclusion}

In this paper, we have completed our previous investigations
\cite{prd1,prd2} concerning  self-gravitating BECs with attractive
self-interaction
($a_s<0$). Axions generally have an attractive self-interaction
so they
enter in the framework of our study. Standard axions can coalesce into
axion stars. Axion stars are stable only if their mass
does 
not exceed a certain maximum mass  [see  Eq. (\ref{intro1})] obtained in 
\cite{prd1,prd2}. In general, this mass is extremely small. The question
then arises to what happens to an axion star with a mass larger than $M_{\rm
max}$? In
this paper, we have considered the possibility that the star collapses and forms
a black hole. We have used a Newtonian model based on the GPP equations and,
using a Gaussian ansatz, we have determined an approximate expression of the
collapse time at which the star collapses to a
point. We have obtained the analytical asymptotic behaviors
\begin{eqnarray}
\frac{t_{\rm coll}}{t_D}\sim \frac{0.688033...}{(M/M_{\rm max})^{1/2}},\qquad
(M\gg M_{\rm max}),
\label{conclusion1}
\end{eqnarray}
\begin{eqnarray}
\frac{t_{\rm coll}}{t_D}\sim \frac{2.90178...}{(M/M_{\rm max}-1)^{1/4}},\qquad
(M\rightarrow  M_{\rm max}^+),
\label{conclusion2}
\end{eqnarray}
where $M_{\rm max}$ is the maximum mass of a stable axion
star and $t_D$ is the dynamical time. For a standard
axion particle with  $m=10^{-4}\,
{\rm eV}/c^2$ and
$a_s=-5.8\, 10^{-53}\, {\rm m}$, we have $M_{\rm max}=6.9\times 10^{-14}\,
M_{\odot}$, $R_*=1.0\times 10^{-4}\, R_{\odot}$ and $t_D=3.4 \,{\rm
hrs}$. From our general
expression of the collapse time [see Eq. (\ref{ct5})],
an axion star with a mass
$M=2M_{\rm max}=1.4\times 10^{-13}\,
M_{\odot}$ and an initial radius $R_0=R_*=1.0\times 10^{-4}\, R_{\odot}$ will
form
a mini black hole of mass $M=1.4\times 10^{-13}\,
M_{\odot}$    in a time $t_{\rm
coll}=0.676301 t_D=2.3\, {\rm hrs}$. This time is very short. Therefore, if dark
matter is made of standard axions,
they could be  in the form of mini black holes instead of mini axion stars.
Of course, our Newtonian treatment is
only
valid as long as the radius of the star is much larger than its Schwarzschild
radius. When the radius of the star approaches its Schwarzschild
radius,  general relativity must be taken into account in order to describe the
formation of a black hole. However, we have shown in Appendix \ref{sec_vna} that
general relativistic effects become important only extraordinarily close to the
collapse time ($\Delta t=1.09\times 10^{-32}\, {\rm s}$ for standard axions) so
our Newtonian treatment is justified. We have also considered the case of
ultralight axions which may cluster into dark matter halos. For  ultralight
axions with
$m=1.93\times
10^{-20}\, {\rm eV}/c^2$ and
$a_s=-8.29\times 10^{-60}\, {\rm fm}$, we
obtain $M_{\rm max}=4.18\times 10^{5}\, M_{\odot}$, $R_*=10.4\, {\rm pc}$, and
$t_D=1.49 \,{\rm
Myrs}$.  An axionic dark
matter halo with a mass
$M=2M_{\rm max}=8.36\times 10^{5}\, M_{\odot}$ and an initial radius
$R_0=R_*=10.4\, {\rm pc}$ will form
a supermassive black hole of mass $M=8.36\times 10^{5}\,
M_{\odot}$ (similar to those reported at the center of
galaxies) in a time $t_{\rm
coll}=0.676301 t_D=1.01  \,{\rm
Myrs}$. General relativistic effects become important only $\Delta t=8.47\times
10^{-8}\, {\rm s}$ before the collapse time.

In addition to the black hole scenario, other possibilities 
can be considered as recently discussed by Braaten {\it et al.} \cite{braaten}
in the case of axion stars. 
The first possibility is that the
collapse of the axion star can be accompanied by a burst of relativistic
axions produced by inelastic reactions when the density reaches high values.
This would lead to a bosenova. A similar phenomenon has been observed in the
collapse of a (nongravitational) BEC of ultracold atoms with an attractive
self-interaction \cite{donley}. The emission of relativistic axions
could decrease the mass of the axion star so that it always remains larger than
its  Schwarzschild radius. Another possibility is that an axion star
with
$M>M_{\rm max}$ loses mass by emitting scalar field radiation through the
process of gravitational cooling \cite{seidel94}. In this manner, its
core
mass $M_{\rm core}$  always remains smaller than $M_{\rm max}$, avoiding its
catastrophic collapse into a black hole. An axion star with $M>M_{\rm max}$
may also fragmentate into several pieces of
mass $M'<M_{\rm max}$. There is also the possibility of
forming dense axion stars.  The stable axions stars that we have considered in 
\cite{prd1,prd2} are dilute axion stars in which the self-gravity and
the attractive
self-interaction are balanced by the quantum force (kinetic energy). As proposed
by Braaten {\it et al.} \cite{braaten}, there may exist
dense axion stars in which
gravity is balanced by the pressure of the axion condensate coming from higher
order terms in the scalar field potential $V(\phi)$ beyond the $\lambda\phi^4$
term. This leads to a
sequence of new
branches of axion stars. The first branch of these dense axion stars has mass
ranging from about $10^{-11}\, M_{\odot}$ to about $M_{\odot}$.

On a technical point of view, our study provides a lot of analytical
results concerning the collapse, the explosion, and the oscillations of
self-gravitating BECs with negative scattering length. As already mentioned, our
analytical results are based
on a Gaussian ansatz that may not always be quantitatively reliable. It
would be therefore interesting to compare our analytical
results with an exact numerical solution of the GPP equations. On the other
hand, we have used a fully Newtonian model although the system becomes
relativistic at the end of the collapse, when it approaches the Schwarzschild
radius. Therefore, one should solve the KGE equations to have a more exact
description of the dynamics. These are interesting problems for the future.
However, the richness of the problem already revealed by our simple analytical
study shows that the full numerical simulation of the GPP and KGE equations, and
their study, will require a considerable amount of work.

\appendix

\section{Derivation of the GP equation from the KG equation}
\label{sec_d}

In this Appendix, we show that the GP equation can be derived from
the KG equation in the nonrelativistic limit $c\rightarrow +\infty$ and we
relate the scattering length $a_s$ appearing in the GP equation to the
self-interaction constant $\lambda$ appearing in the KG
equation. We follow the procedure of \cite{abrilph,playa}. The KG
equation for a complex scalar field $\phi$ writes
\begin{equation}
\frac{1}{c^2}\frac{\partial^2\phi}{\partial
t^2}-\Delta\phi+\frac{m^2c^2}{\hbar^2}\left (1+\frac{2\Phi}{c^2}\right
)\phi+2\frac{dV}{d|\phi|^2}\phi=0,
\label{d5}
\end{equation}
where  $V=V(|\phi|^2)$ is the self-interaction potential of the SF and $\Phi$ is
an
external potential that, in a simplified model, can be identified with the
gravitational potential (see \cite{abrilph,playa} for a fully general
relativistic
treatment). The KG equation without self-interaction can be viewed as the
relativistic generalization of the Schr\"odinger equation. Similarly, the
KG equation with a self-interaction can be viewed as the relativistic
generalization of the GP equation. In order to recover the
Schr\"odinger and GP equations in the nonrelativistic limit
$c\rightarrow +\infty$, we make the Klein
transformation\footnote{The prefactor is derived in \cite{abrilph,playa}.}
\begin{equation}
\phi({\bf r},t)=\frac{\hbar}{m}e^{-im c^2 t/\hbar}\psi({\bf r},t),
\label{d6}
\end{equation}
where $\psi$ is the wave function. The rest-mass density is given by
$\rho=|\psi|^2=(m/\hbar)^2|\phi|^2$. Substituting Eq. (\ref{d6}) into
the KG equation (\ref{d5}), we get
\begin{equation}
\frac{\hbar^2}{2m c^2}\frac{\partial^2\psi}{\partial t^2}-i\hbar
\frac{\partial\psi}{\partial
t}-\frac{\hbar^2}{2m}\Delta\psi+m\Phi\psi+m\frac{dV}{d|\psi|^2}\psi=0.
\label{d7}
\end{equation}
Taking the nonrelativistic limit $c\rightarrow +\infty$ of Eq.
(\ref{d7}), we obtain the nonlinear Schr\"odinger equation
\begin{equation}
i\hbar \frac{\partial\psi}{\partial
t}=-\frac{\hbar^2}{2m}\Delta\psi+m\Phi\psi+m\frac{dV}{d|\psi|^2}\psi.
\label{d8}
\end{equation}
It can be written as a GP equation of the form
\begin{equation}
i\hbar \frac{\partial\psi}{\partial t}=-\frac{\hbar^2}{2m}\Delta\psi+m\Phi\psi+m
h(|\psi|^2)\psi
\label{d9}
\end{equation}
with a nonlinearity
\begin{equation}
h(|\psi|^2)=\frac{dV}{d|\psi|^2}\qquad {\rm i.e.}\qquad h(\rho)=V'(\rho).
\label{d10}
\end{equation}
As we have recalled in footnote 4, the GP equation with a cubic nonlinearity
(corresponding to $h(\rho)\propto \rho$) can be derived from the mean field
Schr\"odinger equation with a pair contact potential. The present
approach shows that the GP equation with an {\it arbitrary} nonlinearity
$h(\rho)$ can be derived from the KG equation with an arbitrary
self-interaction potential $V(\rho)$: the potential $h(\rho)$ determining the
nonlinearity  in the GP equation is equal to the derivative of the potential
$V(\rho)$ in the KG equation; reciprocally, $V(\rho)$ is a primitive of
$h(\rho)$.

The potential $h(\rho)$ determining the nonlinearity in the standard GP equation
(\ref{gpp1}) is
\begin{eqnarray}
\label{d11}
h=\frac{4\pi a_s\hbar^2}{m^3}\rho.
\end{eqnarray}
According to Eq. (\ref{d10}), it corresponds to a  potential of the form
\begin{equation}
V=\frac{2\pi a_s\hbar^2}{m^3}\rho^2=\frac{2\pi a_s m}{\hbar^2}|\phi|^4.
\label{d12}
\end{equation}
With this potential, the KG equation  (\ref{d5})  explicitly writes
\begin{eqnarray}
\frac{1}{c^2}\frac{\partial^2\phi}{\partial
t^2}-\Delta\phi+\frac{m^2c^2}{\hbar^2}\left (1+\frac{2\Phi}{c^2}\right
)\phi
+\frac{8\pi a_s m}{\hbar^2}|\phi|^2\phi=0.\nonumber\\
\label{d13}
\end{eqnarray}
Therefore, a cubic nonlinearity in the GP equation corresponds to a quartic
potential in the KG equation. In general, the quartic potential is written in
the form
\begin{equation}
V(|\phi|^2)=\frac{\lambda}{4\hbar c}|\phi|^4,
\label{d14}
\end{equation}
where $\lambda$ is a dimensionless self-interaction constant. Comparison
between (\ref{d12}) and (\ref{d14}) yields
\begin{eqnarray}
\frac{\lambda}{8\pi}=\frac{a_s m c}{\hbar}.
\label{d15}
\end{eqnarray}
This justifies Eq. (B9) of \cite{prd1}.

The pressure appearing in the quantum Euler equation associated with the GP
equation is due to the self-interaction of the bosons. It is given in terms
of the potential $V$ by \cite{prd1}:
\begin{eqnarray}
P(\rho)=\rho V'(\rho)-V(\rho). 
\label{d16}
\end{eqnarray}
For the quartic potential (\ref{d12}), we get Eq. (\ref{gpp12}), i.e. a
polytropic equation of state of index $n=1$. More generally, for a power-law
potential of the form
\begin{eqnarray}
V(|\phi|^2)=\frac{K}{\gamma-1}\left (\frac{m}{\hbar}\right
)^{2\gamma}|\phi|^{2\gamma},
\label{d17}
\end{eqnarray}
we get the polytropic equation of state
\begin{eqnarray}
P=K\rho^{\gamma}, \qquad \gamma=1+\frac{1}{n}.
\label{d18}
\end{eqnarray}
In particular, for a sextic potential
$V(|\phi|^2)=(K/2)({m}/{\hbar})^6|\phi|^{6}$ (which is the term coming after
the quartic potential), we obtain $P=K\rho^3$, i.e. a polytropic equation of
state of index $n=1/2$.

\section{Lagrangian of a  self-gravitating BEC}
\label{sec_lh}

In this Appendix, we discuss the Lagrangian structure of the GPP
equations and of the corresponding hydrodynamic  equations. The Lagrangian of
the
GPP equation is
\begin{eqnarray}
L=\int \biggl\lbrace\frac{i\hbar}{2m} \left (\psi^*\frac{\partial\psi}{\partial
t}-\psi\frac{\partial\psi^*}{\partial t}\right
)-\frac{\hbar^2}{2m^2}|\nabla\psi|^2\nonumber\\
-\frac{1}{2}\Phi|\psi|^2-\frac{2\pi
a_s\hbar^2}{m^3}|\psi|^4\biggr\rbrace\, d{\bf
r}.
\label{lh1}
\end{eqnarray}
We can view the Lagrangian (\ref{lh1}) as a functional of $\psi$, $\dot\psi$
and $\nabla\psi$. The action is $S=\int L\, dt$. The least action principle
$\delta S=0$, which is equivalent to the Lagrange equation
\begin{eqnarray}
\label{lh2}
\frac{\partial}{\partial t}\left (\frac{\delta
L}{\delta\dot\psi}\right)+\nabla\cdot \left (\frac{\delta
L}{\delta\nabla\psi}\right)-\frac{\delta L}{\delta\psi}=0,
\end{eqnarray}
returns the GP equation (\ref{gpp1}).  The total energy is
obtained from the
transformation
\begin{eqnarray}
E_{\rm tot}=\int \frac{i\hbar}{2m} \left (\psi^*\frac{\partial\psi}{\partial
t}-\psi\frac{\partial\psi^*}{\partial t}\right )\, d{\bf r}-L
\label{lh3}
\end{eqnarray}
leading to
\begin{eqnarray}
E_{\rm tot}=\int\frac{\hbar^2}{2m^2}|\nabla\psi|^2\, d{\bf
r}+\int\frac{1}{2}\Phi|\psi|^2\, d{\bf r}\nonumber\\
+\int\frac{2\pi
a_s\hbar^2}{m^3}|\psi|^4\, d{\bf r}.
\label{lh4}
\end{eqnarray}
The first term is the kinetic energy, the second term
is the gravitational energy and the third term is the self-interaction energy. 
Using the Lagrange equations, one can show that the total energy is conserved.
On
the other hand, the GP equation (\ref{gpp1}) can be written as
\begin{eqnarray}
\label{lh5}
i\hbar\frac{\partial\psi}{\partial t}=m\frac{\delta E_{\rm tot}}{\delta\psi^*}.
\end{eqnarray}
This expression directly implies the conservation of the total energy
$E_{\rm tot}$. From general arguments \cite{holm}, a minimum of energy $E_{\rm
tot}$ under
the normalization condition  $\int |\psi|^2\, d{\bf r}=M$ is a stationary
solution of the GP equation that is formally nonlinearly dynamically stable.
Writing the variational principle as
\begin{eqnarray}
\label{lh5b}
\delta E_{\rm tot}-\frac{\mu}{m}\delta\int |\psi|^2\, d{\bf r}=0,
\end{eqnarray}
where $\mu$
is a Lagrange
multiplier (chemical potential), we recover the time-independent GP equation
(\ref{gpp3}) with $E=\mu$. This shows that the chemical potential $\mu$ can be
identified with the eigenenergy $E$, or reciprocally.

Using the Madelung transformation, we can rewrite the
Lagrangian in terms of hydrodynamic variables. According to
Eqs. (\ref{gpp4}) and (\ref{gpp6}) we have
\begin{eqnarray}
\label{lh6}
\frac{\partial S}{\partial t}=-i\frac{\hbar}{2}\frac{1}{|\psi|^2}\left
(\psi^*\frac{\partial\psi}{\partial t}-\psi\frac{\partial\psi^*}{\partial
t}\right )
\end{eqnarray}
and
\begin{eqnarray}
\label{lh7}
|\nabla\psi|^2=\frac{1}{4\rho}(\nabla\rho)^2+\frac{\rho}{\hbar^2}
(\nabla S)^2.
\end{eqnarray}
Substituting these identities in Eq. (\ref{lh1}) we get
\begin{eqnarray}
\label{lh8}
L=-\int \biggl \lbrace \frac{\rho}{m}\frac{\partial S}{\partial
t}+\frac{\rho}{2m^2}(\nabla S)^2
+\frac{\hbar^2}{8m^2}\frac{(\nabla\rho)^2}{\rho}\nonumber\\
+\frac{1}{2}\rho\Phi+\frac{2\pi a_s\hbar^2}{m^3}\rho^2\biggr\rbrace\, d{\bf r}.
\end{eqnarray}
We can view the Lagrangian (\ref{lh8}) as a functional of $S$, $\dot S$,
$\nabla S$, $\rho$, $\dot\rho$, and $\nabla\rho$. The Lagrange equation for the
action
\begin{eqnarray}
\label{lh9}
\frac{\partial}{\partial t}\left (\frac{\delta L}{\delta\dot
S}\right)+\nabla\cdot \left (\frac{\delta L}{\delta\nabla S}\right)-\frac{\delta
L}{\delta S}=0
\end{eqnarray}
returns the equation of continuity (\ref{gpp8}). The Lagrange equation for the
density
\begin{eqnarray}
\label{lh10}
\frac{\partial}{\partial t}\left (\frac{\delta L}{\delta\dot
\rho}\right)+\nabla\cdot \left (\frac{\delta L}{\delta\nabla
\rho}\right)-\frac{\delta L}{\delta \rho}=0
\end{eqnarray}
returns the quantum Hamilton-Jacobi (or Bernoulli) equation (\ref{gpp8b}) 
leading to the quantum Euler equation (\ref{gpp9}). The total energy is
obtained
from
the transformation
\begin{eqnarray}
\label{lh11}
E_{\rm tot}=-\int \frac{\rho}{m}\frac{\partial S}{\partial t}\, d{\bf r}-L
\end{eqnarray}
leading to
\begin{eqnarray}
\label{lh12}
E_{\rm tot}=\int  \frac{1}{2}\rho {\bf
u}^2\,
d{\bf r}+\int \frac{\hbar^2}{8m^2}\frac{(\nabla\rho)^2}{\rho}\,
d{\bf r}
+\frac{1}{2}\int \rho\Phi\,
d{\bf r}\nonumber\\
+\int \frac{2\pi a_s\hbar^2}{m^3}\rho^2\,
d{\bf r}.\qquad\qquad
\end{eqnarray}
The first term is the classical kinetic energy, the
second term
is the quantum kinetic energy (or quantum potential energy), the third term is
the gravitational energy, and
the fourth term is the self-interaction energy. 
Using the Lagrange equations, one can show that the total energy is conserved.
Therefore, a stable stationary solution of the EP equations is a minimum of
energy $E_{\rm tot}$ under the mass constraint  $M=\int \rho\, d{\bf r}$.
Writing the
variational principle as 
\begin{eqnarray}
\label{lh11b}
\delta E_{\rm tot}-\frac{\mu}{m}\delta M=0,
\end{eqnarray}
where
$\mu$ is a
Lagrange multiplier (chemical potential), we recover Eq. (\ref{gpp8bc}).
Taking its gradient, we recover the condition of hydrostatic equilibrium
(\ref{gpp13}) \cite{prd1}.

With the Gaussian ansatz of Eq. (\ref{ga1}), one has
\begin{eqnarray}
\int \frac{i\hbar}{2m} \left (\psi^*\frac{\partial\psi}{\partial
t}-\psi\frac{\partial\psi^*}{\partial t}\right )\, d{\bf
r}\nonumber\\
=-\int \frac{\rho}{m}\frac{\partial S}{\partial t}\, d{\bf r}
=-\frac{1}{2}\alpha MR^2\dot H
\label{lh13}
\end{eqnarray}
and 
\begin{eqnarray}
E_{\rm tot}=\frac{1}{2}\alpha
M R^2 H^2+\sigma\frac{\hbar^2M}{m^2R^2}\nonumber\\
+\zeta\frac{2\pi
a_s\hbar^2M^2}{m^3R^3}
-\nu\frac{GM^2}{R}.
\label{lh14}
\end{eqnarray}
Substituting these expressions in Eq. (\ref{lh3}) or in Eq. (\ref{lh11}), we
obtain the effective Lagrangian
\begin{eqnarray}
L(H,\dot H,R)=-\frac{1}{2}\alpha
MR^2(\dot H+H^2)-\sigma\frac{\hbar^2M}{m^2R^2}\nonumber\\
-\zeta\frac{2\pi
a_s\hbar^2M^2}{m^3R^3}
+\nu\frac{GM^2}{R}.
\label{lh15}
\end{eqnarray}
We can view the Lagrangian (\ref{lh15}) as a function of $H$, $\dot H$ and $R$.
The Lagrange
equation for $H$ 
\begin{eqnarray}
\label{lh16}
\frac{\partial}{\partial t}\left (\frac{\delta L}{\delta\dot
H}\right)-\frac{\delta
L}{\delta H}=0
\end{eqnarray}
returns Eq. (\ref{ga4}). The Lagrange equation for $R$
\begin{eqnarray}
\label{lh17}
\frac{\delta L}{\delta R}=0
\end{eqnarray}
returns the equation of motion (\ref{gd2}). We also note that, using Eq.
(\ref{ga4}), the total energy
can be written as
\begin{eqnarray}
E_{\rm tot}=\frac{1}{2}\alpha
M\left (\frac{dR}{dt}\right
)^2+\sigma\frac{\hbar^2M}{m^2R^2}\nonumber\\
+\zeta\frac{2\pi
a_s\hbar^2M^2}{m^3R^3}
-\nu\frac{GM^2}{R},
\label{lh18}
\end{eqnarray}
leading to Eqs. (\ref{gd0}) and (\ref{gd1}).

\section{Self-similar solutions close to the
critical point
$M\rightarrow M_{\rm max}^+$ and $R_0\rightarrow R_*$}
\label{sec_sscp}

For $M\rightarrow M_{\rm max}^+$ and $R_0\rightarrow R_*$, the temporal
evolution of the radius $R(t;M,R_0)$  of the BEC and the collapse time $t_{\rm
coll}(M,R_0)$ have a self-similar structure. This self-similar
structure can be directly obtained from the normal form of the effective
potential close to the critical point given by Eq. (\ref{saddle1}). This
corresponds to a saddle node bifurcation.

\subsection{Self-similar solution with a normalization by $R_0>R_*$}
\label{sec_ssu}

The temporal evolution of the radius of the BEC is determined by Eq.
(\ref{saddle5}).
We first consider the case $x_0>0$. Making the change of variables
$y=x/x_0$ and introducing the scaled variables
\begin{eqnarray}
T=\sqrt{\frac{2}{3}}x_0^{1/2}t,\qquad
X(T)=\frac{x(t)}{x_0},\qquad \mu=\frac{6(M-1)}{x_0^2},\nonumber\\
\label{ssu1}
\end{eqnarray}
where $t$, $x(t)$ and $M$ are normalized in terms of $x_0$, we can rewrite
Eq. (\ref{saddle5}) as
\begin{eqnarray}
\int_{X(T)}^{1}
\frac{dy}{\sqrt{\mu(1-y)+1-y^3}}=T.
\label{ssu2}
\end{eqnarray}
The collapse time, corresponding to $X\rightarrow -\infty$, is given by
\begin{eqnarray}
T_{\rm coll}(\mu)=\int_{-\infty}^{1}
\frac{dy}{\sqrt{\mu(1-y)+1-y^3}}.
\label{ssu3}
\end{eqnarray}
With the scaled variables, the collapse dynamics of the BEC close to the
critical point depends on a single
parameter $\mu$ instead of the two parameters $M$ and $x_0$.
For $\mu=0$, we get
\begin{eqnarray}
\int_{X(T)}^{1}
\frac{dy}{\sqrt{1-y^3}}=T
\label{ssu4}
\end{eqnarray}
and 
\begin{eqnarray}
T_{\rm coll}(\mu=0)=\int_{-\infty}^{1}
\frac{dy}{\sqrt{1-y^3}}\nonumber\\
=\sqrt{\pi}\frac{\Gamma(1/3)}{\Gamma(5/6)}
=4.20655....
\label{ssu5}
\end{eqnarray}
In the general case, returning to the original variables,  we obtain for
$R_0>1$:
\begin{eqnarray}
\frac{t_{\rm
coll}(M,R_0)}{t_{\rm coll}(M=1,R_0)}=f_+\left
\lbrack \frac{6(M-1)}{(R_0-1)^2}\right \rbrack
\label{ssu6}
\end{eqnarray}
with
\begin{eqnarray}
f_+(\mu)=\frac{\int_{-\infty}^{1}
\frac{dy}{\sqrt{\mu(1-y)+1-y^3}}}
{\int_{-\infty}^{1}
\frac{dy}{\sqrt{1-y^3}}}
\label{ssu7}
\end{eqnarray}
and 
\begin{eqnarray}
t_{\rm coll}(M=1,R_0)=K_+ (R_0-1)^{-1/2},
\label{ssu8}
\end{eqnarray}
where $K_+$ is given by Eq. (\ref{ru4}). The invariant profile defined by Eq.
(\ref{ssu7}) behaves as
\begin{eqnarray}
f_+(\mu)\sim \frac{6^{1/4}B}{K_+}\mu^{-1/4}\sim 0.881517...\mu^{-1/4}
\label{ssu9}
\end{eqnarray}
for $\mu\rightarrow +\infty$. 

\begin{figure}
\begin{center}
\includegraphics[clip,scale=0.3]{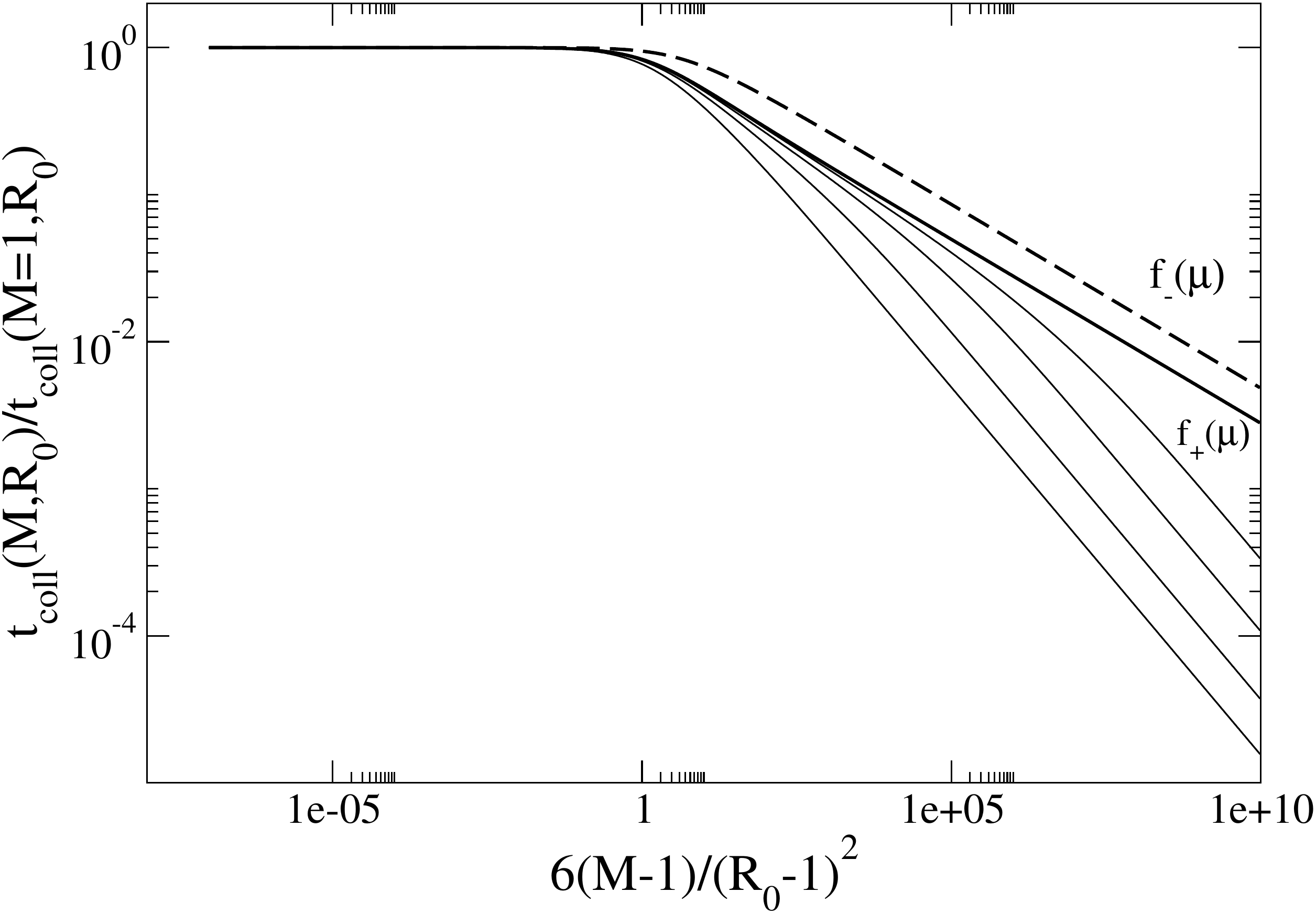}
\caption{Evolution of the collapse time $M \!\mapsto\! t_{\rm
coll}(M,R_0)$ in scaled variables. We have taken $R_0=1.1, 1.01, 1.001, 1.0001$
(bottom to top). The rescaled profile converges towards the invariant profile
$f_+(\mu)$ when $R_0\rightarrow 1^+$ and $M\rightarrow 1$. It has an asymptotic 
logarithmic slope $-1/4$. For $M\gg 1$, the self-similar solution is not valid
anymore and the collapse
time behaves according to Eq. (\ref{la1}) with a logarithmic slope $-1/2$. The
two slopes are clearly
visible on the figure. For comparison, we have also represented the invariant
profile $f_-(\mu)$ corresponding to $R_0<1$ (see Sec.
\ref{sec_ssd}).}
\label{ssfplus}
\end{center}
\end{figure}

\begin{figure}
\begin{center}
\includegraphics[clip,scale=0.3]{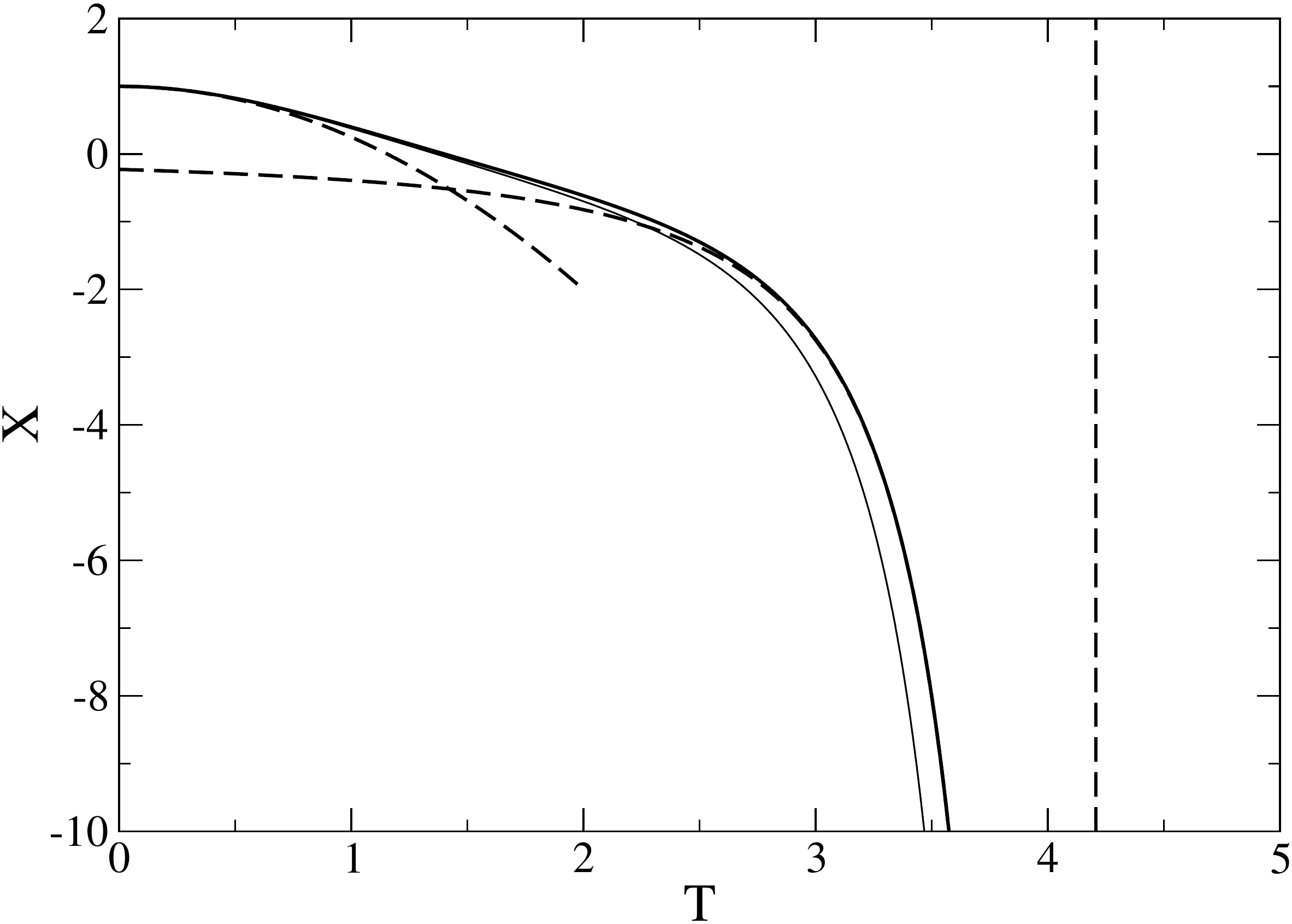}
\caption{Universal evolution of the scaled radius of the BEC $X(T;\mu)$ close
to the critical point corresponding to a saddle-node bifurcation. The thick
curve corresponds to $\mu=0$. According to the results of
Secs. \ref{sec_rtz} and \ref{sec_rti}, we have $X\sim
1-3T^2/4$ for $T\rightarrow 0$ and $X\sim
-4/(T_{\rm coll}-T)^2$ for $T\rightarrow T_{\rm
coll}=4.20655...$. The thin curve corresponds to $\mu=0.1$.
}
\label{txmu}
\end{center}
\end{figure}

The exact collapse time given by Eq. (\ref{ct5}) is plotted  in
scaled variables in Fig. \ref{ssfplus} in order to illustrate its
convergence towards the invariant profile given by Eq. (\ref{ssu7}) as
$R_0\rightarrow 1^+$ and $M\rightarrow 1^+$. The universal  evolution of the
radius of the BEC 
in scaled variables given by Eq. (\ref{ssu2}), valid close to the critical
point,
is plotted in Fig. \ref{txmu}.

{\it Remark:} The integral of Eq. (\ref{ssu4}) has the
analytical expression
\begin{equation}
\int_{X}^{1}
\frac{dy}{\sqrt{1-y^3}}=\frac{\sqrt{\pi}\, \Gamma(4/3)}{\Gamma(5/6)}-X\,
_2F_1\left (\frac{1}{3},\frac{1}{2},\frac{4}{3},X^3\right ).
\label{ssu4ba}
\end{equation}

\subsection{Self-similar solution with a normalization by
$R_0<R_*$}
\label{sec_ssd}

We now consider the case $x_0<0$. Making the change of variables $y=x/|x_0|$,
and introducing the scaled variables
\begin{eqnarray}
T=\sqrt{\frac{2}{3}}|x_0|^{1/2}t,\quad
X(T)=\frac{x(t)}{|x_0|},\quad \mu=\frac{6(M-1)}{x_0^2},\nonumber\\
\label{ssd1}
\end{eqnarray}
where $t$, $x(t)$ and $M$ are normalized in terms of $x_0$, we can rewrite
Eq. (\ref{saddle5}) as
\begin{eqnarray}
\int_{X(T)}^{-1}
\frac{dy}{\sqrt{\mu(-y-1)-y^3-1}}=T.
\label{ssd2}
\end{eqnarray}
The collapse time, corresponding to $X\rightarrow -\infty$, is given by
\begin{eqnarray}
T_{\rm coll}(\mu)=\int_{-\infty}^{-1}
\frac{dy}{\sqrt{\mu(-y-1)-y^3-1}}.
\label{ssd3}
\end{eqnarray}
With the scaled variables, the collapse dynamics of the BEC close to the
critical point depends on a single
parameter $\mu$ instead of the two parameters $M$ and $R_0$. For
$\mu=0$, we get
\begin{eqnarray}
\int_{X(T)}^{-1}
\frac{dy}{\sqrt{-y^3-1}}=T
\label{ssd4}
\end{eqnarray}
and
\begin{eqnarray}
T_{\rm coll}(\mu=0)=\int_{-\infty}^{-1}
\frac{dy}{\sqrt{-y^3-1}}\nonumber\\
=2\sqrt{\pi}\frac{\Gamma(7/6)}{\Gamma(2/3)}
=2.42865...
\label{ssd5}
\end{eqnarray}
In the general case, returning to the original variables, we obtain for $R_0<1$:
\begin{eqnarray}
\frac{t_{\rm
coll}(M,R_0)}{t_{\rm coll}(M=1,R_0)}=f_-\left
\lbrack \frac{6(M-1)}{(1-R_0)^2}\right \rbrack
\label{ssd6}
\end{eqnarray}
with
\begin{eqnarray}
f_{-}(\mu)=\frac{\int_{-\infty}^{-1}
\frac{dy}{\sqrt{\mu(-y-1)-y^3-1}}}
{\int_{-\infty}^{-1}
\frac{dy}{\sqrt{-y^3-1}}}
\label{ssd7}
\end{eqnarray}
and 
\begin{eqnarray}
t_{\rm coll}(M=1,R_0)=K_- (1-R_0)^{-1/2},
\label{ssd8}
\end{eqnarray}
where $K_-$ is given by Eq. (\ref{ru7}). The invariant profile defined by Eq.
(\ref{ssd7})
behaves as
\begin{eqnarray}
f_-(\mu)\sim \frac{6^{1/4}B}{K_-}\mu^{-1/4}\sim 1.52683...\mu^{-1/4}
\label{ssd9}
\end{eqnarray}
for $\mu\rightarrow +\infty$. It is
plotted in Fig. \ref{ssfplus}.

\begin{figure}
\begin{center}
\includegraphics[clip,scale=0.3]{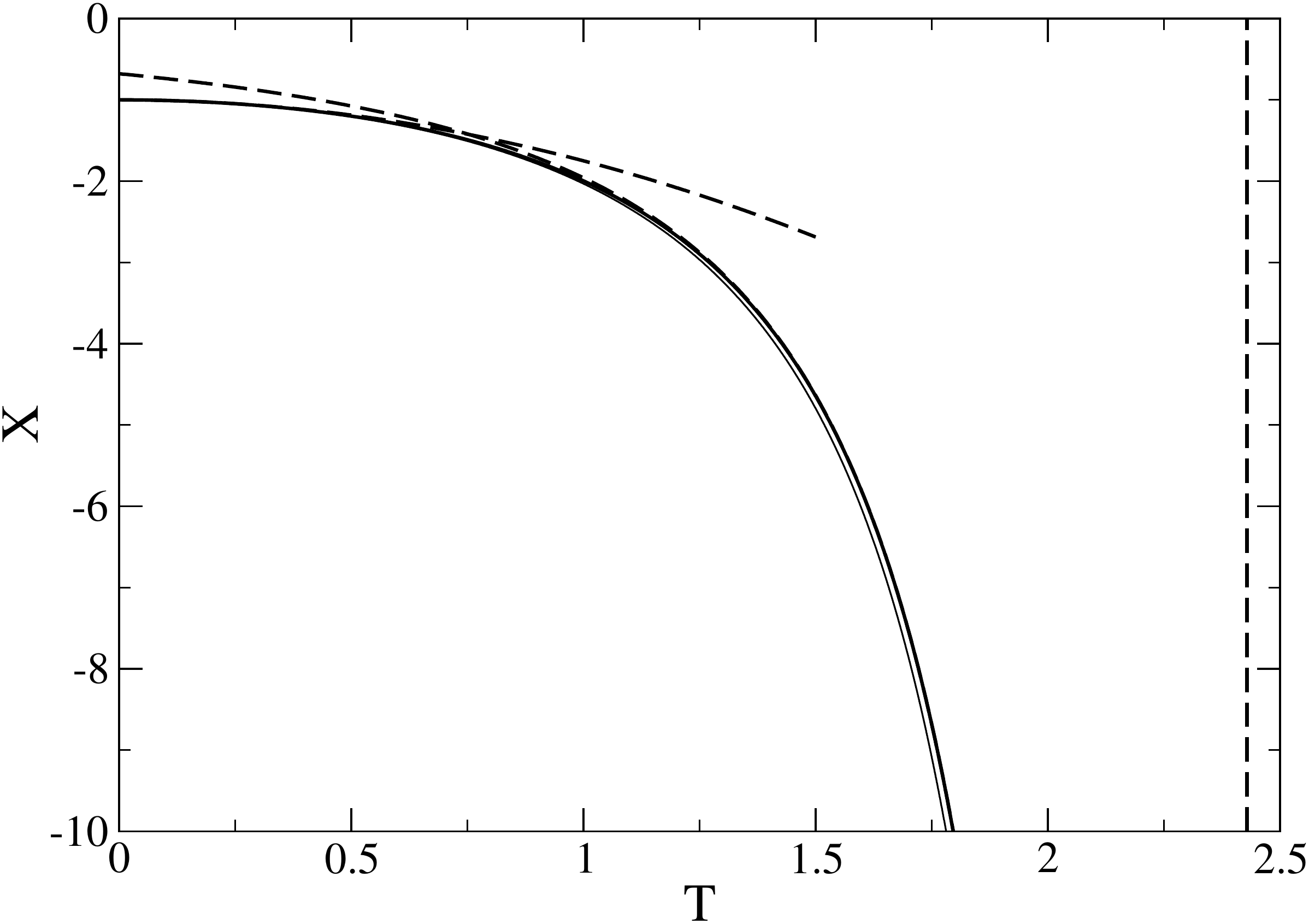}
\caption{Universal evolution of the scaled radius of the BEC $X(T;\mu)$ close
to the critical point corresponding to a saddle-node bifurcation. The thick
curve corresponds to $\mu=0$. According to the results of
Secs. \ref{sec_rtz} and \ref{sec_rti}, we have $X\sim
-1-3T^2/4$ for $T\rightarrow 0$ and $X\sim
-4/(T_{\rm coll}-T)^2$ for $T\rightarrow T_{\rm
coll}=2.42865...$.  The thin curve corresponds to $\mu=0.1$.
}
\label{txnegmu}
\end{center}
\end{figure}

The universal  evolution of the radius of the BEC 
in scaled variables given by  Eq. (\ref{ssd2}), valid close to the critical
point,
is plotted in Fig. \ref{txnegmu}.

{\it Remark:} The integral of Eq. (\ref{ssd4}) has the
analytical expression
\begin{eqnarray}
\int_{X}^{-1}
\frac{dy}{\sqrt{-y^3-1}}=\frac{2\sqrt{\pi}\,
\Gamma(7/6)}{\Gamma(2/3)}\nonumber\\
-\frac{2}{\sqrt{-X}}\,
_2F_1\left (\frac{1}{6},\frac{1}{2},\frac{7}{6},-\frac{1}{X^3}\right ).
\label{ssd4ba}
\end{eqnarray}

\subsection{Self-similar solution with a normalization by 
$M>1$}
\label{sec_sst}

Making the change of variables
$y=x/\sqrt{6(M-1)}$ and introducing the scaled variables
\begin{eqnarray}
T=\left (\frac{8}{3}\right )^{1/4}(M-1)^{1/4}t,
\label{sst1}
\end{eqnarray}
\begin{eqnarray}
X(T)=\frac{x(t)}{\sqrt{6(M-1)}},\qquad X_0=\frac{x_0}{\sqrt{6(M-1)}},
\label{sst2}
\end{eqnarray}
where $t$, $x(t)$ and $x_0$ are normalized in terms  of $M$, we can
rewrite Eq. (\ref{saddle5}) as
\begin{eqnarray}
\int_{X(T)}^{X_0}\frac{dy}{\sqrt{X_0-y+X_0^3-y^3}}=T.
\label{sst3}
\end{eqnarray}
The collapse time, corresponding to $X\rightarrow -\infty$, is given by
\begin{eqnarray}
T_{\rm coll}(X_0)=\int_{-\infty}^{X_0}\frac{dy}{\sqrt{X_0-y+X_0^3-y^3}}.
\label{sst4}
\end{eqnarray}
With the scaled variables, the collapse dynamics of the BEC close to the
critical point depends on a single
parameter $X_0$ instead of the two parameters $M$ and $R_0$. For
$X_0=0$, we get
\begin{eqnarray}
\int_{X(T)}^{0}\frac{dy}{\sqrt{-y-y^3}}=T
\label{sst5}
\end{eqnarray}
and
\begin{eqnarray}
T_{\rm
coll}(X_0=0)=\int_{-\infty}^{0}\frac{dy}{\sqrt{-y-y^3}}\nonumber\\
=\frac{8}{
\sqrt { \pi } } \Gamma(5/4)^2=3.70815... .
\label{sst6}
\end{eqnarray}
\begin{figure}
\begin{center}
\includegraphics[clip,scale=0.3]{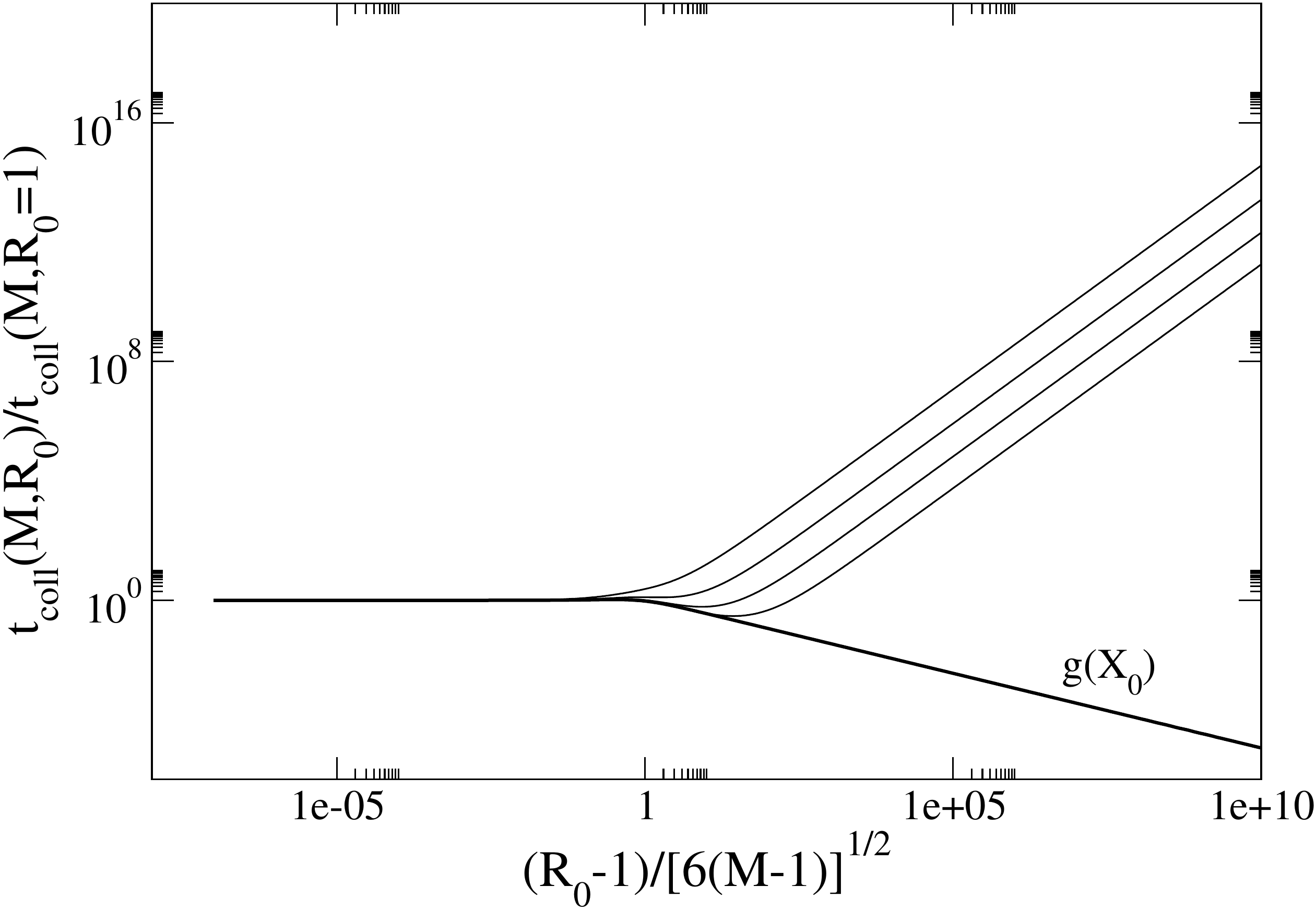}
\caption{Evolution of the collapse time $R_0 \!\mapsto\!
t_{\rm coll}(M,R_0)$ with $R_0>1$ in
scaled variables. We have taken $M=1.1, 1.01, 1.001, 1.0001$ (top to bottom).
The rescaled profile converges towards the invariant profile
$g(X_0)$ when
$M\rightarrow 1^+$ and $R_0\rightarrow 1^+$. It has an asymptotic  logarithmic
slope $-1/2$.
For $R_0\gg 1$, the self-similar solution is not valid anymore and the collapse
time
behaves according to Eq. (\ref{rl4}) with a logaritmic slope $+3/2$.
The two slopes, of different sign, are clearly
visible on the figure.}
\label{ssg}
\end{center}
\end{figure}
\begin{figure}
\begin{center}
\includegraphics[clip,scale=0.3]{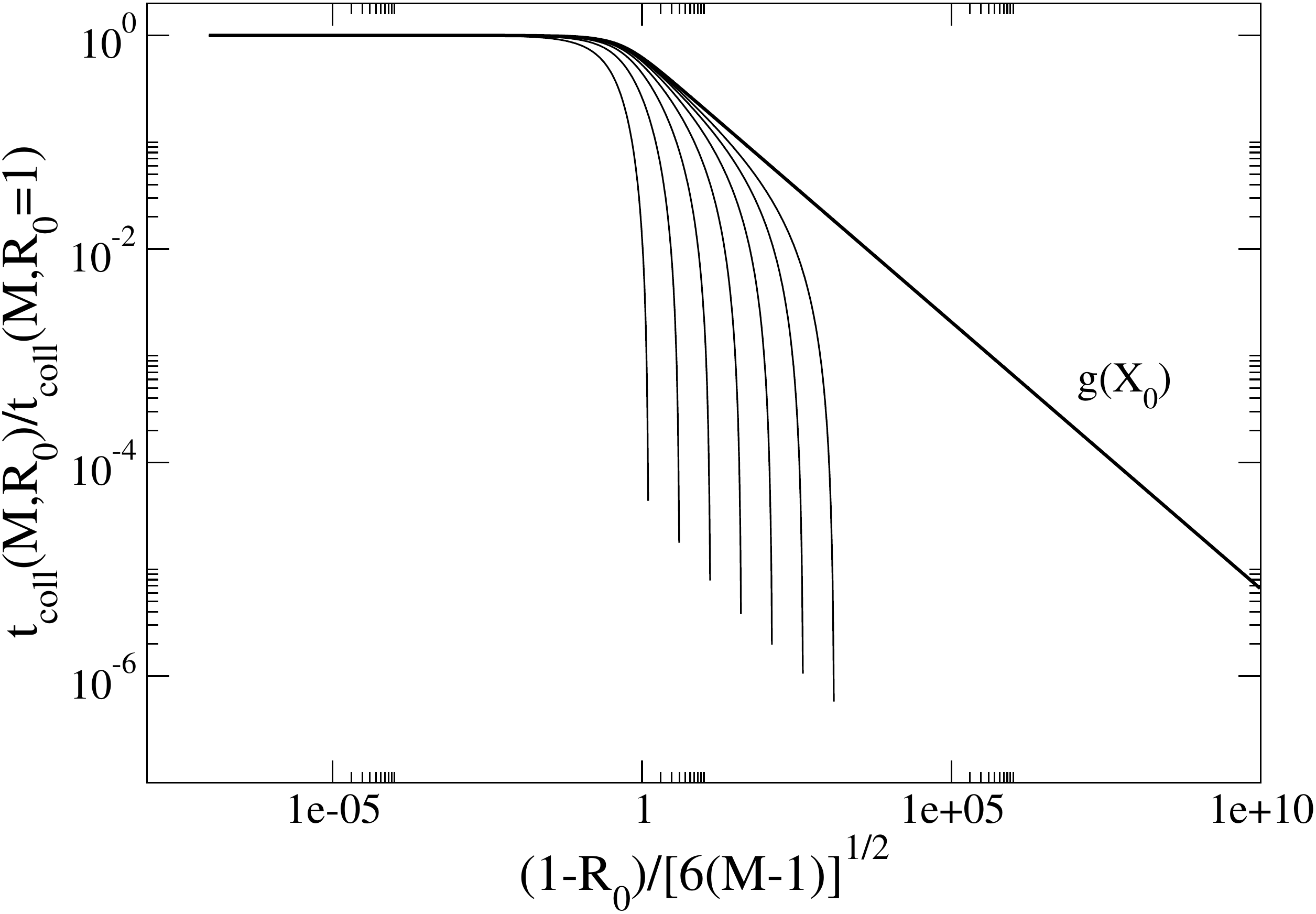}
\caption{Evolution of the collapse time $R_0 \!\mapsto\!
t_{\rm coll}(M,R_0)$ with $R_0<1$ in scaled variables. We have taken $M=1.1,
1.01, 1.001, 1.0001, 1.00001, 1.000001, 1.0000001$ (left to right). The rescaled
profile converges towards the invariant profile $g(X_0)$ when $M\rightarrow 1^+$
and $R_0\rightarrow 1^-$. It has an asymptotic  logarithmic slope $-1/2$.
For $R_0\rightarrow 0$, the self-similar solution is not valid anymore and the
collapse time behaves according to Eq. (\ref{rs4}) leading to a divergence in
log-log plot.}
\label{ssgR0neg}
\end{center}
\end{figure}
In the general case, returning to the original variables, we obtain
\begin{eqnarray}
\frac{t_{\rm coll}(M,R_0)}{t_{\rm coll}(M,R_0=1)}=g\left\lbrack 
\frac{R_0-1}{\sqrt{6(M-1)}}   \right\rbrack
\label{sst7}
\end{eqnarray}
where
\begin{eqnarray}
g(X_0)=\frac{\int_{-\infty}^{X_0}\frac{dy}{\sqrt{X_0-y+X_0^3-y^3}}}
{\int_{-\infty}^{0}\frac{dy}{\sqrt{-y-y^3}}}
\label{sst8}
\end{eqnarray}
and
\begin{eqnarray}
t_{\rm
coll}(M,R_0=1)=B(M-1)^{-1/4},
\label{sst9}
\end{eqnarray}
where $B$ is given by Eq. (\ref{saddle11}). The invariant profile defined by Eq.
(\ref{sst8})
behaves as
\begin{eqnarray}
g(X_0)\sim \frac{K_{\pm}}{6^{1/4}B|X_0|^{1/2}}
\label{sst10}
\end{eqnarray}
for $X_0\rightarrow \pm\infty$. The coefficients are $1.134408...$ (+) 
and $0.6549509...$ (-).

\begin{figure}
\begin{center}
\includegraphics[clip,scale=0.3]{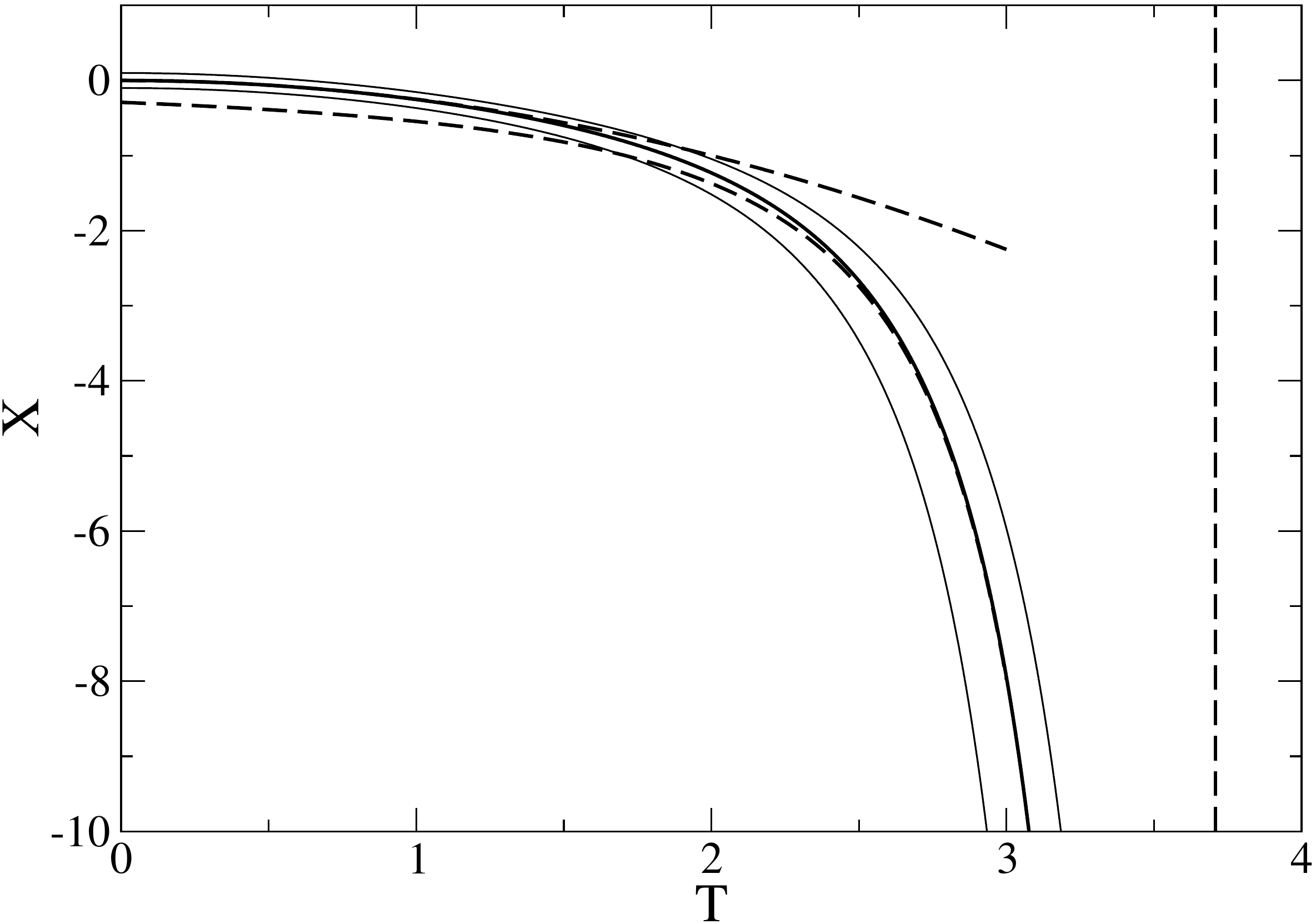}
\caption{Universal evolution of the scaled radius of the BEC $X(T;X_0)$ close
to the critical point corresponding to a saddle-node bifurcation. The thick
curve corresponds to $X_0=0$.  According to the results of
Secs. \ref{sec_rtz} and \ref{sec_rti}, we have $X\sim
-T^2/4$ for $T\rightarrow 0$ and $X\sim
-4/(T_{\rm coll}-T)^2$ for $T\rightarrow T_{\rm
coll}=3.70815...$. The thin curves correspond to $X_0=0.1$ and
$X_0=-0.1$.
}
\label{txmore}
\end{center}
\end{figure}

The exact collapse time given by Eq. (\ref{ct5}) is plotted  in
scaled variables in Figs. \ref{ssg} and \ref{ssgR0neg} in order to illustrate
its convergence towards the invariant profile given by Eq. (\ref{sst8}) as
$M\rightarrow 1^+$ and $R_0\rightarrow 1$. The universal  evolution of the
radius of the BEC 
in scaled variables given by Eq. (\ref{sst3}), valid close to the critical
point,
is plotted in Fig. \ref{txmore}.

{\it Remark:} making the change of variables $y=\sinh(u)$ we also have
\begin{eqnarray}
\int_{\sinh^{-1}(X(T))}^0
\frac{du}{\sqrt{-\sinh(u)}}=T
\label{rem1}
\end{eqnarray}
and
\begin{eqnarray}
\int_{-\infty}^0
\frac{du}{\sqrt{-\sinh(u)}}=T_{\rm coll}(X_0=0).
\label{rem2}
\end{eqnarray}

\section{Particular regimes of interest}
\label{sec_sum}

In this Appendix, we highlight  particular regimes of interest considered in
our paper and provide the sections where they are discussed in detail. 

{\it (i) Self-gravity alone:} The case $E_{\rm tot}<0$, corresponding to the
collapse of the BEC, is treated at the end of Secs. \ref{sec_tfa}
and \ref{sec_mi}. This solution is similar to the Mestel solution describing
the gravitational collapse of a homogeneous sphere. It is also similar to a FLRW
cosmological model of pressureless universe with curvature $k=+1$ exhibiting a
phase of
contraction (Big Crunch) after an initial phase of expansion (Big Bang). The
case $E_{\rm tot}>0$, corresponding to the explosion of the BEC, is treated in
Secs. \ref{sec_gppos}. This solution is similar to a FLRW
cosmological model of pressureless universe with curvature $k=-1$ that is
expanding. The
case $E_{\rm tot}=0$, corresponding to a marginal case of explosion, is
treated in Sec. \ref{sec_gpnul}.  This solution is similar to a FLRW
cosmological model  of pressureless universe without curvature
($k=0$) known as the EdS model.

{\it (ii) Quantum potential alone:} The case $E_{\rm tot}>0$ (the only possible
one), corresponding to the explosion of the BEC, is treated at the end of Sec.
\ref{sec_niq}.

{\it (iii) Self-interaction alone:} The case $E_{\rm tot}<0$, corresponding to
the collapse of the BEC, is treated at the begining of  Secs.  \ref{sec_tfa}
and \ref{sec_mi}. The case
$E_{\rm tot}=0$, corresponding to the marginal collapse or explosion of the BEC,
is treated
at the end of Sec.
\ref{sec_mzng}. 

{\it (iv) Self-gravity and quantum potential:} The general case is treated in
Sec. \ref{sec_niq} (the specific case of an energy $E_{\rm tot}=0$ is treated
in Sec. \ref{sec_gpnul}).

{\it (v) Self-interaction and quantum potential:}  The general case is
treated in Sec. \ref{sec_mzng}.

{\it (vi) Self-gravity and self-interaction:} The case $E_{\rm tot}<0$ is
treated in Sec. \ref{sec_tf}.

\section{Validity of the Newtonian approximation}
\label{sec_vna}

\begin{figure}
\begin{center}
\includegraphics[clip,scale=0.3]{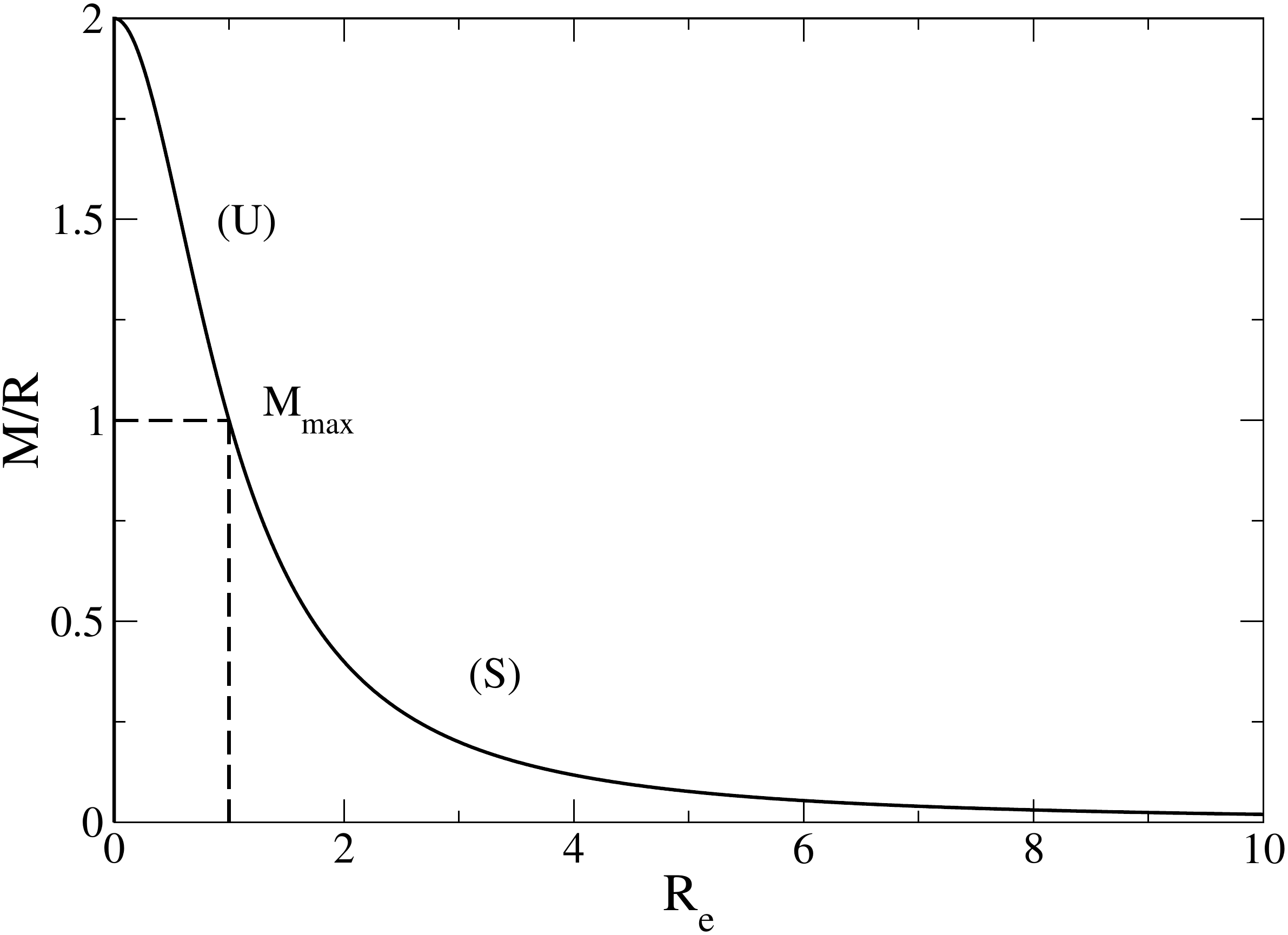}
\caption{Mass-radius ratio along the series of equilibria of a
self-gravitating
BEC with attractive self-interaction (we have used the dimensionless variables
of Sec. \ref{sec_de}).
}
\label{msrr}
\end{center}
\end{figure}

In this Appendix, we study the validity of the Newtonian
approximation used in this paper. General relativistic effects become important
when the radius $R$ of the system is of the order of the Schwarzschild
radius $R_S=2GM/c^2$. It is convenient to introduce the mass-radius ratio
\begin{eqnarray}
\chi=\frac{R_S}{R}=\frac{2GM}{Rc^2}.
\label{vna0}
\end{eqnarray}
The Newtonian approximation is valid when $\chi\ll 1$ and
general relativistic effects come into play when $\chi\sim 1$. Considering the
mass-radius relation of Eq. (\ref{gd5}), it is easy to see that the mass-radius
ratio is maximum on the stable branch (S) at the point $(R_*,M_{\rm max})$.
Therefore, we define
\begin{eqnarray}
\chi_{\rm max}=\frac{2GM_{\rm max}}{R_*c^2}.
\label{vna1}
\end{eqnarray}
The mass-radius ratio can then be written as
\begin{eqnarray}
\frac{\chi}{\chi_{\rm max}}=\frac{M}{M_{\rm max}}\,\frac{R_*}{R}.
\label{vna2}
\end{eqnarray}
From the mass-radius relation of Eq. (\ref{mr1}), we get 
\begin{eqnarray}
\frac{\chi_e}{\chi_{\rm
max}}=\frac{2}{1+(R_e/R_*)^2}.
\label{vna3}
\end{eqnarray}
This function decreases monotonically with $R_e$ (see Fig. \ref{msrr}). It
starts from $2$ at $R_e=0$, reaches the value $1$ at $R_e=R_*$ and tends to $0$
as $2/(R_e/R_*)^2$ when $R_e\rightarrow +\infty$.

Using Eqs. (\ref{nsl1}) and (\ref{nsl2}), we can write the maximum mass-radius
ratio as
\begin{eqnarray}
\chi_{\rm
max}=\frac{\sigma}{6\pi\zeta}\frac{2Gm}{|a_s|c^2}=\frac{\sigma}{6\pi\zeta}\frac{
r_S}{|a_s|}.
\label{vna4}
\end{eqnarray}
In the last term, we have introduced the Schwarzschild radius $r_S=2Gm/c^2$  of
the bosons.  Therefore, $\chi_{\rm max}$ is of the
order of the  ratio between the Schwarzschild
radius of the bosons and their scattering length.  In general, this ratio is
very
small ($r_S\ll |a_s|$) so that
$\chi_{\rm max}\ll 1$, ensuring the validity of the Newtonian
approximation.

To be specific, let us make a numerical application. For a standard
axion particle with  $m=10^{-4}\,
{\rm eV}/c^2$ and
$a_s=-5.8\, 10^{-53}\, {\rm m}$, we have $M_{\rm max}=6.9\times 10^{-14}\,
M_{\odot}$ and $R_*=1.0\times 10^{-4}\, R_{\odot}$, implying $R_S=2.93\times
10^{-19}\, R_{\odot}$,
$r_S=2.65\times 10^{-67}\, {\rm m}$ and $\chi_{\rm max}=2.9\times 10^{-15}$.
For ultralight axions with $m=1.93\times
10^{-20}\, {\rm eV}/c^2$ and
$a_s=-8.29\times 10^{-60}\, {\rm fm}$, we
have $M_{\rm max}=4.18\times 10^{5}\, M_{\odot}$ and $R_*=10.4\, {\rm
pc}$, implying $R_S=4.00\times
10^{-8}\, {\rm pc}$, $r_S=5.11\times 10^{-83}\, {\rm m}$ and $\chi_{\rm
max}=3.85\times
10^{-9}$. These two systems, which correspond respectively to axion stars and
axionic dark matter halos, can be treated by Newtonian gravity since $\chi_{\rm
max}\ll 1$.

When $M>M_{\rm max}$, the system collapses and ultimately forms a black hole.
Since $R(t)\rightarrow 0$ as $t\rightarrow t_{\rm coll}$, there exists a time
$t_{\rm GR}$ at which general relativity has to be taken into account. We
define it such that $R(t_{\rm GR})=R_S$ or, equivalently, $\chi(t_{\rm GR})=1$.
Using Eq. (\ref{vna2}), it corresponds to $[R(t_{\rm
GR})/R_*]/[M/M_{\rm max}]=\chi_{\rm max}$. When $t_{\rm GR}$ is
close to $t_{\rm coll}$, we can use the asymptotic formula of Eq. (\ref{bco3}).
This gives 
\begin{eqnarray}
\Delta t=t_{\rm coll}-t_{\rm GR}=\left (\frac{6}{25}\right )^{1/2}\chi_{\rm
max}^{5/2}\left (\frac{M}{M_{\rm max}}\right )^2 t_D.
\label{vna5}
\end{eqnarray}
For standard axions  with  $m=10^{-4}\, {\rm eV}/c^2$ and $a_s=-5.8\,
10^{-53}\, {\rm m}$, assuming $M=2M_{\rm
max}=1.4\times 10^{-13}\,
M_{\odot}$ and $R_0=R_*=1.0\times 10^{-4}\, R_{\odot}$, we find
$t_{D}=3.4\, {\rm hrs}$, $t_{\rm
coll}=2.3\, {\rm hrs}$, and $\Delta t=1.09\times 10^{-32}\, {\rm s}$.
For ultralight axions with
$m=1.93\times
10^{-20}\, {\rm eV}/c^2$ and
$a_s=-8.29\times 10^{-60}\, {\rm fm}$,
assuming $M=2M_{\rm max}=8.36\times 10^{5}\, M_{\odot}$ and 
$R_0=R_*=10.4\, {\rm pc}$, we find $t_D=1.49 \,{\rm Myrs}$, $t_{\rm
coll}=1.01  \,{\rm
Myrs}$ and  $\Delta t=8.47\times 10^{-8}\, {\rm s}$. In each case, general
relativistic effects become important only extraordinarily close to the
collapse time.

{\it Remark:} For consistency, we have used the result
of Eq. (\ref{bco3}) based on the Gaussian ansatz. However, we have
mentioned at the end of Sec \ref{sec_rti} that
this result may not be accurate. Using the exact result of \cite{sulem} does not
change our conclusion that  general relativistic effects become important only
extraordinarily close to the collapse time. This is intrinsically due to the
fact that $R\rightarrow 0$ as $t\rightarrow t_{\rm coll}$ with an infinite
slope.


\begin{thebibliography}{99}



\bibitem{baldeschi}{\small M.R. Baldeschi, G.B. Gelmini, R. Ruffini, Phys. Lett.
B {\bf  122}, 221 (1983)}
\bibitem{khlopov}{\small M.Yu. Khlopov, B.A. Malomed, Ya.B. Zeldovich, Mon. Not.
R. astr. Soc. {\bf  215}, 575 (1985) }
\bibitem{membrado}{\small M. Membrado, A.F. Pacheco, J. Sanudo, Phys. Rev. A
{\bf  39}, 4207 (1989)}
\bibitem{sin}{\small S.J. Sin, Phys. Rev. D {\bf  50}, 3650 (1994)}
\bibitem{jisin}{\small S.U. Ji, S.J. Sin, Phys. Rev. D {\bf  50}, 3655 (1994)}
\bibitem{leekoh}{\small J.W. Lee, I. Koh, Phys. Rev. D {\bf  53}, 2236 (1996)}
\bibitem{schunckpreprint}{\small F.E. Schunck, [astro-ph/9802258]}
\bibitem{matosguzman}{\small T. Matos,
F.S. Guzm\'an, F. Astron. Nachr. {\bf 320}, 97 (1999)}
\bibitem{guzmanmatos}{\small F.S. Guzm\'an,
T. Matos, Class. Quantum Grav.  {\bf 17}, L9 (2000)}
\bibitem{hu}{\small W. Hu, R. Barkana, A. Gruzinov, Phys. Rev. Lett. {\bf  85},
1158 (2000)}
\bibitem{peebles}{\small P.J.E. Peebles, Astrophys. J. {\bf 534}, L127 (2000)}
\bibitem{goodman}{\small J. Goodman, New Astronomy {\bf 5}, 103 (2000)}
\bibitem{mu}{\small T. Matos, L.A. Ure\~na-L\'opez,
Phys. Rev. D {\bf 63}, 063506 (2001)}
\bibitem{arbey1}{\small A. Arbey, J. Lesgourgues, P. Salati, Phys. Rev. D {\bf
64}, 123528 (2001)}
\bibitem{silverman1}{\small M.P. Silverman, R.L. Mallett, Class. Quantum Grav.
{\bf  18}, L103 (2001)}
\bibitem{matosall}{\small M. Alcubierre, F.S. Guzm\'an, T. Matos, D. N\'u\~nez,
L.A. Ure\~na-L\'opez, P.  Wiederhold, Class. Quantum. Grav. {\bf 19}, 5017
(2002)}
\bibitem{silverman}{\small M.P. Silverman, R.L. Mallett, Gen. Rel. Grav. {\bf
34}, 633 (2002)}
\bibitem{lesgourgues}{\small J. Lesgourgues,  A. Arbey, P. Salati, New Astron.
Rev. {\bf 46}, 791 (2002)}
\bibitem{arbey}{\small A. Arbey, J. Lesgourgues, P. Salati, Phys. Rev. D {\bf
68}, 023511 (2003)}
\bibitem{bohmer}{\small C.G. B\"ohmer, T. Harko, J. Cosmol. Astropart. Phys.
{\bf 06}, 025 (2007)}
\bibitem{bmn}{\small A. Bernal, T. Matos, D. N\'u\~nez, Rev. Mex. Astron.
Astrofis.  {\bf 44}, 149 (2008)}
\bibitem{sikivie}{\small P. Sikivie, Q. Yang, Phys. Rev. Lett. {\bf  103},
111301 (2009)}
\bibitem{mvm}{\small T. Matos, A. V\'azquez-Gonz\'alez,
J. Maga\~na, Mon. Not. R. Astron. Soc. {\bf 393}, 1359 (2009)}
\bibitem{lee09}{\small J.W. Lee, Phys. Lett. B {\bf 681}, 118 (2009)}
\bibitem{ch1}{\small T.P. Woo, T. Chiueh, Astrophys. J. {\bf 697}, 850 (2009)}
\bibitem{lee}{\small J.W. Lee, S. Lim, J. Cosmol. Astropart. Phys.  {\bf 01},
007 (2010)}
\bibitem{prd1}{\small P.H. Chavanis, Phys. Rev. D {\bf 84}, 043531 (2011)}
\bibitem{prd2}{\small P.H. Chavanis, L. Delfini, Phys. Rev. D {\bf 84}, 043532
(2011)}
\bibitem{prd3}{\small P.H. Chavanis, Phys. Rev. D {\bf 84}, 063518 (2011)}
\bibitem{briscese}{\small F. Briscese, Phys. Lett. B
{\bf 696}, 315 (2011)}
\bibitem{harkocosmo}{\small T. Harko, Mon. Not. R. Astron. Soc. {\bf 413}, 3095
(2011)}
\bibitem{harko}{\small T. Harko, J. Cosmol. Astropart. Phys. {\bf 05}, 022
(2011)}
\bibitem{abrilMNRAS}{\small A. Su\'arez, T. Matos, Mon. Not. R. Astron. Soc.
{\bf 416}, 87 (2011)}
\bibitem{aacosmo}{\small P.H. Chavanis, Astron. Astrophys. {\bf 537}, A127
(2012)}
\bibitem{velten}{\small H. Velten, E. Wamba, Phys. Lett. B {\bf 709}, 1
(2012)}
\bibitem{pires}{\small M.O.C. Pires, J.C.C. de Souza, J. Cosmol. Astropart.
Phys. {\bf 11} (2012) 024}
\bibitem{park}{\small C.-G. Park, J.-C. Hwang, H. Noh, Phys. Rev. D {\bf 86},
083535 (2012)}
\bibitem{rmbec}{\small V.H. Robles, T. Matos, Monthly Not. Roy. Astron. Soc.
{\bf 422},
282 (2012)}
\bibitem{rindler}{\small T. Rindler-Daller, P. R. Shapiro, Monthly Not. Roy.
Astron.
Soc. {\bf 422}, 135 (2012)}
\bibitem{lora}{\small V. Lora, J. Maga\~na, A. Bernal, F.J. S\'anchez-Salcedo,
E.K.
Grebel, J. Cosmol. Astropart. Phys.  {\bf  02}, 011 (2012)}
\bibitem{abrilJCAP}{\small J. Maga\~na, T. Matos, A. Su\'arez,
F. J. S\'anchez-Salcedo, JCAP {\bf 10}, 003 (2012)}
\bibitem{mhh}{\small G. Manfredi, P.A. Hervieux, F. Haas, Class. Quantum Grav.
{\bf 30}, 075006 (2013)}
\bibitem{lensing}{\small A.X. Gonz\'alez-Morales, A. Diez-Tejedor, L.A.
Ure\~na-L\'opez, O. Valenzuela, Phys. Rev. D {\bf 87}, 021301(R) (2013)}
\bibitem{glgr1}{\small F.S. Guzm\'an, F.D. Lora-Clavijo, J.J.
Gonz\'alez-Avil\'es, F.J. Rivera-Paleo, J. Cosmol. Astropart. Phys. {\bf 09}
(2013) 034}
\bibitem{ch2}{\small H.Y. Schive, T. Chiueh, T. Broadhurst, Nature Physics {\bf
10}, 496 (2014)}
\bibitem{ch3}{\small H.Y. Schive {\it et al.}, Phys. Rev. Lett. {\bf 113},
261302 (2014)}
\bibitem{shapiro}{\small B. Li, T. Rindler-Daller, P.R. Shapiro, Phys. Rev. D
{\bf 89}, 083536 (2014)}
\bibitem{bettoni}{\small D. Bettoni, M. Colombo, S. Liberati, JCAP
{\bf 02}, 004 (2014)}
\bibitem{lora2}{\small V. Lora, J. Maga\~na, JCAP
{\bf 09}, 011 (2014)}
\bibitem{mlbec}{\small P.H. Chavanis,  Eur. Phys. J. Plus {\bf 130}, 180 (2015)}
\bibitem{madarassy}{\small E.J.M. Madarassy, V.T. Toth,  Phys. Rev. D {\bf 91},
044041 (2015)}
\bibitem{abrilph}{\small A. Su\'arez, P.H. Chavanis,  Phys. Rev. D {\bf 92},
023510 (2015)}
\bibitem{playa}{\small A. Su\'arez, P.H. Chavanis, J. Phys.: Conf. Series {\bf
654}, 012088 (2015)}
\bibitem{stiff}{\small P.H. Chavanis,  Phys. Rev. D {\bf 92},
103004 (2015)}
\bibitem{guth}{\small A.H. Guth, M.P. Hertzberg, C. Prescod-Weinstein,  Phys.
Rev. D {\bf 92},
103513 (2015)}
\bibitem{souza}{\small J.C.C. de Souza, M. Ujevic, Gen. Relat. Grav. {\bf 47},
100 (2015)}
\bibitem{freitas}{\small R.C. de Freitas, H. Velten, Eur. Phys. J. C {\bf 75},
597 (2015)}
\bibitem{alexandre}{\small J. Alexandre,  Phys. Rev. D {\bf 92},
123524 (2015)}
\bibitem{schroven}{\small K. Schroven, M. List, C. L\"ammerzahl,  Phys. Rev. D
{\bf 92}, 124008 (2015)}
\bibitem{eby}{\small J. Eby, C. Kouvaris, N.G. Nielsen, L.C.R. Wijewardhana, 
JHEP {\bf 02}, 028 (2016)}
\bibitem{cembranos}{\small J.A.R. Cembranos, A.L. Maroto, S.J. N\'u\~nez
Jare\~no, JHEP {\bf 03}, 013 (2016)}
\bibitem{braaten}{\small E. Braaten, A. Mohapatra, H. Zhang, [arXiv:1502.00108]}
\bibitem{davidson}{\small S. Davidson, T. Schwetz, [arXiv:1603.04249]}
\bibitem{fan}{\small J. Fan, [arXiv:1603.06580]}
\bibitem{calabrese}{\small E. Calabrese, D.N. Spergel [arXiv:1603.07321]}
\bibitem{revueabril}{\small  A. Su\'arez, V.H. Robles, T. Matos, Astrophys.
Space Sci. Proc. {\bf 38}, 107 (2014)}
\bibitem{revueshapiro}{\small T. Rindler-Daller, P.R. Shapiro, Astrophys.
Space Sci. Proc. {\bf 38}, 163 (2014)}
\bibitem{bookspringer}{\small P.H. Chavanis, {\it
Self-gravitating
Bose-Einstein condensates}, in Quantum Aspects of Black Holes, edited by X.
Calmet (Springer, 2015)}
\bibitem{madelung}{\small E. Madelung, Zeit. F. Phys. {\bf 40}, 322 (1927)}
\bibitem{kaup}{\small D.J. Kaup, Phys. Rev. {\bf  172}, 1331 (1968)}
\bibitem{rb}{\small R. Ruffini, S. Bonazzola, Phys. Rev. {\bf  187}, 1767
(1969)}
\bibitem{colpi}{\small M. Colpi, S.L. Shapiro, I. Wasserman, Phys. Rev. Lett.
{\bf  57}, 2485 (1986)}
\bibitem{chavharko}{\small P.H. Chavanis, T. Harko, Phys. Rev. D {\bf 86},
064011 (2012)}
\bibitem{Lat}{\small J. M. Lattimer and M. Prakash, in From
Nuclei to Stars, Ed: S. Lee (Singapore, World
Scientific, 2011), [arXiv:1012.3208]}
\bibitem{Dem}{\small P. B. Demorest, T. Pennucci, S. M. Ransom, M. S. E. Roberts
and J. W. T. Hessels,
Nature {\bf 467}, 1081 (2010)}
\bibitem{black1}{\small O. Barziv, L. Karper, M. H. van Kerkwijk, J. H. Telging,
and J. van Paradijs, Astron. Astrophys.{\bf  377}, 925 (2001)}
\bibitem{black2}{\small  H. Quaintrell, A. J. Norton, T. D. C. Ash, P. Roche, B.
Willems, T. R. Bedding, I. K. Baldry, and R. P. Fender, Astron. Astrophys. {\bf
401}, 303 (2003)}
\bibitem{black3}{\small  M. H. van Kerkwijk, R. Breton, and S. R. Kulkarni,
Astrophys. J. {\bf 728}, 95 (2011)}
\bibitem{antoniadis}{\small J. Antoniadis {\it et al.}, Science {\bf 340},
6131 (2013)}
\bibitem{ov}{\small J.R. Oppenheimer, G.M. Volkoff, Phys. Rev. {\bf 55}, 374
(1939)}
\bibitem{kc}{\small J.E. Kim, G. Carosi, Rev. Mod. Phys. {\bf 82}, 557
(2010)}
\bibitem{marsh}{\small D. Marsh, [arXiv:1510.07633]}
\bibitem{pq}{\small R.D. Peccei, H.R. Quinn, Phys. Rev. Lett. {\bf 38}, 1440
(1977)}
\bibitem{axiverse}{\small A. Arvanitaki, S. Dimopoulos, S. Dubovsky, N. Kaloper,
J.  March-Russell,  Phys. Rev. D {\bf 81}, 123530 (2010)}
\bibitem{sikivie1}{\small P. Sikivie, Q. Yang, Phys. Rev. Lett. {\bf 103},
111301 (2009)}
\bibitem{sikivie2}{\small O. Erken, P. Sikivie, H. Tam, Q. Yang,
 Phys. Rev. D {\bf 85}, 063520 (2012)}
\bibitem{kingfermionic}{\small P.H. Chavanis, M. Lemou, F. M\'ehats, Phys.
Rev. D {\bf 92}, 123527 (2015)}
\bibitem{nfw}{\small J.F. Navarro, C.S. Frenk, S.D.M. White, Astrophys. J.
{\bf 462}, 563 (1996)}
\bibitem{seidel94}{\small E. Seidel, W.M. Suen, Phys. Rev. Lett.
{\bf  72}, 2516 (1994)}
\bibitem{prep}{\small P.H. Chavanis, in preparation}
\bibitem{gross}{\small E.P. Gross, Ann. of Phys. {\bf 4}, 57 (1958); Nuovo
Cimento {\bf 20}, 454 (1961); J. Math. Phys. {\bf 4}, 195 (1963)}
\bibitem{pitaevskii}{\small L.P. Pitaevskii, Sov. Phys. JETP {\bf 9}, 830
(1959); ibid {\bf 13}, 451 (1961)}
\bibitem{bogoliubov}{\small N. Bogoliubov, J. Phys. {\bf 11}, 23
(1947)}
\bibitem{revuebec}{\small F. Dalfovo, S. Giorgini, L.P. Pitaevskii, S.
Stringari, Rev. Mod. Phys. {\bf 71}, 463 (1999)}
\bibitem{chandra}{\small S. Chandrasekhar, An Introduction to the Study of
Stellar Structure (Dover, 1958)}
\bibitem{holm}{\small D.D. Holm, J.E. Marsden, T. Ratiu, A. Weinstein, Phys.
Rep. {\bf 123}, 1 (1985)}
\bibitem{weinbergbook}{\small S. Weinberg, Gravitation and Cosmology (John
Wiley, 2002)}
\bibitem{cosmopoly1}{\small P.H. Chavanis, Eur. Phys. J. Plus {\bf 129}, 38
(2014)}
\bibitem{poincare}  {\small H. Poincar\'e, Acta Math. {\bf 7}, 259 (1885)}
\bibitem{katzpoincare}  {\small J. Katz,  Mon. Not. R. Astron. Soc. {\bf 183},
765 (1978)}
\bibitem{ijmpb}{\small P.H. Chavanis, Int. J. Mod. Phys. B {\bf 20}, 3113
(2006)}
\bibitem{sulem}{\small C. Sulem, P.L. Sulem, The Nonlinear Schr\"odinger
Equation (Springer, 1999)}
\bibitem{mestel}{\small L. Mestel, Q. Jl R. Astr. Soc. {\bf 6}, 161 (1965)}
\bibitem{pomeau}{\small Y. Pomeau, M. Le Berre, P.H. Chavanis, B. Denet, Eur.
Phys. J. E {\bf 37}, 26 (2014)}
\bibitem{donley}{\small E.A. Donley, N.R. Claussen, S.L. Cornish, J.L. Roberts,
E.A. Cornell, C.E. Wieman, Nature {\bf 412}, 295
(2001)}













\end{thebibliography}
\end{document}